\pdfoutput=1
\documentclass[aip,reprint,twoside]{revtex4-1} 

\def\ispreprint{1}

\usepackage{graphicx}
\usepackage{dcolumn}
\usepackage{placeins}
\usepackage[mathlines]{lineno}

\usepackage[version=3]{mhchem} 
\usepackage{amssymb}
\usepackage{multirow}
\usepackage{graphicx}
\usepackage{amsmath}
\usepackage{threeparttable}
\usepackage{float}
\usepackage{hyperref}
\usepackage{caption}
\DeclareCaptionFormat{withunderline}{#1#2#3\par\vspace{2pt}\hrule\vspace{4pt}}
\usepackage{ragged2e}
\usepackage{listings}
\usepackage[table]{xcolor}
\usepackage{todonotes}
\usepackage{makecell}

\usepackage[utf8]{inputenc} 
\usepackage[T1]{fontenc}    

\usepackage{diagbox}
\usepackage{booktabs}
\renewcommand{\thetable}{\arabic{table}}
\usepackage[a4paper,margin=2cm]{geometry} 
\usepackage{colortbl}

\definecolor{fessa1}{RGB}{ 31, 59,  115}
\definecolor{fessa2}{RGB}{ 47, 146, 148}
\definecolor{fessa3}{RGB}{ 80, 178, 141}
\definecolor{fessa4}{RGB}{167, 214,  85}
\definecolor{fessa5}{RGB}{255, 224,  62}
\definecolor{fessa6}{RGB}{255, 169,  85}
\definecolor{fessa7}{RGB}{214,  87,  59}

\newcommand{\software}[1]{{\small\texttt{\textbf{\color{teal}#1}}}\normalfont}
\newcommand{\method}[1]{{\small\textsf{\textbf{\color{purple}#1}}}\normalfont}
\newcommand{\paragraphtitle}[1]{\textsf{\textbf{\small {#1}}}}

\usepackage{titlesec}
\renewcommand{\thesection}{\arabic{section}}
\renewcommand{\thesubsection}{\arabic{subsection}}
\renewcommand{\thesubsubsection}{\arabic{subsubsection}}

\titleformat{\section}
  {\sffamily\bfseries} 
  {\arabic{section}}                            
  {1em}                                   
  {}
  
\titleformat{\subsection}
  {\sffamily\bfseries\small} 
  {\arabic{section}.\arabic{subsection}}                            
  {1em}                                   
  {}
  
\titleformat{\subsubsection}
  {\sffamily\bfseries\itshape\small} 
  {\arabic{section}.\arabic{subsection}.\arabic{subsubsection}}                            
  {1em}                                   
  {}                                      

\makeatletter
\renewcommand\@seccntformat[1]{\csname the#1\endcsname\quad}

\renewcommand{\@currentlabel}{\csname the\@currenvir\endcsname}
\makeatother



\usepackage{fancyhdr}
\fancypagestyle{plain}{%
    \fancyhead{} 
    \fancyfoot{} 
    \fancyfoot[RO,LE]{\footnotesize \thepage}
}

\begin{document}

\title{Enhanced Sampling in the Age of Machine Learning: \\Algorithms and Applications}

\author{Kai Zhu}
\thanks{Contributed equally to this work}
\affiliation{\small College of Pharmaceutical Sciences, Zhejiang University, Hangzhou, 310058, Zhejiang, China}
\author{Enrico Trizio}
\thanks{Contributed equally to this work}
\affiliation{\small Atomistic Simulations, Italian Institute of Technology, Genova 16152, Italy}
\author{Jintu Zhang}
\affiliation{\small College of Pharmaceutical Sciences, Zhejiang University, Hangzhou, 310058, Zhejiang, China}
\author{Renling Hu}
\affiliation{\small College of Pharmaceutical Sciences, Zhejiang University, Hangzhou, 310058, Zhejiang, China}
\author{Linlong Jiang}
\affiliation{\small College of Pharmaceutical Sciences, Zhejiang University, Hangzhou, 310058, Zhejiang, China}
\author{Tingjun Hou}
\email{tingjunhou@zju.edu.cn}
\affiliation{\small College of Pharmaceutical Sciences, Zhejiang University, Hangzhou, 310058, Zhejiang, China}
\affiliation{\small Zhejiang Provincial Key Laboratory for Intelligent Drug Discovery and Development, Jinhua 321016, China}
\author{Luigi Bonati}
\email{luigi.bonati@iit.it}
\affiliation{\small Atomistic Simulations, Italian Institute of Technology, Genova 16152, Italy}

\begin{abstract}
Molecular dynamics simulations hold great promise for providing insight into the microscopic behavior of complex molecular systems. However, their effectiveness is often constrained by long timescales associated with rare events. Enhanced sampling methods have been developed to address these challenges, and recent years have seen a growing integration with machine learning techniques.
This review provides a comprehensive overview of how they are reshaping the field, with a particular focus on the data-driven construction of collective variables. Furthermore, these techniques have also improved biasing schemes and unlocked novel strategies via reinforcement learning and generative approaches.
In addition to methodological advances, we highlight applications spanning different areas such as biomolecular processes, ligand binding, catalytic reactions, and phase transitions. We conclude by outlining future directions aimed at enabling more automated strategies for rare-event sampling.

\end{abstract}

\maketitle


\tableofcontents

\clearpage
\section{Introduction}
Molecular dynamics (MD) simulations have become an indispensable tool for understanding physical, chemical, and biological processes at the molecular scale~\cite{frenkel2023understanding}. Their value is that they can be thought of as a computational microscope, allowing us to zoom in on the molecular motions that underpin these processes. By integrating Newton’s equations of motion, MD generates trajectories that reveal the dynamic evolution of atomic configurations, providing a detailed and time-resolved view of complex systems and enabling direct calculation of thermodynamic and kinetic properties. Over the past decades, advances in algorithms and computational power have extended the reach of this computational microscope. Yet, significant challenges remain.

Two of the most pressing challenges in atomistic simulations are (i) constructing accurate yet efficient models for describing atomic interactions and (ii) overcoming the so-called rare events problem. The \textit{accuracy} of a simulation is fundamentally determined by the quality of the underlying potential energy surface (PES). Ab initio methods, such as Car–Parrinello~\cite{car1985unified} and Born–Oppenheimer MD~\cite{marx2009ab}, employ highly accurate descriptions of the PES derived from quantum mechanics but are computationally expensive, restricting simulations to small systems and short timescales. At the other end of the spectrum, (semi)empirical force fields~\cite{ponder2003force} enable simulations of larger systems but often lack the fidelity required to capture complex chemical processes and reactive events. Bridging this gap, machine-learning potentials~\cite{unke2021machine} have emerged over the past decade as a transformative solution, offering near-ab initio accuracy at a fraction of the cost and accelerating first-principles simulations by several orders of magnitude.

\begin{figure*}[ht!]
  \centering
  \includegraphics[width=0.95\linewidth]{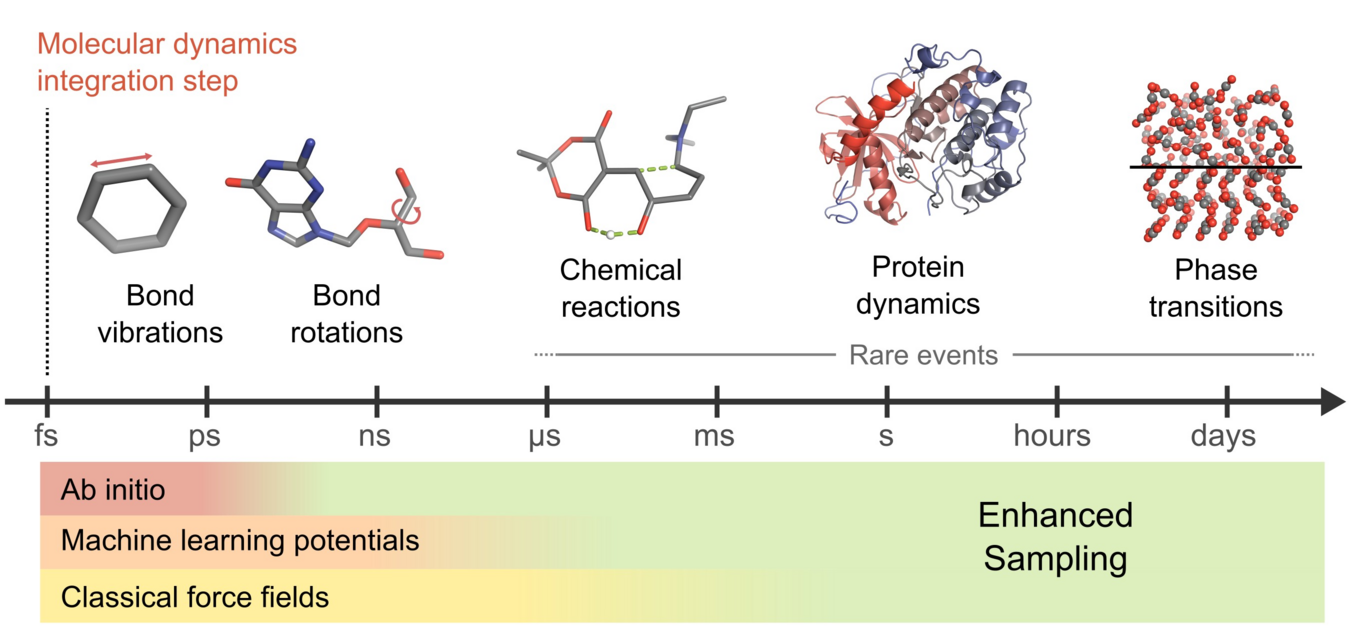}
  \caption{\justifying Timescales in MD simulations. MD simulations capture atomic motions across a broad range of timescales. Rare events such as chemical reactions, large-scale conformational changes in proteins, and phase transitions occur on timescales from microseconds (\text{\textmu s}) to days, far beyond the reach of standard MD. To efficiently access these rare events, enhanced sampling techniques are indispensable. The bottom panel illustrates the approximate timescales accessible by different classes of potential energy models.}
  \label{timescale}
  \end{figure*}
  
The second major challenge lies in the \textit{timescales} accessible by MD. In principle, atomistic simulations hold the potential to reveal how a protein folds into its native state, how a drug binds to its target, or how a material undergoes a phase transition. However, these processes often unfold on timescales, from milliseconds to seconds or even hours, that far exceed the reach of conventional MD, even with powerful supercomputers~\cite{shaw2021anton} (see Fig.~\ref{timescale}). This limitation arises from the intrinsic serial nature of molecular dynamics and the necessity of using an integration timestep smaller than the fastest molecular motions, typically on the femtoseconds scale~\cite{frenkel2023understanding}. As a result, many processes of chemical and biological relevance remain inaccessible without additional methodological advances.

To overcome this barrier, diverse enhanced sampling methods have been developed~\cite{henin2022Enhanced}. These approaches accelerate the exploration of the configurational space by various means, such as by biasing the dynamics along selected \textit{collective variables (CVs)}~\cite{valsson2016enhancing} or increasing the likelihood of rare events~\cite{bolhuis2002transition}, thereby enabling efficient sampling of transitions that would otherwise remain elusive. Nevertheless, the high dimensionality of the configurational space and the large number of degrees of freedom involved make this task still quite challenging. This complexity naturally calls for data-driven approaches that can integrate physical intuition with powerful statistical tools to efficiently explore and understand the relevant regions of phase space.

In recent years, \textit{machine learning (ML)} has emerged as a transformative technology for many fields, and atomistic simulations are no exception, see for instance the review by Noe \textit{et al.}~\cite{noe2020machine} and the Chemical Reviews special issue \textit{"Machine Learning at the Atomic Scale"}~\cite{ceriotti2021introduction}. ML has significantly impacted several aspects of atomistic modeling. These tools are indeed particularly useful for learning structural representations~\cite{musil2021physics} and uncovering meaningful patterns from a large amount of data~\cite{glielmo2021unsupervised}. Beyond constructing accurate PESs~\cite{behler2021four,unke2021machine}, ML has enabled large-scale computational discovery~\cite{nandy2021computational} and exploration of chemical compound space~\cite{huang2021ab}.

The field of enhanced sampling has likewise been profoundly influenced by ML~\cite{sidky2020machine,wang2020machine,chen2021collective,mehdi2024enhanced}, from the data-driven identification of CVs to the development of novel biasing schemes and advanced post-processing tools. On one hand, this review aims to provide a comprehensive methodological overview of the integration of ML and enhanced sampling techniques. On the other hand, we also seek to offer a perspective to readers more interested in applying this computational microscope to their own problems of interest. To this end, we will present applications across diverse areas, highlighting the requirements and challenges involved in deploying such models in practice. Relevant areas include the study of biological conformational changes, such as protein folding and the thermodynamics and kinetics of ligand binding. Other important fields of application are chemical and catalytic reactions, as well as structural phase transformations. In all these domains, the integration of ML and enhanced sampling has provided crucial insights into atomistic mechanisms, effectively focusing the lens of our computational microscope on rare events. The period covered by this review is approximately from 2018 to 2025.

The structure of the manuscript is as follows: Sec.~\ref{sec:fund} provides a brief overview of the fundamentals of enhanced sampling and a glossary of ML. Sec.~\ref{sec:mlcvs_methods} focuses on the construction of machine learning collective variables, while Sec.~\ref{sec:mlcvs_applications} illustrates relevant applications to different areas of molecular simulations. In Sec.~\ref{sec:ml_bias}, we discuss how ML methodologies have been integrated to improve the construction of biasing schemes. Sec.~\ref{sec:generative} highlights the emerging role of generative models in improving sampling efficiency. Finally, we offer our perspectives on current challenges and future research directions in the field of enhanced sampling and its integration with ML.

\section{Fundamentals of ML-based enhanced sampling}
\label{sec:fund}

In this section, we present some fundamental elements of the fields of enhanced sampling simulations and ML.
Such information is reported in a very concise way, more intended to refresh some key concepts that will be recurrent throughout the review rather than formally discussing them at length, as more detailed information is already provided in some recent reviews.
In particular, for a comprehensive review of enhanced sampling in atomistic simulations, we refer the reader to Refs.\citenum{valsson2016enhancing, yang2019enhanced, henin2022Enhanced}, whereas for an overview of ML methods and their application to science, we refer to Refs.\citenum{bishop2025deep, wang2023scientific, mahesh2020machine,noe2020machine,carleo2019machine}.

\subsection{Atomistic simulations}

Atomistic simulations, such as MD, allow to study physical, chemical, and biological systems at the atomic scale, offering microscopic insight into their behavior and enabling the computation of physical and chemical properties\cite{frenkel2023understanding}.
Central to these simulations is the PES \( U(\mathbf{R}) \), which governs the interactions between atoms as a function of the atomic coordinates \( \mathbf{R} \). This quantity can be described using a variety of models, ranging from quantum-mechanical \textit{ab initio} methods and empirical force fields to ML potentials or coarse-grained models.

Given a model for the PES, the equilibrium properties of a system in the canonical ensemble (constant number of particles N, volume V, and temperature T) are described within the framework of statistical mechanics by the Boltzmann distribution:
\begin{equation}
    p(\mathbf{R}) = \frac{1}{Z} e^{-\beta U(\mathbf{R})}
\end{equation}
where \( \beta = 1/(k_B T) \) is the inverse temperature, \( k_B \) is the Boltzmann constant, and the partition function \( Z = \int d\mathbf{R} \, e^{-\beta U(\mathbf{R})} \) ensures normalization. Sampling this distribution is central to atomistic simulations, as it enables the computation of equilibrium properties as ensemble averages:
\begin{equation}
    \langle O(\mathbf{R}) \rangle = \int d\mathbf{R} \,O(\mathbf{R}) p(\mathbf{R})
\end{equation}
This can be accomplished through computational approaches such as Monte Carlo or MD simulations. In this review, we mostly focus on the latter, which not only enable sampling from equilibrium distributions but also provide access to time-dependent dynamical information by integrating Newton’s equations of motion.

\begin{figure}[htbp]
  \centering
  \includegraphics[width=\linewidth]{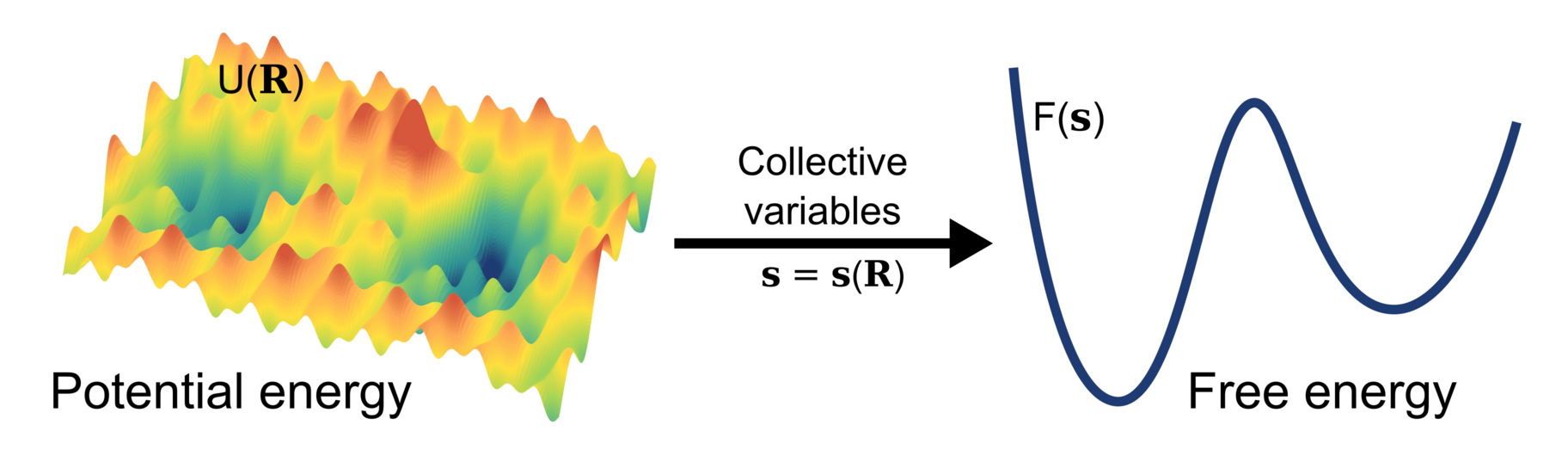}
  \caption{\justifying The transformation from the high-dimensional and possibly rugged PES to the low-dimensional and smooth FES projected along some CVs.}
  \label{fig:pes_vs_fes}
\end{figure}

However, sampling the Boltzmann distribution for complex systems is highly challenging. For a system of \( N \) atoms, the configuration space has \( 3N-1 \) degrees of freedom, making a direct exploration of \( p(\mathbf{R}) \) intractable.
To mitigate this complexity, it is common to reduce the dimensionality of the problem by introducing a set of CVs, \( \mathbf{s} = \mathbf{s}(\mathbf{R}) \), which are functions of the atomic coordinates. These CVs are often designed to capture the slow and thermodynamically relevant modes of the system, conceptually similar to reaction coordinates in chemistry or order parameters in statistical physics. The equilibrium distribution along the CVs is obtained by marginalizing the full distribution:
\begin{equation}
    p(s) = \int d\mathbf{R} \, \delta[s - s(\mathbf{R})] p(\mathbf{R}) = \langle \delta[s - s(\mathbf{R})] \rangle
\end{equation}
which in turn defines the \textit{free energy surface} (FES):
\begin{equation}
    F(\mathbf{s}) = -\frac{1}{\beta} \log p(\mathbf{s})
\end{equation}
The FES provides a low-dimensional - and typically smoother - thermodynamic landscape of the system, which also accounts for entropic contributions, with metastable states corresponding to local minima and reaction pathways to transitions between them (Fig.~\ref{fig:pes_vs_fes}). 

The other element of complexity stems from the fact that transitions between metastable states typically involve crossing large free energy barriers and are rarely observed in standard MD simulations, leading to inefficient sampling. This issue is particularly pronounced in systems where the relevant transitions are \textit{rare events}, occurring on timescales many orders of magnitude longer than those accessible by conventional simulations. A prototypical example is the folding of a protein from an extended to a native conformation, which, despite being thermodynamically favorable, often occurs over milliseconds or longer timescales, far beyond those typically accessible with standard MD.

\subsection{Enhanced sampling}
\begin{figure*}[htbp]
  \centering
  \includegraphics[width=0.85\linewidth]{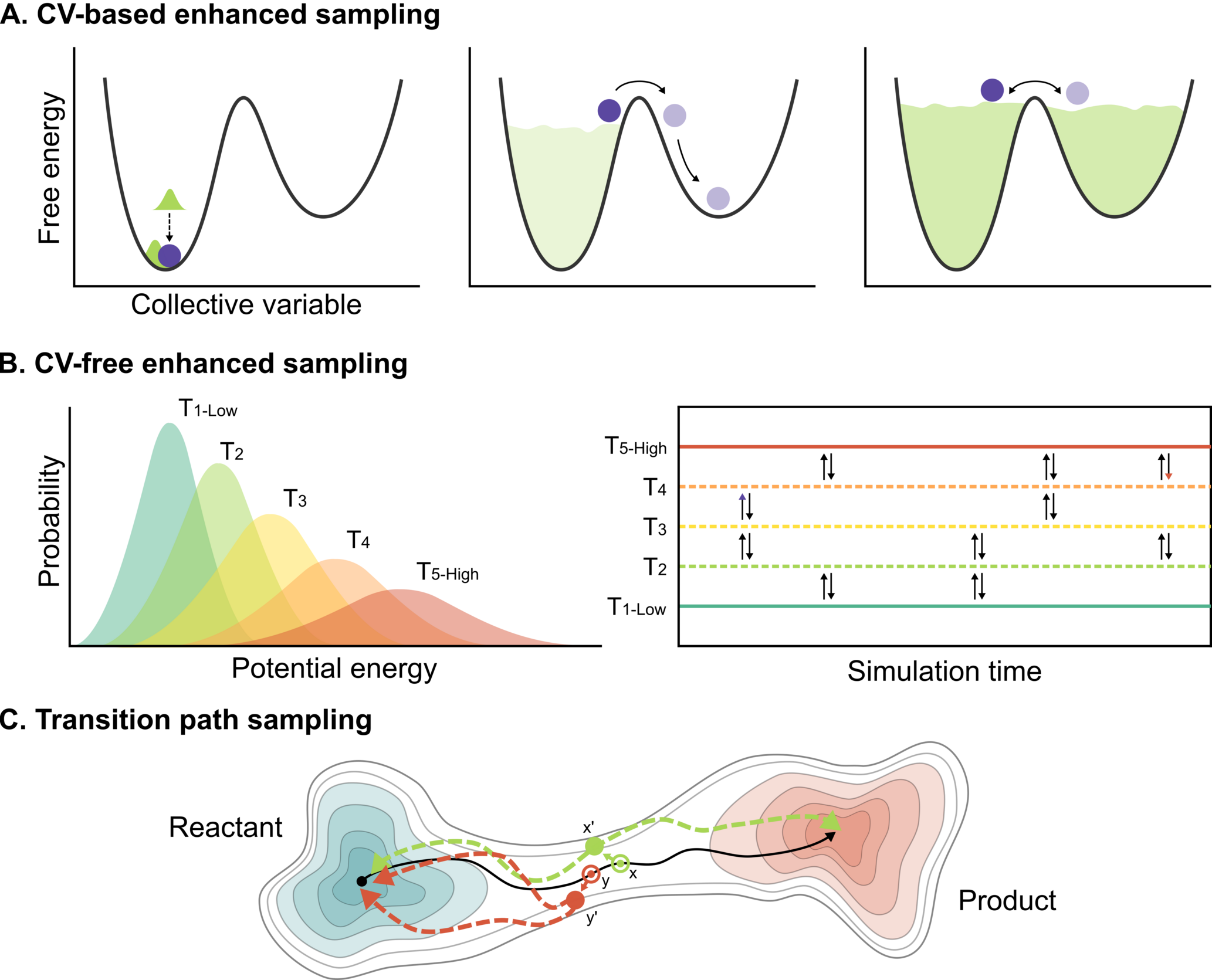}
  \caption{\justifying
  (A) Schematic representation of CV-based enhanced sampling methods, exemplified by metadynamics. Initially, the system is confined in a local free energy minimum. As the bias accumulates, it reduces energy barriers and promotes transitions between metastable states. Eventually, the FES is flattened, allowing uniform exploration. (B) An example of CV-free enhanced sampling with REMD. Multiple simulations are run in parallel with different parameters (e.g., temperatures) and exchanges between them are attempted according to the Metropolis criterion. (C) Schematic representation of TPS, exemplified by the shooting method. Starting from an initial reactive trajectory (solid black line), a configuration is randomly selected and slightly perturbed to create new initial conditions (e.g., $x'$, green, or $y'$, red). Two MD simulations are launched forward and backward in time. Trajectories connecting distinct stable states (dashed green line) are accepted, while those returning to the same basin (dashed red line) are discarded. }
  \label{fig:CV-ES}
\end{figure*}

To address the challenge of rare events, several \textit{enhanced sampling} methods have been developed. These techniques aim to accelerate the exploration of configuration space, enabling efficient sampling of rare transitions. Below, we limit ourselves to outlining the three main families, examples of which are depicted in Fig.~\ref{fig:CV-ES}, and their characteristics to better understand how ML techniques have been integrated with them. For a high-level overview of the different approaches, see, for instance, the review by Pietrucci \cite{pietrucci2017strategies}, while for a more detailed discussion, see the review by Henin \textit{et al.}~\cite{henin2022Enhanced}.

\paragraphtitle{CV-based enhanced sampling}. In the first family of methods, a bias potential \( V(\mathbf{s}) \) is introduced to modify the effective PES experienced by the system in the space of a few selected CVs. The goal of this bias is to facilitate the exploration of rarely visited regions, which are typically separated by high free energy barriers, while preserving the ability to reconstruct the unbiased thermodynamics through reweighting. For an introduction, see, for instance, the review by Valsson \textit{et al.}~\cite{valsson2016enhancing}.

One of the earliest strategies to enhance the sampling along a CV is \method{umbrella sampling}\cite{torrie1977nonphysical}. In this method, the system is simulated under a set of fixed external (harmonic) bias potentials centered at different CV values. These simulations, referred to as windows or umbrellas, collectively span the relevant region of the CV space. The data from different windows are then combined using the weighted histogram analysis method (WHAM) \cite{kumar1992weighted, souaille2001extension} or umbrella integration \cite{kastner2005bridging} to reconstruct the global FES.
While effective, standard umbrella sampling requires \textit{a priori} selection of bias centers and force constants, often involving trial-and-error. To address this limitation, adaptive umbrella sampling \cite{mezei1987adaptive} updates the bias iteratively based on the sampled distribution. Related approaches, such as self-healing umbrella sampling \cite{marsili2006self} and local elevation \cite{dickson2010free} also dynamically modify the bias to improve exploration of poorly sampled regions.

Ideally, if the exact FES \( F(\mathbf{s}) \) was known, one could apply a bias potential equal to its negative, \( V(\mathbf{s}) = -F(\mathbf{s}) \), which would \textit{flatten the free energy profile} and lead to uniform sampling in the CV space. While this is not feasible in practice, many enhanced sampling methods are based on approximating or iteratively constructing such a bias during the course of the simulation.
The most prominent among these is \method{metadynamics} \cite{laio2002escaping}, schematically depicted in Fig.~\ref{fig:CV-ES}A, in which repulsive Gaussians are periodically deposited in the CV space, progressively \textit{filling} and flattening the free energy landscape. Variants such as well-tempered metadynamics\cite{barducci2008well} introduce a tempering factor to ensure convergence. The free energy profile can then be reconstructed from the asymptotic profile of the bias, or via time-dependent reweighting schemes \cite{tiwary2015time}. 

Another class of approaches directly estimates the \textit{gradient of the free energy} surface from simulations. \method{Adaptive biasing force (ABF)} \cite{darve2001calculating} computes the average force acting along a CV and uses it to counteract the underlying free energy gradient. This approach avoids constructing the free energy explicitly, though multidimensional generalizations require numerical integration of the sampled gradient field \cite{alrachid2015long}.

Finally, other methods focus on the \textit{target distribution }\( p_{\mathrm{tg}}(\mathbf{s}) \) to be sampled, and then construct a bias potential that drives the system toward this distribution. Examples are the \method{variationally enhanced sampling (VES)} \cite{valsson2014variational} and the recent \method{on-the-fly probability enhanced sampling (OPES)} \cite{invernizzi2020rethinking}. In the latter, the bias is defined as
$  V(\mathbf{s}) = \frac{1}{\beta} \log\left(p(\mathbf{s})/p_{\mathrm{tg}}(\mathbf{s})\right),
$
where \( p(\mathbf{s}) \) is the equilibrium distribution estimated during the simulation via an on-the-fly reweighting of the trajectory data. The flexibility in choosing \( p_{\mathrm{tg}}(\mathbf{s}) \) makes this approach highly versatile: with suitable choices, it can recover the same sampling distribution as well-tempered metadynamics, adaptive umbrella sampling, or generalized ensembles such as multithermal or multibaric ensembles~\cite{invernizzi2020unified}. Moreover, the reweighting procedure is greatly facilitated by the rapid convergence of the bias toward a quasi-static regime. For a practical overview on the different OPES variants, we refer the reader to the review by Trizio \textit{et al.}~\cite{trizio2024advanced}.

\paragraphtitle{CV-free enhanced sampling}. 
Instead of focusing on the identification of appropriate variables, other methods aim to enhance the exploration of configuration space more generally, often by altering the thermodynamic ensemble or employing multiple replicas.
A prominent class of such methods is based on generalized ensembles, where the system is allowed to sample from a more general probability distribution, such as the one obtained by combining multiple overlapping probability distributions. These are typically constructed to bridge an easily sampled distribution (such as one at high temperature), with the target distribution of interest (low temperature). Enhanced sampling is then achieved by allowing coordinated exchanges between replicas simulated under different conditions, which promotes transitions across energy barriers that would otherwise be rarely crossed in standard simulations.
Examples of these methods include parallel tempering or \method{replica exchange (REX)} \cite{swendsen1986replica,hansmann1997parallel,sugita2000multidimensional}, as well as solute tempering approaches \cite{liu2005replica,wang2011replica}, where only part of the system (e.g., the solute) is tempered, allowing more focused acceleration of relevant degrees of freedom.

Another class of CV-free methods enhances sampling by adding a boost potential that effectively smooths the PES. Notable examples include \method{accelerated molecular dynamics (aMD)} \cite{hamelberg2004accelerated} and \method{Gaussian accelerated molecular dynamics (GaMD)} \cite{miao2015gaussian}. In GaMD, a boost potential with a near-Gaussian distribution is applied whenever the system’s potential energy falls below a predefined threshold. At the end of the simulation, a cumulant expansion is used to reconstruct unbiased thermodynamic averages, a procedure referred to as ``Gaussian approximation''.

\paragraphtitle{Path sampling.} A third category of methods  
 enhances the exploration of rare events by performing a Monte Carlo simulation in path space rather than configuration space, such as \method{transition path sampling (TPS)}\cite{dellago1998efficient}. For more details, see also the recent perspective by Bolhouis and Swenson\cite{bolhuis2021transition}. At variance with previous methods, which modify the PES to accelerate the sampling of rare events, TPS focuses on generating an ensemble of unbiased reactive trajectories, which is known as the transition path ensemble. In fact, this can provide insight into the unbiased mechanisms underlying rare events.
To achieve this goal, one is required to define Monte Carlo moves to create a new pathway from a previous one. A typical TPS move, called \textit{shooting}, perturbs a configuration along a reactive trajectory and integrates the dynamics forward and backward in time to generate a new path \cite{dellago1998efficient}. The new path is then accepted or rejected based on whether it connects distinct metastable states.
Further extensions also allow for computing kinetic rates\cite{van2005elaborating,allen2006simulating} as well as to reconstruct the free energy profiles~\cite{lazzeri2023molecular}.

\paragraphtitle{Enhanced sampling software}. Here we highlight the main software packages for performing enhanced sampling simulations, with growing support for integration with ML libraries such as \texttt{PyTorch} and \texttt{TensorFlow}.

\software{PLUMED} \cite{bonomi2009plumed,tribello2014plumed} is a widely used open-source plugin for enhanced sampling and free energy calculations that can be interfaced with most classical and \textit{ab initio} MD engines, including \texttt{AMBER}, \texttt{GROMACS}, \texttt{LAMMPS}, \texttt{NAMD}, \texttt{OpenMM}, \texttt{CP2K}, \texttt{Quantum Espresso}. It supports a broad range of methods, including metadynamics, VES, and OPES, and provides an extensive library of CVs. In addition to enhanced sampling, PLUMED offers standalone tools for post-processing and trajectory analysis. It is a community-driven project that promotes reproducibility through \texttt{PLUMED-NEST} \cite{plumed2019promoting}, a repository for input files, and supports learning via the user-contributed \texttt{PLUMED Tutorials} \cite{tribello2025plumed}.
Conveniently, it also provides a native interface for PyTorch-based MLCVs through the additional \texttt{pytorch} module\cite{bonati2021deep, bonati2023unified}.

\software{Colvars} \cite{fiorin2013using} is directly integrated into several widely used classical MD engines, including \texttt{NAMD}, \texttt{LAMMPS}, and \texttt{GROMACS}. It allows users to define a wide range of CVs and apply enhanced sampling methods such as adaptive biasing force, metadynamics, and umbrella sampling.

\software{SSAGES} \cite{sidky2018ssages} is a modular and extensible framework for enhanced sampling simulations. It interfaces with classical MD engines like \texttt{LAMMPS}, \texttt{GROMACS}, and \texttt{OpenMD} and supports both CV-based methods and path-based techniques such as the string method and forward flux sampling. 

Finally, to perform transition path sampling simulations, Python libraries such as \software{OpenPathSampling} \cite{swenson2018openpathsampling,swenson2018openpathsampling2} and \software{PyRETIS} \cite{lervik2017pyretis,riccardi2020pyretis,vervust2024pyretis} provide tools to construct and analyze ensembles of reactive trajectories.

\subsection{Glossary of machine learning}
\label{sec:ml}
ML is a broad field encompassing computational and statistical techniques designed to automatically extract patterns and learn from data, which has become ubiquitous in recent years.
In this section, we provide a brief and essential overview, contextualized to the field of atomistic simulations, of some of the key concepts that will be recurrent in the rest of the review: learning approaches, data types, architectures, and loss functions.
For Readers seeking a more comprehensive introduction to ML,  we refer them to specialized literature, for example, the recent book by Bishop and Bishop\cite{bishop2025deep} or the introduction by Mehta \textit{et al.} \cite{mehta2019high}. 

\paragraphtitle{Types of data.}
At the core of any ML approach lies the data, which can be used for training (i.e., optimizing the model on available information) or for inference (i.e., making predictions with a trained model on new inputs).
Broadly speaking, datasets can be categorized based on the amount and type of information provided, which in turn determines the appropriate learning strategy (see below).
In the most general case, the dataset consists of a collection of raw samples, such as images, atomic configurations, or scalar properties, without additional annotations (\textit{unlabeled} datasets). 
In contrast, \textit{labeled} datasets associate each sample with one or more \textit{labels} that encode target properties the model is expected to learn. For example, in a set of animal pictures, each one could be labeled with the corresponding species, or, in the case of an atomic system, a given configuration can be labeled with the corresponding energy value.
A relevant subclass of labeled data is that of \textit{time series} or sequences, where each sample is accompanied by a timestamp or ordering index. This temporal structure enables the learning of sequential or dynamic relationships. Examples include sequences of atomic configurations collected during a simulation or word tokens in a sentence.

\paragraphtitle{Learning approaches.}
ML models can be trained using different learning paradigms, each suited to specific types of data and tasks. These paradigms also dictate the form of the loss function used during optimization.
In \textit{supervised learning}, the model learns from labeled data by minimizing a loss that quantifies the mismatch between predictions and known labels. This setting is typical for tasks such as classification (e.g., image recognition) or regression (e.g., predicting the energy of molecular structures).
In \textit{unsupervised learning}, the model is trained without labeled inputs and instead seeks to discover hidden structures in the data, such as clusters, manifolds, or latent variables. Typical algorithms include clustering and dimensionality reduction; see the reviews by Glielmo et al. \cite{glielmo2021unsupervised} and by Ceriotti \cite{ceriotti2019unsupervised}. 
A third paradigm is \textit{reinforcement learning}, where learning is driven by interactions between an \textit{agent} (the model) and an \textit{environment} (data or simulation). The agent makes decisions and receives feedback in the form of \textit{rewards} or \textit{penalties}. The model is optimized to maximize the cumulative reward, allowing it to improve its behavior over time through trial and error.

\paragraphtitle{Loss functions.}
Training a ML model requires formalizing its learning objective as a {loss function}, which quantifies how far the model's predictions deviate from the desired outcomes. The optimization then proceeds by adjusting model parameters to minimize this loss, for instance, using gradient-based methods such as stochastic gradient descent in the case of neural networks. In addition, multiple loss functions aiming at different learning objectives can also be combined and minimized simultaneously to enforce different properties into a single model.

In the following, we describe some of the commonly used loss functions in ML as well as in the physical sciences. 
In regression tasks, the \textit{mean squared error (MSE)} is frequently used. Given a set of $N$ predicted values \( \textbf{x}_i \) and target values \( \textbf{x}'_i \), the MSE is defined as:
\begin{equation}
    \mathcal{L}_{\mathrm{MSE}} = \frac{1}{N} \sum_{i=1}^N \left| \textbf{x}_i - \textbf{x}'_i \right|^2
    \label{eq:MSE}
\end{equation}

When comparing predicted and reference {probability distributions}, the \textit{Kullback-Leibler (KL) divergence} is commonly used. Given two distributions \( P(\textbf{x}) \) and \( Q(\textbf{x}) \), the KL divergence is defined as
\begin{equation}
    \mathcal{D}_{\mathrm{KL}}(P||Q) = \sum_\textbf{x} P(\textbf{x}) \log{\frac{P(\textbf{x})}{Q(\textbf{x})}}
    \label{eq:KL_divergence}
\end{equation}
It quantifies how much information is lost when using \( Q \) to approximate \( P \), and it is widely used in variational inference and generative modeling.

Another important principle is that of \textit{maximum likelihood estimation (MLE)}, which aims to find model parameters \( \theta \) that maximize the likelihood of observing the training data under a model distribution \( Q_\theta(\textbf{x}) \), defined as
\begin{equation}
    p(\textbf{x}|\theta) = \prod_{i=1}^N Q_\theta(\textbf{x}_i)
\end{equation}
In practice, more commonly the log-likelihood is minimized, since it is numerically more stable:
\begin{equation}
    -\log p(\textbf{x}|\theta) = -\sum_{i=1}^N \log Q_\theta(\textbf{x}_i)
\end{equation}

\paragraphtitle{Architectures.}
The architecture of a ML model defines the structure of the function \( f_\theta \), parameterized by trainable weights \( \theta \), used to map inputs to outputs. The required complexity of the architecture is not solely determined by the difficulty of the task, but also, crucially, by the quality and expressiveness of the input features. If the input representation already encodes relevant symmetries, invariances, or physically meaningful correlations, even relatively simple models may suffice. Conversely, when using raw or generic features, the architecture must compensate by being more expressive, often at the cost of interpretability, computational efficiency, or data efficiency. In the following, we briefly discuss some important families.

\paragraphtitle{Kernel-based models} compute similarities between inputs using a kernel function \( K(\textbf{x}_i, \textbf{x}_j) \), which implicitly maps data to a high-dimensional feature space:
\begin{equation}
    K(\textbf{x}_i, \textbf{x}_j) = \langle \phi(\textbf{x}_i), \phi(\textbf{x}_j) \rangle
\end{equation}
where \( \phi \) is an implicit (and typically infinite-dimensional) feature map. These methods are typically data-efficient and offer strong theoretical guarantees.

\paragraphtitle{Feed-forward neural networks (FNNs)} represent functions as compositions of simpler transformations $f_i$, typically involving linear layers, characterized by weights $\textbf{W}$ and biases $b$ and nonlinear activation functions $\sigma$:
\begin{equation}
\begin{aligned}
f_\theta(\textbf{x}) 
&= f_L \circ f_{L-1} \circ \dots \circ f_1(\textbf{x}), \\
f_i(\textbf{x}) 
&= \sigma(\textbf{W}_i \textbf{x} + b_i)
\end{aligned}
\end{equation}
Their compositional nature enables them to learn complex hierarchical representations from data, and they are widely used in regression, classification, and representation learning tasks.

\paragraphtitle{Graph neural networks (GNNs)} are tailored for structured data, such as molecular graphs, where atoms and bonds are naturally represented as nodes and edges. These models iteratively update node features by exchanging messages with neighboring nodes:
\begin{equation}
    \textbf{h}_i^{(t+1)} = U \left(\textbf{h}_i^{(t)}, \sum_{j \in \mathcal{N}(i)} M\left(\textbf{h}_i^{(t)}, \textbf{h}_j^{(t)}, \textbf{e}_{ij}\right) \right)
\end{equation}
where \( \textbf{h}_i^{(t)} \) is the feature vector of node \( i \) at iteration \( t \), \( \textbf{e}_{ij} \) encodes edge attributes, and \( M \), \( U \) are learnable functions. This formulation allows GNNs to incorporate both the connectivity and geometry of atomic systems.

More advanced and specialized architectures, such as those used in generative models, will be discussed in detail in the following sections.

\section{Data-driven learning of collective variables }
\label{sec:mlcvs_methods}
Key to the success of many enhanced sampling methods is the identification of suitable CVs. 
Traditionally, CVs have been constructed based on physical intuition by choosing quantities that are experimentally measurable or directly related to the nature of the process.
Examples include torsional angles for conformational changes in molecules and proteins, distances associated with bond formation or breakage for chemical reactions, coordination numbers to describe solvent interaction, or angular order parameters to describe short-range order variation in a phase transition. 
However, these simple CVs can typically account only for a few specific degrees of freedom each, thus making it very likely to overlook important modes of the system.
As a consequence, for a thorough description of complex processes, such as the conformational changes in large biological systems, one may need to use many such CVs to completely describe the relevant modes of the system that are related to the transitions between its long-lived metastable states. 
However, as the computational cost of many enhanced sampling techniques scales highly unfavorably with the number of CVs, this approach is bound to fail as the complexity of the studied process increases.

Over the past decade, it has been widely proposed to improve the CV design process with the help of ML, that is, to learn the CVs directly from a given dataset, optimizing a model with learnable parameters following a suitable learning objective. 
These approaches have already proven effective on a variety of challenging systems, as we will see in Sec.~\ref{sec:mlcvs_applications}.
Many ways of expressing the CVs have been explored, ranging from linear combinations of primitive descriptors to using more complex approaches based on geometric GNNs, which operate directly on the atomic coordinates.
Similarly, many different criteria for optimizing CV models have been proposed, from those derived from ML (e.g., supervised or unsupervised techniques) to physics-informed approaches based on learning dynamic operators or committor probabilities.

The following section aims to provide an organic overview of such methods, trying to group them based on the spirit of their working principles.
To this aim, we first discuss what good CVs are (Sec.~\ref{sec:good_cvs}), presenting this topic from different theoretical and practical points of view. 
Then, we provide an overview of the key ingredients of MLCV models (Sec.~\ref{sec:mlcvs_ingredients}). 
Finally, we illustrate relevant methods proposed so far, grouped into two broad categories. 
First, we discuss approaches that exploit ML-derived techniques to obtain CV surrogates based on geometrical (i.e., structural) information, using techniques such as classification or dimensionality reduction (Sec.~\ref{sec:mlcvs_structural}).
Next, we will present methods in which ML is used as a tool to encode well-defined physical principles into CV models, such as parametrization of dynamic operators (Sec.~\ref{sec:mlcvs_physical_slow}) and committor functions (Sec.~\ref{sec:mlcvs_physical_committor}).

\subsection{What are good collective variables?}
\label{sec:good_cvs}

As introduced earlier, the concept of CVs is closely related to order parameters in physics and reaction coordinates in chemistry. 
CVs are mathematical functions of atomic coordinates, expressed as $\mathbf{s} = \mathbf{s}(\mathbf{R})$, designed to provide a compact and meaningful representation of a reactive process. 
These variables play a crucial role both in data analysis and in enhanced sampling simulations.

CVs should \textit{respect the intrinsic symmetries} of the system, meaning they must be invariant under global rotations and translations, and sometimes also permutation of identical atoms. 
In the context of enhanced sampling, they must satisfy an additional requirement: they should be \textit{continuous and differentiable} to ensure the smooth propagation of biasing forces. 
Indeed, for a one-dimensional CV $s$, the effective potential is expressed as:
    \begin{equation}
        U_{biased}(\mathbf{R}) = U(\mathbf{R}) + V(s(\mathbf{R})) 
    \end{equation}
and hence the force acting on atom $i$ will be:
        \begin{equation}
    f_{biased}^{(i)} = -\nabla_i U(\textbf{R}) - \frac{\partial V}{\partial s} \nabla_{i} s.
    \end{equation}

A key characteristic of a good CV is its ability to achieve \textit{dimensionality reduction}. 
Since molecular systems with $N$ atoms exist in a high-dimensional phase space of 3$N$ dimensions, CVs should provide a low-dimensional representation, ideally in one or two dimensions, while still capturing the essential information about the process of interest. 
Without such a reduction, most analyses would become impractical or difficult to interpret, and CV-based enhanced sampling techniques would be infeasible. 
However, not every low-dimensional representation qualifies as a good CV.
The representation must indeed encode the relevant physical or chemical information that characterizes the reactive process. 
One fundamental requirement is that a CV should be able to \textit{distinguish} between different metastable and transition states, ensuring that configurations from distinct basins are mapped to separate regions of CV space and that transition pathways are clearly represented. 
This property is indeed crucial both for analysis and for applying enhanced sampling methods effectively. 
The latter scenario, to be effective, also requires the ability of CVs to \textit{capture the slowest modes} of the system’s dynamics. 
These modes correspond to rare transitions between long-lived metastable states, which typically involve overcoming significant free energy barriers that hinder sampling. 
Identifying and representing these slowest modes is essential for constructing effective CVs, as they dictate the fundamental kinetics of the system.

From a theoretical standpoint, different approaches have been developed to rigorously define the slowest modes and establish criteria for selecting CVs. 
One widely used perspective is based on the committor function, which describes the probability that a given configuration will evolve toward a particular metastable state. 
A good CV should exhibit a strong correlation with this function, as the committor effectively encodes the progress of a transition (see also Sec.~\ref{sec:mlcvs_physical_committor}). 
Another perspective comes from spectral analysis, where CVs are chosen to approximate the eigenvectors of dynamical operators that govern system evolution. 
In particular, the first non-trivial eigenvectors of the transfer operator correspond to the slowest dynamical modes, making them valuable candidates for CV construction (see also Sec.~\ref{sec:mlcvs_physical_slow}).
It is worth noting that these definitions have somewhat different scopes of applicability (such as a two-state scenario in the case of the committor or many states for the dynamical operators) and also requirements (such as the presence of a spectral gap for learning the dominant eigenfunctions, or data from the transition state region for the committor function).

\subsection{Ingredients of machine learning CVs}
\label{sec:mlcvs_ingredients}

\begin{figure*}[htbp]
   \centering
   \includegraphics[width=1\linewidth]{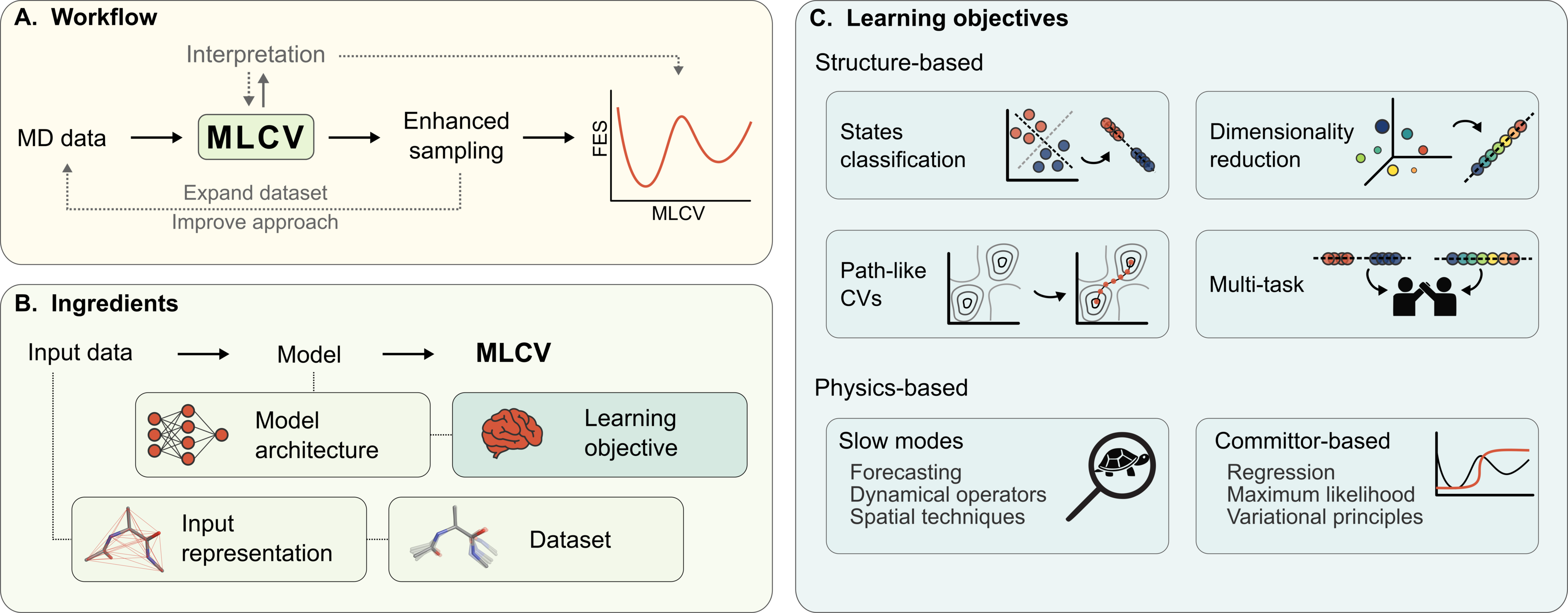}
   \caption{\justifying Typical workflow of machine learned CVs. 
   (A) Starting from data collected with MD simulations, the MLCV is trained and used to drive enhanced sampling simulations, for example, to compute free energy estimates. The procedure can often be improved in an iterative way by expanding the training set with the newly collected configurations and eventually exploiting such data to enforce a more refined learning criterion. In addition, an analysis of the CV model can both help improve the design process and the interpretation of the results.
   (B) Ingredients of MLCVs. The input data from MD simulations is encoded through a representation of the system (e.g., physical descriptors or atomic coordinates) and stored in a dataset. The functional form of the CV model is determined by its architecture (e.g., a neural network), and its optimization is driven by the learning criterion that characterizes the adopted CV method. The MLCV value for a given input is returned as the output of the CV model.
   (C) Types of CVs learning objectives. Structure-based methods exploit structural and topological features, with criteria such as classification of states, dimensionality reduction, approximation of path-CVs, or combination of such approaches in a multi-task framework. Physics-based methods aim at encoding specific physical properties into the CV model, for instance by targeting slow modes or by leveraging properties of the committor function.}
   \label{fig:ML-CV}
\end{figure*}

Here, we briefly describe the three main ingredients that define a data-driven approach: the representation of the system (input features), the choice of model architecture, and the construction of the dataset, which are schematically depicted in Fig.~\ref{fig:ML-CV}. 

\paragraphtitle{Input representation}. The first ingredient is the choice of how to represent the system, that is, what constitutes the input of our ML model. 
A natural choice would be to use raw atomic coordinates; however, they do not inherently respect the relevant physical symmetries, such as rotational and translational invariance. 
Hence, some additional pre-processing step is required, such as \textit{aligning} the system's coordinate to a template structure. 
This option can be exploited when there are rigid motifs in the system and/or one is interested in conformational changes, while care should be taken in the case of reactive events. 
Alternatively, the invariance under the roto-translational symmetry can be implicitly learned with a \textit{data-augmentation }scheme, in which the training dataset is augmented by randomly rotating and translating the input coordinate structures while keeping the same target. 
Finally, \textit{geometric GNNs }provide a more elegant (and expensive) solution by representing atoms as graph nodes, naturally encoding relational information while maintaining symmetry invariance or equivariance.
An alternative way to encode the physical symmetries is to construct descriptors to represent the system (\textit{featurization}). 
Simple physical quantities, such as interatomic distances and torsional angles, have indeed long been used to promote sampling of reactions and conformational changes. 
They could also be more complex descriptions of the local environments, such as the Steinhardt parameters to measure the orientational order in crystals or symmetry functions \cite{behler2007generalized} and the smooth overlap of atomic positions (SOAP) \cite{bartok2013representing} descriptors, which are commonly employed in ML potentials. 
These offer richer representations of local environments but come at a higher computational cost. 
Other domain-specific features, such as structure factor peaks for crystallization \cite{karmakar2021collective} or graph-based descriptors for chemical reactions and phase transitions \cite{rahimi2023comparison, pietrucci2011graph, raucci2022discover, yang2024structure}, have also been successfully applied.

\paragraphtitle{Model architectures}. Different architectures have been used to construct CVs starting from their representation, each offering distinct trade-offs between expressiveness and computational cost. 
Early approaches relied on linear models, optimizing linear combinations of predefined descriptors.
These were later extended using kernel methods and, more prominently, FNNs, which provide greater flexibility in learning nonlinear transformations. 
More recently, geometric GNNs have been exploited, offering richer representations of molecular systems by treating atomic environments as graph structures, although with a higher computational cost.

\paragraphtitle{Datasets}. It is important to note that the choice of dataset typically depends not only on the chosen representation and model architecture, but also on the learning objective, as different MLCV methods require different types of data. 
For example, unsupervised learning approaches for dimensionality reduction can use raw MD trajectories without labels, making them broadly applicable. 
In contrast, supervised learning methods rely on labeled data, such as configurations classified by the metastable states or transition states. 
Physics-informed approaches that aim to extract the slow modes of the system often require ergodic simulations or biased simulations in a stationary limit. 
Ensuring that the dataset adequately represents relevant system configurations and, possibly, transitions is essential for training reliable CVs.

\subsection{Structure-based approaches}
\label{sec:mlcvs_structural}

In the first broad category, we discuss structure (or geometry)-based methods for CV optimization, which assume that relevant transitions can be captured by analyzing geometric or topological features. 
These methods include classification-based CVs, which rely on supervised learning to distinguish between metastable states, dimensionality reduction techniques, which extract low-dimensional representations without prior labeling, and path-like CVs, which approximate transition pathways to describe molecular processes. 
At the end of this section, we also discuss the possibility of combining different criteria into a multi-task framework. Table~\ref{tb:structural_mlcv} provides an overview of the methods discussed in the following sections, together with a concise summary of their reported applications, with the aim of highlighting the areas in which they have been applied.

\subsubsection{Metastable states classification}

Since one of the basic requirements of CVs is to be able to distinguish metastable states, several methods have been proposed to construct CVs from classifiers optimized to discriminate between different states.
This applies to the situation in which (some) metastable states of a system are known, such as the reactants and products of a chemical reaction, or the folded and unfolded states of proteins, or the bound and unbound states of a host-guest system. 
For instance, information on the native states of proteins or the bound state could come from experimental data such as x-ray crystallography, while other states, such as the unfolded or unbound ones, can be rather easily obtained through MD simulations at higher temperatures. 
Once we have a set of states, we can create a dataset of configurations with labels indicating which state they belong to, for example, by running a series of short MD trajectories for each metastable state. 
Sultan and Pande \cite{sultan2018automated} have explored the use of various classifier outputs, such as the distance to the decision hyperplane of a support vector machine, logistic regression probability estimates, and classifier outputs from deep or shallow neural networks, to build CVs and accelerate molecular simulations, demonstrating the feasibility of this approach.

Because the main goal is to obtain a variable whose values are able to discriminate between different states, many methods for constructing CVs have been based on \method{linear discriminant analysis (LDA)}. 
This is a supervised learning algorithm that separates the different classes by maximizing the \textit{inter-}class variance \( \mathbf{S}_b \) while minimizing the \textit{intra-}class variance \( \mathbf{S}_w \), by solving the generalized eigenvalue problem:
\begin{equation}
    \mathbf{S}_b \mathbf{w} = \lambda \mathbf{S}_w \mathbf{w}
\end{equation}
Here, the eigenvectors \( \mathbf{w} \) define the directions in the feature space \( \mathbf{x} \) (e.g., interatomic distances, dihedrals) that best separate the predefined states, and the eigenvalues $\lambda=\frac{\mathbf{w}^T \mathbf{S}_b \mathbf{w}}{\mathbf{w}^T \mathbf{S}_w \mathbf{w}}$ measure the degree of separation. This can be seen as a similar operation to principal component analysis (PCA), but where the principal discriminant components are the linear projections that distinguish the states the most. 
Note that the number of nonzero eigenvalues (and thus usable CVs) is \( N_S - 1 \), where \( N_S \) is the number of metastable states.

Mendels \textit{et al.} proposed \method{Harmonic-LDA (HLDA)}\cite{mendels2018collective}, a variant that computes covariance matrices using a harmonic mean, in order to address the problem that the LDA method assigns high variance weights to CVs, resulting in suboptimal sampling for the more stable states characterized by smaller fluctuations. 
They showed that this approach can be successfully used in numerous cases from biology~\cite{mendels2018folding} to chemistry~\cite{piccini2018metadynamics}.
Recently, Sasmal \textit{et al.} used standard LDA to learn CVs directly from atomic positions\cite{sasmal2023reaction}. 
To this end, they treated a molecular configuration as a member of an equivalence class in size-and-shape space, containing all molecular configurations that can be optimally translated and rotated to align with a reference distribution. 
Furthermore, they reformulated the LDA eigenvalue problem in terms of generalized singular value decomposition (SVD) to extend the applicability of the method in this setting. This way, they were able to study the folding and the right-left helix transition in small proteins.

\begin{figure*}[htbp]
\centering
\includegraphics[width=0.75\linewidth]{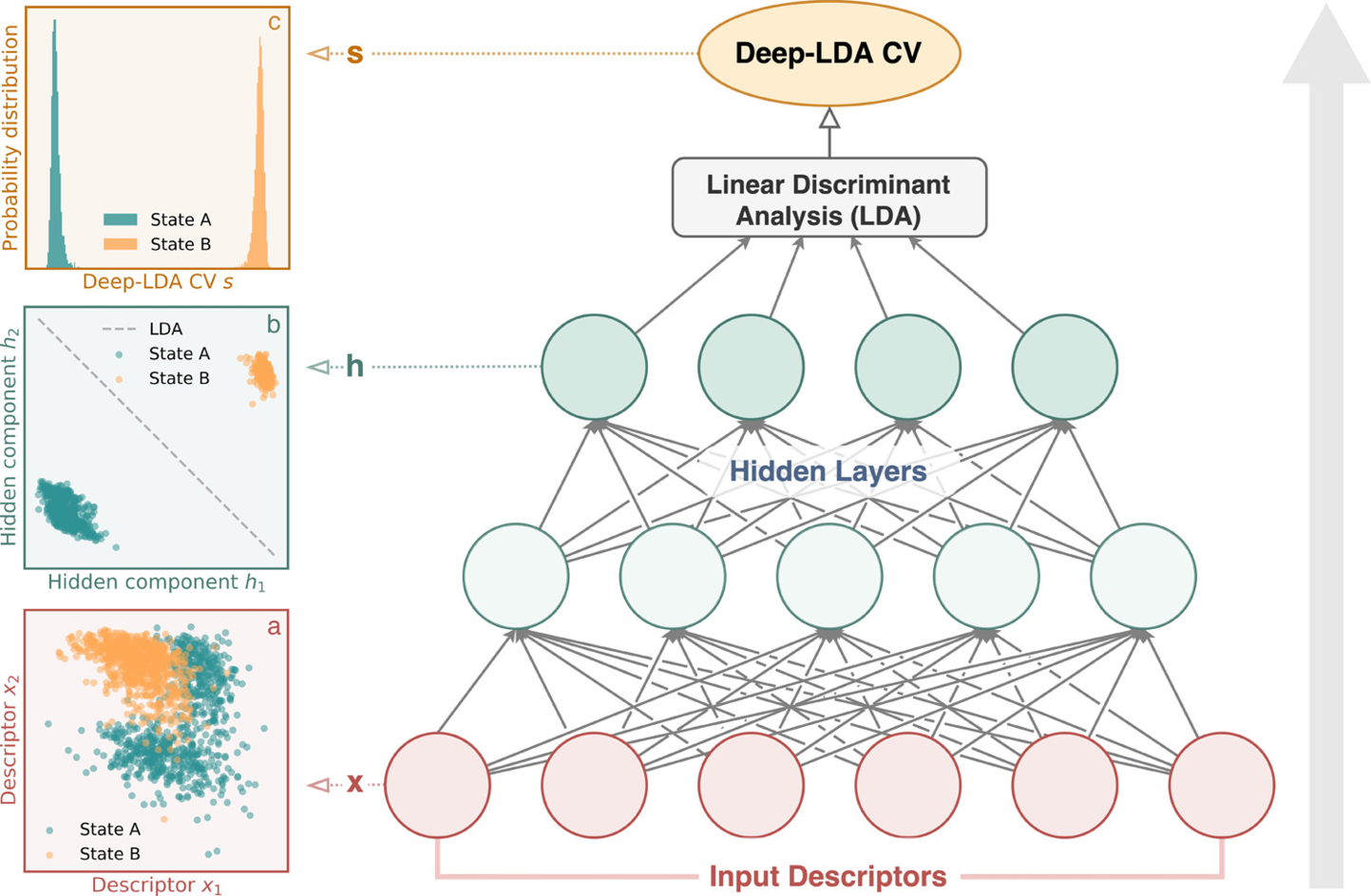}
\caption{\justifying
Schematic of Deep-LDA, an example of a supervised, classifier-based CV. A set of physical descriptors serves as input to a feedforward neural network, which performs a nonlinear transformation to a feature space where the separation between metastable states is maximized. In the final layer, Fisher's discriminant analysis is applied to identify directions that best discriminate between the predefined classes, yielding the CVs. The network is optimized to enhance this discriminative power by maximizing the LDA separation score. The panels illustrate this process: (a) distribution of input descriptors for two metastable states, showing partial overlap; (b) transformed variables in the neural network’s feature space with the LDA boundary, where the states become linearly separable; and (c) probability distribution of the resulting CV, demonstrating clear discrimination between states.
Image reproduced from Ref.~\citenum{bonati2020data}. Copyright 2020 American Chemical Society.}
\label{fig:deeplda}
\end{figure*}

However, the main limitation of LDA/HLDA is the linearity of the projection, and thus the need to identify a (small) set of descriptors where the states are already linearly separable. 
To address this, Bonati \textit{et al.} proposed to use a nonlinear extension called \method{Deep-LDA}\cite{bonati2020data}. In this method, the original inputs \( \mathbf{x} \) are first transformed via a neural network into a latent space of hidden features  \( \mathbf{h}_\theta = f_\theta(\mathbf{x}) \) (see Fig.~\ref{fig:deeplda}). Then, the CVs are obtained by performing LDA in the transformed space $\mathbf{h}_\theta$. The network's parameters are optimized to maximize the LDA discrimination score, or in other words, its generalized eigenvalues. In the case of two states, this corresponds to using as loss function $\mathcal{L}_{Deep-LDA}=-\lambda$. This process corresponds to transforming the feature space to maximize the ability to discriminate between states. 
After training, Deep-LDA CVs can be used to drive enhanced sampling simulations between metastable states and reconstruct the free energy profile.

Generalizing this to more than two states requires, as in the case of LDA, $N_{S-1}$ CVs.
To ease this requirement, Trizio and Parrinello proposed the \method{deep targeted discriminant analysis (Deep-TDA)} method, where the discrimination criterion is obtained by a distribution regression procedure.\cite{trizio2021enhanced} 
In this case, the neural network outputs are used directly as CVs, which are optimized by imposing a target distribution on the projected training data.
This distribution is defined as a linear combination of $N_S$ multivariate Gaussian distributions with diagonal covariances, one associated with each state. 
Each Gaussian is defined by $N_{\rho} = N_S - 1$ CV positions and covariances, so the loss function is as follows:
    \begin{equation}
      \mathcal{L}_{\mathrm{Deep-TDA}} = \sum_{k}^{N_S} \sum_{\rho}^{N_{\rho}} \left( \alpha L_{k,\rho}^{\mu} + \beta L_{k,\rho}^{\sigma} \right)
    \end{equation}
where $\rho$ are the components of the CV space. 
While the results are similar to Deep-LDA for a two-state scenario, Deep-TDA can reduce the dimensionality of the CV space (i.e., $N_\rho < N_S - 1$) in the case where one has additional information, such as having a set of ordered states (e.g. intermediate steps) or mutually exclusive reactants and products, circumstances in which a one-dimensional variable is sufficient~\cite{das2023and, das2024correlating}.

\paragraphtitle{Augmenting classifier-based CVs.} Since these methods are trained exclusively to distinguish metastable states, their performance can be suboptimal in the transition state region, resulting in a limited sampling.  
For this reason, it has been proposed to improve them by adding data belonging to the transition region, which can be accomplished in different ways. 
For instance, Ray \textit{et al.} proposed incorporating data from the transitions path ensemble obtained from reactive trajectories \cite{ray2023deep} as an additional state in Deep-TDA CVs. Specifically, they performed OPES-flooding \cite{ray2022rare} simulations based on the Deep-TDA CVs to obtain several reactive trajectories, and the configurations located outside of the metastable basins are collected and assigned to a new state, characterized by a broader distribution. 
As an alternative, Yang \textit{et al.} proposed a simulation-free data augmentation strategy for CV learning in protein folding environments\cite{yang2024learning}. 
They used geodesic interpolation on Riemannian conformational manifolds of proteins, as proposed by Diepeveen \textit{et al.} \cite{diepeveen2024riemannian}, which faithfully models protein folding transitions. 
Although these are not true realizations of transition states, augmenting the training data with these interpolations can improve the quality of the CVs and thus sampling. 
Furthermore, since the interpolation parameter $t \in [0,1]$ represents the progress of the transition, they also proposed to train CVs by performing regression on this parameter. 
Finally, multi-task approaches can be used to enhance classifier-based CVs by augmenting them with more data outside the metastable states (see Sec.~\ref{sec:mlcvs_multitask}). 

\subsubsection{ Dimensionality reduction}
While classification-based CVs require a labeled dataset, another large family of CV optimization methods is based on unsupervised learning strategies. 
In this case, the goal of ML approaches is to extract meaningful information from simulations without providing explicit targets, but rather by exploiting their ability to identify meaningful low-dimensional representations. 
Note that not all unsupervised techniques can be applied to CV discovery, as, if the purpose is to find variables to perform enhanced sampling, we need continuous and differentiable functions of atomic positions (or descriptor functions of them). 
For additional information on unsupervised approaches, we also refer to the recent review from Glielmo \textit{et al.}\cite{glielmo2021unsupervised}.

The most famous example from this family is \method{principal component analysis (PCA)}\cite{sittel2018perspective,amadei1993essential,garcia1992large,hegger2007complex,sittel2017principal}. 
The purpose of this method is to reduce the number of variables describing a given dataset while retaining most of the original information. 
To this aim, PCA diagonalizes the covariance matrix of a set of features and projects the data onto its leading eigenvectors (called principal components). 
These represent the linear combinations of the input features that encode as much of the variance as possible. 
For this reason, PCA is often used as a dimensionality reduction algorithm to preprocess the dataset before feeding it to other algorithms. 
It has also been used by Spiwok \textit{et al.} to directly learn a set of CVs to understand the system and enhance the sampling via biased simulations\cite{spiwok2007metadynamics}.
Of course, this approach is suitable when the transition we are interested in is associated with a large structural change in the system, and is thus related to principal components. 
Also, the projection operated by PCA acts on a linear subspace of the original space, which may not be adequate when the relationship between the relevant degrees of freedom is nonlinear. 

\begin{figure}[t!]
\centering
\includegraphics[width=1.0\linewidth]{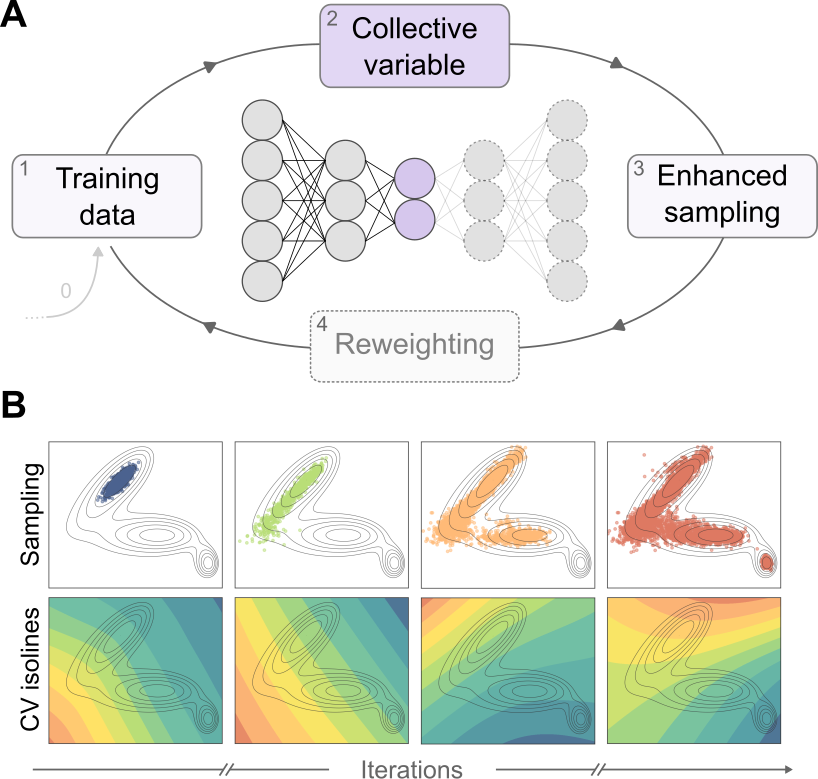}
\caption{\justifying
(A) Many unsupervised approaches to CV discovery are based on autoencoders, which learn low-dimensional representations directly from unlabeled simulation data. These methods are typically used in an exploratory fashion, interleaving rounds of CV learning with free energy biasing. In each iteration, the learned CVs are used to bias the system and promote the exploration of new configurations, which are then added to the training set for the next round. In some variants, statistical reweighting of the sampled configurations is applied before proceeding to the next iteration.
(B) Example of progress of an iterative approach used for exploration on a simple 2D potential surface. As the number of iterations increases, a larger portion of the phase space is sampled (explored points in the 2D space, upper row) and better CVs are learned (CV isolines, lower row).}
\label{fig:autoencoders}
\end{figure}

Among the non-linear unsupervised methods used for CV discovery, many of them rely on \method{autoencoders (AE)}. 
An AE is an artificial neural network consisting of two parts: the first part (encoder) $E$ maps the high-dimensional input space to a low-dimensional latent space, often referred to as the bottleneck of the model (see Fig.~\ref{fig:autoencoders}). 
The second part (decoder) $D$ simultaneously learns to reconstruct the input data by mapping the latent space back to the high-dimensional space of the inputs. 
The parameters of the encoder and decoder are optimized to minimize the discrepancy between the reconstructed output and the original input features $\mathbf{x}_i$, typically by using MSE as a loss function:
    \begin{equation}
      \mathcal{L}_{\mathrm{AE}} = \sum_{i=1}^{N} \left| \mathbf{x}_i - \mathrm{D} \circ \mathrm{E} \left( \mathbf{x}_i \right) \right|^2
    \end{equation}
Through this process, the model learns to recover the original data from the low-dimensional representation of the bottleneck, with the latent space often capturing key features of the data, thus providing a sort of non-linear generalization of PCA. 
In the context of enhanced sampling, the latent space is typically used as the CV for analysis and biasing, while the decoder is only used during training.

The earliest adoption of AEs for enhanced sampling is the \method{molecular enhanced sampling with autoencoders (MESA)} method proposed by Ferguson and coworkers\cite{chen2018molecular,chen2018collective}, which uses AEs to learn nonlinear low-dimensional CVs describing the important configurational motions of biomolecules from atomic coordinates, as demonstrated on small test proteins. 
In addition, MESA also uses a data augmentation approach to resolve internal structural reconfigurations and exclude trivial changes in rotational orientation and alternates between CV learning and free energy biasing (umbrella sampling) along these CVs. 
Similarly, Belkacemi \textit{et al.} developed an iterative algorithm for CV learning with AEs, named \method{free energy biasing and iterative learning with autoencoders (FEBILAE)} \cite{belkacemi2021chasing}. 
Contrary to MESA, when learning from biased samples, FEBILAE reweights the configurations sampled from a biased distribution $\tilde{\mu}$ by a factor $w(x) = \frac{\mu(\mathbf{x})}{\tilde{\mu}(\mathbf{x})}$ to target the unbiased one $\mu$, corresponding to the Boltzmann distribution. 
Moreover, FEBILAE relies on adaptive techniques to sample configurations and compute the free energy by reweighting them. Beyond the differences, the iterative aspect that alternates between optimization and biasing is a recurring feature of these and many other AE-based methods. This enables these methods to be used in an exploratory way, without knowing beforehand which the relevant metastable states are.

Different types of systems and processes can be addressed by combining AEs with suitable sets of descriptors. 
In the context of chemical reactions, Ketkaew \textit{et al.} developed a non-instructor-led \method{deep autoencoder neural network (DAENN)} to discover CVs from unbiased MD of the reactants' state of chemical reactions\cite{ketkaew2022machine}. 
To this end, the Authors introduced an unsupervised training descriptor (xSPRINT) which extends the original SPRINT \cite{pietrucci2011graph} variables by including information on distant atoms not directly involved in the reaction. 
The authors then used AEs to reduce the dimensionality of these descriptors into a small set of CVs, employing, in addition to the reconstruction loss, also a penalty function based on root mean squared deviation (RMSD) of atomic positions to promote exploration of the free energy landscape.

To facilitate the sampling of systems involving indistinguishable particles, which are commonly encountered in self-assembly and solvation systems, Ferguson and coworkers proposed an approach called \method{permutationally invariant networks for enhanced sampling (PINES)} \cite{herringer2023permutationally}. 
PINES combines permutation-invariant vector (PIV) descriptors\cite{pipolo2017navigating,gallet2013structural} with AEs to learn nonlinear CVs that are invariant not only to translational and rotational symmetry but also to the permutational one.
The methods integrate PIV characterization with MESA\cite{chen2018molecular, chen2018collective}, iteratively training the CVs and performing enhanced sampling to achieve converged thermodynamic averages. 

One general aspect of AEs is that they only implicitly optimize the latent space as an intermediate step in the reconstruction of the inputs, without imposing any particular structure on the CV space. 
To improve this aspect, different strategies can be applied to enforce specific properties. 
Lemke and Peter introduced a dimensionality reduction algorithm called \method{EncoderMap} \cite{lemke2019encodermap}, which combines an AE with the cost function of sketch-map \cite{tribello2012using}. 
Sketch-map is a multidimensional scaling-like algorithm that aims to reproduce in a low-dimensional space the distances between points in a high-dimensional space, thus enforcing a metric on the latent space. 
In a similar spirit, regularization techniques \cite{lelievre2024analyzing} and multi-task approaches~\cite{bonati2023unified} (see Sec.~\ref{sec:mlcvs_multitask})) 
can be used to enforce desired properties in the CV space of the AEs.

Besides standard AEs, other methods rely on the so-called \method{variational autoencoders (VAEs)} \cite{kingma2014auto}, which are a particular class of AEs based on Bayesian theory. 
In VAEs,  the data in the latent space is enforced to follow a prior distribution, commonly chosen to be a multivariate Gaussian distribution. 
First, the encoder learns to output the Gaussian distribution's mean and variance, and the decoder's sample is drawn from this distribution. 
Second, the encoder/decoder parameters are optimized to maximize the evidence lower bound (ELBO) \cite{blei2017variational}, consisting of two terms: the reconstruction loss measuring how well the VAE can reconstruct the input data and the KL divergence between the approximate posterior and the prior distributions. 

Among the applications of VAEs to the CV discovery problem, Ribeiro \textit{et al.} proposed \method{reweighted autoencoded variational bayes for enhanced sampling (RAVE)} \cite{ribeiro2018reweighted}. 
RAVE is based on the idea that the probability distribution of the latent space can be taken as the most relevant feature learned from the VAE as opposed to the latent variable itself. 
Then, a physical proxy variable is obtained from a linear combination of a set of descriptors, optimized to have the same probability distribution as the latent space. 
Later, Vani \textit{et al.} also proposed to integrate the RAVE algorithm with AlphaFold to sample Boltzmann ensembles starting from protein sequences\cite {vani2023exploring}, showing applications to challenging proteins such as G-protein coupled receptors (GPCRs).
From a different perspective, Schober \textit{et al.} employed VAEs to frame the construction of CVs as a Bayesian inference problem \cite{schoberl2019predictive}. 
In this framework, CVs are considered as low-dimensional hidden generators \cite{jordan1999learning} of all-atom trajectories. The identification of CVs is thus formulated as a Bayesian inference task, where the posterior distribution of the latent CVs is inferred given fine-scale atomic training data. 
The Bayesian latent variable model for CV discovery also incorporates uncertainty quantification to provide confidence in the discovered CVs, which is particularly useful when the training data is sparse or noisy. 

Following a different strategy, Sipka \textit{et al.} used a variational autoencoder to construct CV by compressing a pre-trained representation obtained from an ML potential\cite{sipka2023constructing}. 
Specifically, they extracted an intermediate representation of a graph network based on SchNet\cite{schutt2017schnet} architecture, which intrinsically respects rotational, translational, and permutational invariance. 
Moreover, this approach is an example of transfer learning, in which the representation learned to construct the potential is also used to learn another task (the CV) with little effort. 

\subsubsection{Path-like collective variables}
In this section, we describe another approach to CV construction, which is based on approximating a given path in the (atomic or collective) space. 
This idea has indeed inspired numerous ML approaches.

We start by recalling the original formulation of the so-called \textit{CV pathways} from A to B.
This method requires a reference pathway, given by a sequence of $D$ intermediate molecular structures $S_0(\mathbf{R}) = \left(S_1(\mathbf{R}), S_2(\mathbf{R}), \dots, S_D(\mathbf{R})\right)$. 
These configurations can be represented by either their atomic positions or a set of CVs. 
From these configurations, we can define the progress along the reference path using the following expression:
    \begin{equation}
        s(\mathbf{R}) = \lim_{\lambda \to \infty} \frac{\int_{0}^{1} dt \, t \, e^{-\lambda \| S(\mathbf{R}) - S_0(t) \|^2}}{\int_{0}^{1} dt \, e^{-\lambda \| S(\mathbf{R}) - S_0(t) \|^2}}
    \end{equation}
where the parameter $\lambda$ ensures localization around the closest point in the path, as it can be interpreted as the inverse of a Gaussian variance, and $t \in [0,1]$ parametrizes the position along the reference path \(S_0(t)\). 
Formally, the $s$ variable induces a foliation in the $S$ space and, near the reference path $S_0(t)$, the foliating surfaces become flat and orthogonal to $S_0(t)$.
The distance from the reference path $S_0(t)$ is defined as:
    \begin{equation}
        z(\mathbf{R}) = \lim_{\lambda \to \infty} \left( -\frac{1}{\lambda} \log \int_{0}^{1} dt \, e^{-\lambda \| S(\mathbf{R}) - S_0(t) \|^2} \right)
    \end{equation}
Moreover, the FES $F(s, z)$ as a function of the path CVs can reveal other qualitatively distinct pathways that may be separated from the reference path by significant energy barriers.
Other formulations of path CVs have been proposed, such as in path-metadynamics (PMD) \cite{diaz2012path,perez2018advances}, where the objective is to reconstruct (and optimize) the average transition path connecting two states in the space spanned by the CVs.

One of the problems with conventional path CVs is related to the definition of an optimal similarity measure to describe the process of interest in a high-dimensional space \cite{hovan2018defining}, which is well suited for ML approaches. 
For example, Rogal \textit{et al.} proposed a  \method{path CV based on neural networks}\cite{rogal2019neural} designed to enhance sampling of solid-solid phase transformations in molecular simulations. 
Instead of relying on manually selected reaction coordinates, they employed a neural network classifier to identify local structural environments, which are then used to define a global reaction path in a low-dimensional feature space. 
The path CV is constructed by first classifying atomic environments using the neural network, which assigns a structural label to each local atomic configuration, and then using such classification as global structural descriptors, allowing the definition of a one-dimensional continuous reaction path that captures the transition between phases. 

France-Lanord \textit{et al.}\cite{france2024data} recognized a formal connection between path CVs and kernel methods, interpreting the variable which describes the progress along a reference path as a similarity measure between a configuration $S(\mathbf{R})$ and a set of reference frames $S_0(t)$, typically via a Gaussian kernel. They proposed a \method{data-driven generalization of path CVs} using kernel ridge regression (KRR), enabling the model to accommodate a larger set of reference configurations and to use higher-dimensional, structured inputs such as SOAP descriptors.
In their approach, the KRR model is trained to predict committor probabilities directly from structural descriptors, effectively learning a smooth and differentiable approximation of the progress variable $s(\mathbf{R})$. To construct the training set, committor estimates for selected configurations are obtained from a combination of biased simulations and TPS, ensuring accurate coverage across the transition region (see Sec.~\ref{sec:mlcvs_physical_committor}).

Frohlking \textit{et al.} proposed a method called \method{deep-locally non-linear embedding (DeepLNE)} \cite{frohlking2024deep}, that aims at constructing a directional CV which can describe the progress of the transition through a nonlinear combination of feature vectors inspired by the locally linear embedding method \cite{roweis2000nonlinear}.
Such an architecture is a generalized AE that performs a continuous k-nearest neighbors (k-NN) step on each data point before reducing the dimensionality through the encoder to the bottleneck representing the 1D CV ($s$), whereas the decoder is used to compute the perpendicular distance ($z$) CV. 
One of the main advantages of DeepLNE is its ability to automatically select the metric used for neighbor searches and learn the path from state A to state B without the need for hand-picking landmark selection in advance. 
However, the nearest neighbor step in DeepLNE resulted in a substantial computational cost that the authors later addressed with the revised DeepLNE\texttt{++} \cite{frohlking2024deeplne++} strategy, which uses knowledge distillation to construct a more computationally efficient CV by labeling the training data to guide directionality and employing an ANN student model to represent the DeepLNE variables $s$ and $z$.

\subsubsection{Multitask learning}
\label{sec:mlcvs_multitask}
While many methods are optimized with a single objective, it is often desirable for the CVs to obey multiple requirements. 
This can be accomplished within a multi-task framework~\cite{caruana1997multitask,zhang2018overview,zhang2021survey}.  
This is an umbrella term to describe methods in which a single model is optimized using \textit{multiple learning objectives}, and is generally achieved by including multiple terms in the loss function (e.g., via a sum of them). 
This can be useful also to regularize the learning~\cite{lelievre2024analyzing} and to exploit complementary information across different datasets~\cite{bonati2023unified}.

One way this can be implemented is to learn a single CV that is then able to perform \method{multiple downstream tasks}. 
Kozinsky and collaborators~\cite{sun2022multitask} framed CV learning as a dimensionality reduction that must be able to both separate basins and predict potential energy. In their scheme, the multitask CV consists of a common encoder that performs dimensionality reduction together with multiple decoders that perform separate downstream tasks (potential energy predictor and basin classifier). 

\begin{figure}[htbp]
\centering
\includegraphics[width=1\linewidth]{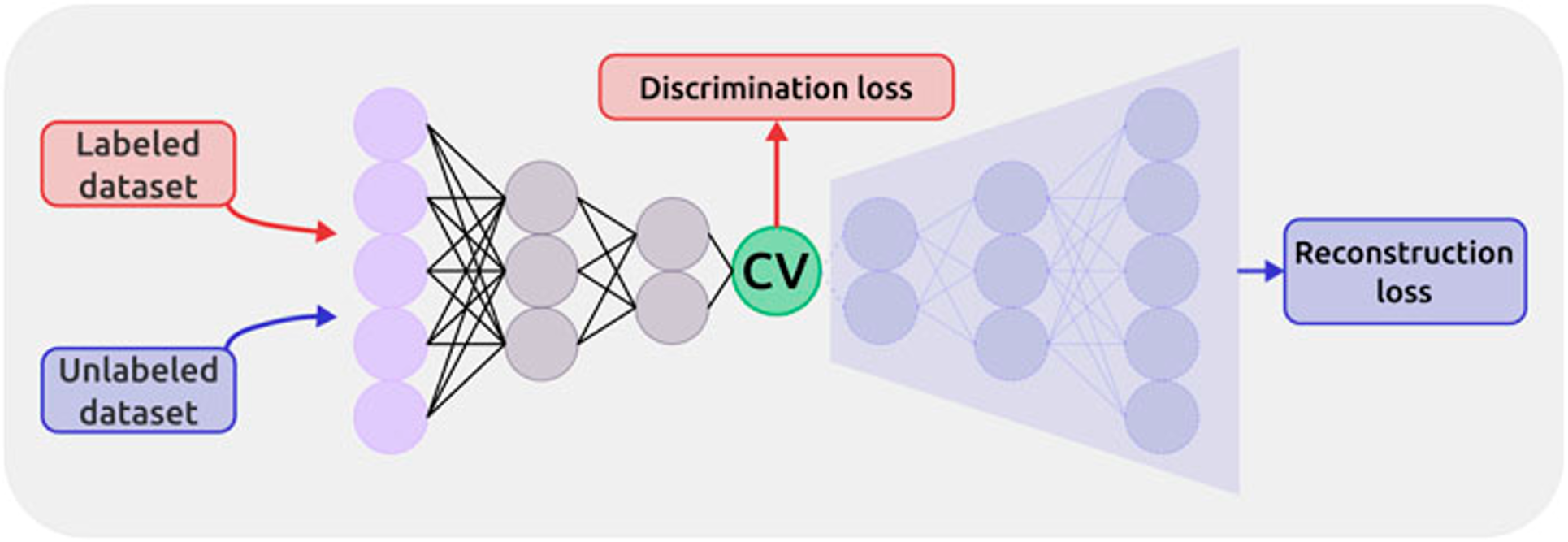}
\caption{\justifying
Multi-task CV optimized on different datasets. This approach combines multiple objectives into a single CV model. In the semi-supervised setup, an autoencoder is used to process data from an unlabeled dataset (blue path) with an unsupervised loss (e.g., reconstruction MSE) computed from the decoder’s output, while labeled data (red path) contribute to a supervised loss applied directly in the CV space (e.g., TDA loss).
Image reproduced from Ref.~\citenum{bonati2023unified}. Copyright 2023 AIP Publishing LLC.} 
\label{fig:Multitask}
\end{figure}

Bonati \textit{et al.} proposed a more general multi-task learning framework, which enables the optimization of a single model via a combination of loss functions evaluated on \textit{different dataset types}~\cite{bonati2023unified}. 
Indeed, often one is faced with many datasets for the same system that are different in nature, for example, because they are sampled using different approaches. These may include, for instance, a subset of labeled configurations coming from unbiased simulations in the different metastable states and unlabeled configurations obtained in biased simulations. 
To effectively integrate information from these different datasets, a multitask learning structure can be adopted. 
In particular, Bonati \textit{et al.}~\cite{bonati2023unified} proposed a \method{semi-supervised multitask CV} that uses an autoencoder-like architecture combined with the Deep-TDA objective (see Fig.~\ref{fig:Multitask}). 
The loss function for the multitask CV is given by a linear combination of the reconstruction loss (calculated on the unlabeled dataset $D_1=\{x_i\}$) and the Deep-TDA loss (calculated on the labeled dataset $D_2=\{(x_i,y_i)\}$) acting on the CVs $s$ 
    \begin{equation}
        \mathcal{L}_{\mathrm{multitask}} = \mathcal{L}_{\mathrm{AE}}|_{D_1} + \alpha \mathcal{L}_{\mathrm{Deep-TDA}}|_{D_2}
    \end{equation}
where $\alpha$ is a hyperparameter that scales the relative weight of the two losses. 
This means that the resulting CV is optimized to reconstruct the data as in a standard autoencoder, but also to discriminate between states. 
This approach can be used, for example, to combine equilibrium data with data from biased simulations, but it is not limited to that. 

Indeed, such a multitask approach was later employed by Zhang \textit{et al.} \cite{zhang2024combining} to learn CVs from TPS simulations. 
Specifically, a semi-supervised autoencoder was trained on TPS trajectories using a reconstruction loss, whereas the classification loss was enforced using a labeled dataset collected with unbiased MD in the initial and end states.  
Furthermore, this CV was also used to bias the shooting point selection towards the region of high reactivity (i.e., close to the transition region), identified by fitting the density of shooting points in the low-dimensional space identified by the multitask CV. 
The algorithm then proceeds iteratively by refining both the CV and the shooting range, yielding both the transition path ensemble and the free energy profiles obtained via biased simulations using the optimized CV, showing the strength of the multitask approach in deriving high-quality CVs by combining multiple simple objectives.

\setlength{\arrayrulewidth}{0.0mm}
\setlength{\tabcolsep}{-1pt}
\renewcommand{\arraystretch}{1.5}

\newcolumntype{C}[1]{>{\centering\arraybackslash}p{#1}}

\begin{table*}[htbp]
\centering
\footnotesize{
\begin{tabular}{lC{5.5cm}C{2.2cm}C{2.2cm}C{2.2cm}C{2.2cm}}
& \shortstack{Notes} & \shortstack{Conformational\\Biophysics} & \shortstack{Ligand\\Binding} & \shortstack{Phase\\Transformations} & \shortstack{Chemical\\Systems} \\
\hline
\rowcolor{fessa3!20} \paragraphtitle{Classifier-based CVs}& &  &  &  & \\
\rowcolor{fessa3!20} SVM~\cite{sultan2018automated} & Support vector machines& ~\citenum{sultan2018automated}  &  &  & \\
\rowcolor{fessa3!20} (H)LDA\cite{mendels2018collective} & Linear Discriminant Analysis & \citenum{mendels2018collective},\citenum{sasmal2023reaction} &  & \citenum{zhang2019improving} & \citenum{piccini2018metadynamics}\\
\rowcolor{fessa3!20} Deep-LDA \cite{bonati2020data} & NN extension of LDA & \citenum{ansari2021water}, \citenum{ruiz2023cation},\citenum{visigalli2025coordinated} & \citenum{rizzi2021role}, \citenum{ansari2022water}, \citenum{siddiqui2023application}  & \citenum{karmakar2021collective} &  \\
\rowcolor{fessa3!20} Deep-TDA\cite{trizio2021enhanced, ray2023deep} & Targeted discrimination & \citenum{ray2023deep} & \citenum{trizio2021enhanced},\citenum{das2023and},\citenum{das2024correlating} \citenum{kumari2025enhanced} & \citenum{yang2024structure} &  \\

\hline
\hline
\rowcolor{fessa4!20} \paragraphtitle{Autoencoders}& &  &  &  & \\
\rowcolor{fessa4!20} MESA\cite{chen2018molecular,chen2018collective} & Iterative autoencoder (AE) & \citenum{chen2018molecular,chen2018collective} &  &  & \\
\rowcolor{fessa4!20} RAVE\cite{ribeiro2018reweighted} & Linear CV from variational AE & \citenum{vani2023exploring} & \citenum{ribeiro2019toward}, \citenum{wang2022interrogating}, \citenum{lamim2020combination} &  & \\
\rowcolor{fessa4!20} FEBILAE \cite{belkacemi2021chasing} & AE + data reweighting & \citenum{belkacemi2023autoencoders} &  &  & \\
\rowcolor{fessa4!20} EncoderMap\cite{lemke2019encodermap} & AE + Sketch-Map loss  & \citenum{lemke2019encodermap} &  &  & \\
\rowcolor{fessa4!20} DAENN\cite{ketkaew2022machine} & AE + xSPRINT inputs (topology changes)&  &  &  &  \citenum{ketkaew2022machine},\citenum{ketkaew2024metadynamics}\\
\rowcolor{fessa4!20} PINES\cite{herringer2023permutationally} & AE + PIV inputs (solvent)&  &  &  & \citenum{herringer2023permutationally}\\

\hline
\hline
\rowcolor{fessa5!20} \paragraphtitle{Path-like CVs}& &  &  &  & \\
\rowcolor{fessa5!20} NN PCV\cite{rogal2019neural} & NN local classifier to build path CV&  &  & \citenum{rogal2019neural} & \\
\rowcolor{fessa5!20} KRR PCV\cite{france2024data} & Kernel ridge regression of committor probabilities &  &  &  &\citenum{france2024data}\\
\rowcolor{fessa5!20} Deep-LNE \cite{frohlking2024deep, frohlking2024deeplne++} & Path-like CV via AE and nearest neighbor & \citenum{frohlking2024deep}, \citenum{frohlking2024deeplne++} &  &  &  \\

\hline
\hline
\rowcolor{fessa6!20} \paragraphtitle{Multi-task learning}& &  &  &  & \\
\rowcolor{fessa6!20} Multiple tasks\cite{sun2022multitask} & Encoder with multiple downstream decoder &  &  & \citenum{sun2022multitask} & \\
\rowcolor{fessa6!20} Multiple properties\cite{bonati2023unified} & Semi-supervised AE optimized on different datasets & \citenum{visigalli2025coordinated} &  &  & \citenum{zhang2024combining} \\

\hline
\end{tabular}

}

\caption{Overview of structure-based machine learning collective variables and their applications.}
\label{tb:structural_mlcv}
\end{table*}

\setlength{\arrayrulewidth}{0.1mm}

\subsection{Physics-based approaches: slow modes}
\label{sec:mlcvs_physical_slow}
In this section, we examine physics-based approaches that seek to identify CVs by focusing on the slow modes that govern rare transitions. These include unsupervised techniques that predict future configurations, dynamical operator learning, which designs CVs as eigenfunctions of the relevant operators, and techniques based on the transition matrix, such as diffusion maps and spectral methods.

\subsubsection{Forecasting the dynamics}
\begin{figure}[htbp]
\centering
\includegraphics[
  width=\if\ispreprint1
    1\linewidth
\else
    0.8\linewidth
\fi]{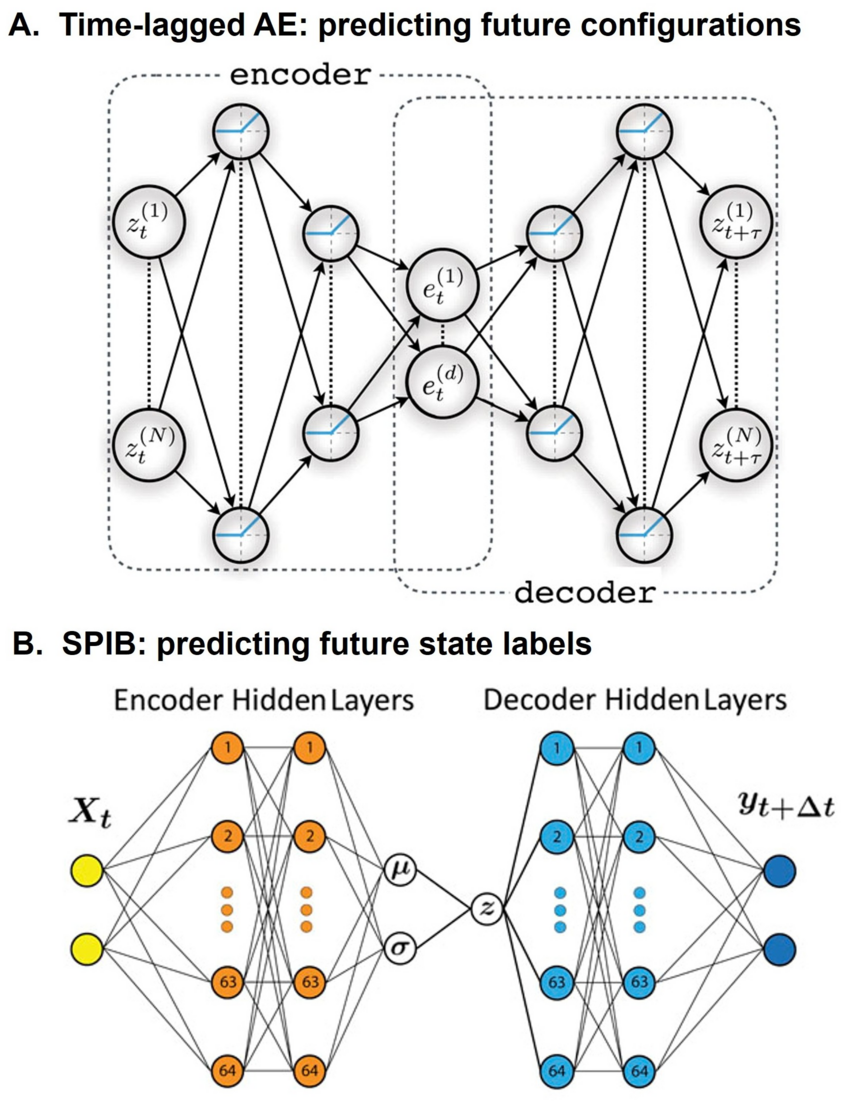}
\caption{\justifying
\textbf{Autoencoder-based frameworks for forecasting dynamics.} (A) Time-lagged autoencoders (TAEs) learn a latent representation that predicts configurations at a future time $t+\tau$. Image reproduced from Ref.~\citenum{wehmeyer2018time}. Copyright 2018 AIP Publishing LLC. (B) State Predictive Information Bottleneck (SPIB) encodes configurations to predict future state labels, enabling automatic identification of metastable states. Image reproduced from Ref.~\citenum{wang2021state}. Copyright 2021 AIP Publishing LLC.
}
\label{fig:future}
\end{figure}

Unsupervised methods can be extended to search for a representation capable not only of compressing the data without losing information, but also of describing the temporal evolution of the data.
One example of such an approach is the \method{time-lagged autoencoders (TAEs)} proposed by Wehmeyer and Noè, which optimize the parameters of the encoder and decoder to predict a configuration observed after a given lag-time $\tau$ (see Fig.~\ref{fig:future}A) \cite{wehmeyer2018time}.
In particular, the encoder of TAEs compresses the information from configurations at time $t$ into the latent space, which represents the CV space as well, and the decoder reconstructs from the bottleneck time-lagged configurations at $t + \tau$. 
Hernandez \textit{et al.} proposed the \method{variational dynamics encoder (VDE)} method, employing variational autoencoders instead, optimized with a time-lagged reconstruction and trained to maximize the autocorrelation of the latent space to resemble the properties of the transfer operator.\cite{hernandez2018variational} Sultan \textit{et al.} \cite{sultan2018transferable} applied the VDE framework to encode information about the slow modes of the systems into CVs for enhanced sampling. 
As a general tendency, however, it has been shown that both TAEs and VDEs tend to learn a mixture of slow and maximum variance modes\cite{chen2019capabilities}.

Similar to TAE and VDE, a time lag can also be introduced in the RAVE approach.\cite{ribeiro2018reweighted} 
Tiwary \textit{et al.} used a \method{past-future information bottleneck (PIB)} optimization scheme and modified the objective function of RAVE to \( L = I(\chi, \mathbf{x}_{\Delta t}) - \gamma I(\mathbf{x}, \chi) \)\cite{wang2019past}. The mutual information \( I(\chi, \mathbf{x}_{\Delta t}) \) quantifies how much an observation at one instant of time \( t \) can tell us about an observation at another instant of time \( t + \Delta t \), while \( I(\mathbf{x},\chi) \) represents the mutual information between input and latent representation $\chi$ at time \( t \). Wang \textit{et al.} discuss the role of predictive time delay in RAVE and further introduce a correction for the objective function to take into account the effect of the biasing potential on the dynamical propagator of the system \cite{wang2020understanding}. Later, Wang \textit{et al.} introduced \method{state predictive information bottleneck (SPIB)} \cite{wang2021state}, which constructs a compressed representation able to predict the future state label.   Once a time delay \( \Delta t \) is selected, SPIB can automatically index the high-dimensional state space into metastable states through an iterative retraining algorithm. Additionally, SPIB tries to carry the maximum information of the state-transition density, which, in principle, can be equivalent to the traditional committor function if there is a timescale separation between the state-to-state transitions and the fluctuations within metastable states. 

\subsubsection{Dynamical operator learning}

Another broad class of approaches for identifying slow CVs in molecular simulations is based on the idea of learning the dynamical operator that governs the time evolution of the system, such as the Koopman or transfer operator\cite{noe2017collective}. Learning these operators offers a description of the system's dynamical modes, which can be obtained from their spectral decomposition, namely, their eigenfunctions \( \psi_i \) and eigenvalues \( \lambda_i \). Of particular interest are the eigenfunctions associated with the largest eigenvalues, which describe the slowest evolving components of the dynamics and often are associated with the rare transitions between metastable states. These eigenfunctions thus offer a natural, low-dimensional representation of the system’s long-time behavior and arguably serve as ideal candidates for CVs in enhanced sampling methods~\cite{mcgibbon2017identification}.

While these operators can't be determined analytically, they can be approximated from time series data. This has led to the development of a family of approaches known as \textit{dynamical operator learning}, which also spans different communities, in which one seeks to recover the dominant spectral components of the underlying operator directly from trajectories. Most of these methods rely on variational principles to construct optimal finite-dimensional approximations of the operator’s eigenfunctions within a chosen function space, such as in (extended) dynamic mode decomposition \cite{schmid2010dynamic,williams2015data} and the \textit{variational approach for conformation dynamics} (\textit{VAC})~\cite{prinz2011markov,perez2013identification,nuske2014variational}. Here we focus on the latter, which has been developed in the context of atomistic simulations and can be seen as a specific instance of Koopman operator learning under the assumptions of equilibrium and time-reversible dynamics~\cite{klus2018data}. VAC relies on a variational principle that allows the eigenfunctions to be approximated using a set of trial functions $\tilde{\psi}_i$. The idea is to find functions that maximize their time-autocorrelation:
\begin{equation}
    \tilde{\lambda}_i = 
    \frac{
      \left\langle \tilde{\psi}_i(\mathbf{R}_t)\tilde{\psi}_i(\mathbf{R}_{t+\tau}) \right\rangle
    }{
      \left\langle \tilde{\psi}_i^2(\mathbf{R}_t) \right\rangle
    }
\end{equation}
The optimal trial functions approximate the true eigenfunctions of the transfer operator, and the corresponding values \( \tilde{\lambda}_i \ge{\lambda}_i \) reflect how slowly these modes decay over time. This variational formulation connects directly to quantities accessible from trajectory data, enabling the extraction of slow CVs from molecular simulations in a statistically grounded way.

In practice, the trial functions can be expressed as linear or nonlinear combinations of features, with parameters optimized to maximize the autocorrelation score. A widely used implementation of the VAC principle is \method{time-lagged independent component analysis (TICA)}~\cite{molgedey1994separation,naritomi2011slow,perez2013identification,schwantes2015modeling}. Originally introduced as a signal processing technique to extract slowly decorrelating components from multivariate time series~\cite{molgedey1994separation}, TICA has been shown to be equivalent to VAC when the trial functions are restricted to linear combinations of input features~\cite{perez2013identification}:
\begin{equation}
\tilde{\psi}_i(\mathbf{R}_t) = \mathbf{w}_i^\top \mathbf{x}_t
\end{equation}
where \( \mathbf{x}_t \) is the features vector (e.g., distances, angles) at time \( t \), and \( \mathbf{w}_i \) are the coefficients. These are obtained by solving the generalized eigenvalue problem:
\begin{equation}
C_{\tau} \, \mathbf{w}_i = \tilde{\lambda}_i \, C_0 \, \mathbf{w}_i
\end{equation}
where \( C_0 = \langle \mathbf{x}_t \mathbf{d}_t^\top \rangle \) is the covariance matrix and \( C_{\tau} = \langle \mathbf{d}_t \mathbf{d}_{t+\tau}^\top \rangle \) is the time-lagged covariance matrix. In this way, TICA identifies orthogonal directions in feature space that maximize autocorrelation at a chosen lag time, thereby capturing the slowest dynamical processes in the data. 
Unlike methods such as PCA and LDA, which focus on maximizing structural variance, TICA is explicitly designed to extract slow modes and is particularly useful for identifying reaction coordinates and kinetic bottlenecks in complex molecular systems. It has been applied to analyze molecular trajectories~\cite{mccarty2017variational,yang2018refining,m2017tica} and to derive CVs for enhanced sampling~\cite{m2017tica}. McCarty and Parrinello further expanded this idea by learning effective CVs from biased metadynamics trajectories using TICA in combination with reweighting techniques~\cite{mccarty2017variational}.

\begin{figure}[htbp]
\centering
\includegraphics[width=1.0\linewidth]{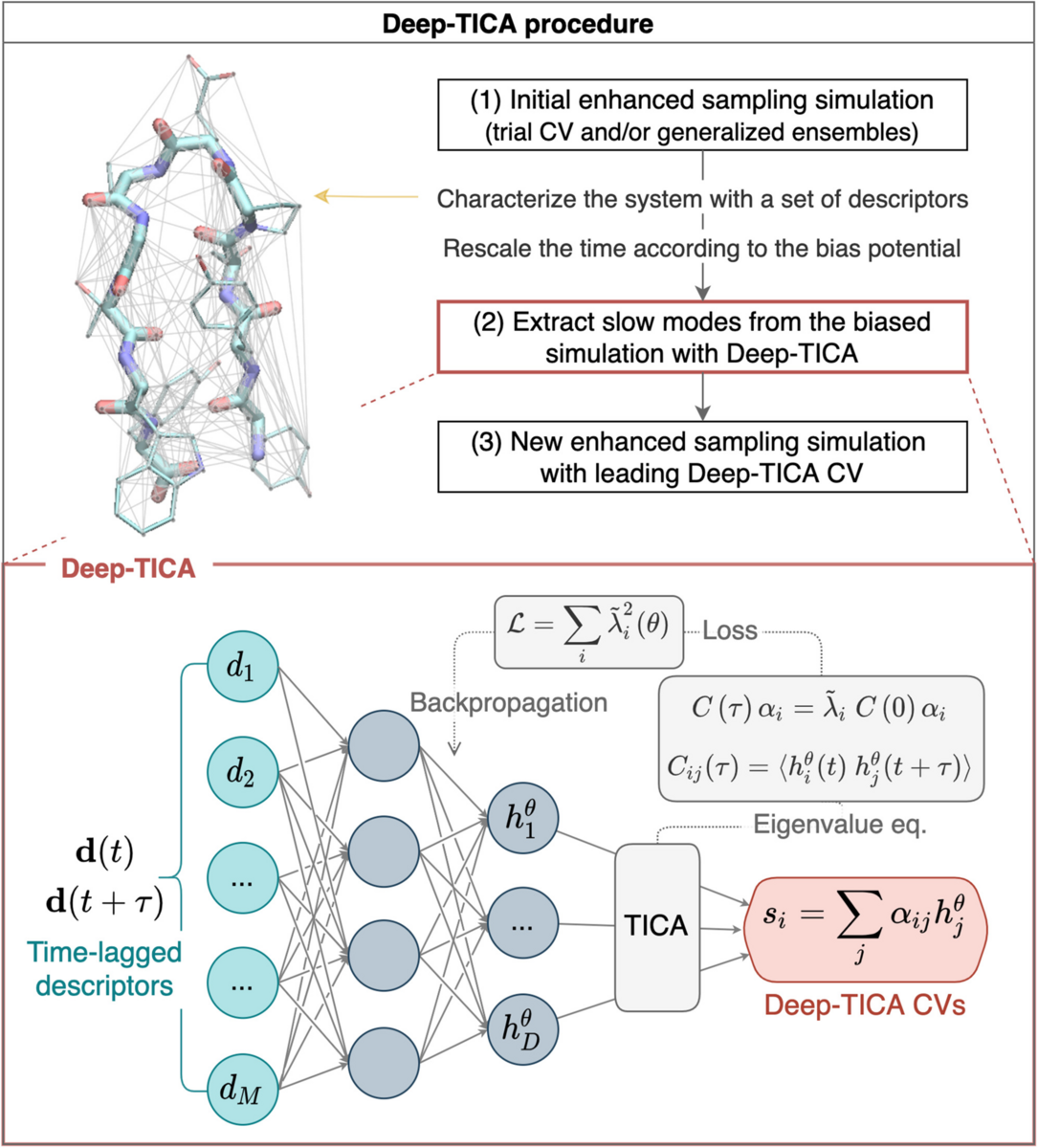}
\caption{\justifying
Deep-TICA for learning slow CVs. This method uses the transfer operator framework to learn CVs that capture the system’s slow modes and remove dynamical bottlenecks in simulations.  (Top) The protocol used: an initial enhanced sampling simulation is performed using a trial CV or generalized ensemble; time is rescaled to account for the bias potential; slow modes are extracted and used as CVs to drive a new enhanced sampling simulation. (Bottom) Neural network architecture: pairs of time-lagged descriptors are mapped into a latent space, where TICA is applied to compute eigenvalues and eigenfunctions. The NN transformation is then optimized to maximize the TICA score (eigenvalues). Image reproduced from Ref.~\citenum{bonati2021deep}. Copyright 2021 National Academy of Science.}
\label{fig:deeptica}
\end{figure}

Several nonlinear extensions to TICA have been proposed to increase its representational power and better capture complex dynamical modes, including kernel methods~\cite{schwantes2015modeling} and neural networks~\cite{bonati2021deep,spiwok2020time}. Here, we focus in particular on those relevant to enhanced sampling simulations. Bonati \textit{et al.} introduced \method{Deep-TICA}~\cite{bonati2021deep} which uses neural networks trial functions in the VAC framework by applying TICA to the output of a learned nonlinear transformation (see Fig.~\ref{fig:deeptica}).
The original inputs \( \mathbf{d}_t \) (e.g., atomic coordinates or structural features) are transformed by a neural network into hidden features  \( \mathbf{h}_\theta = f_\theta(\mathbf{d}_t) \), where \( \theta \) denotes the trainable parameters. TICA is then applied in the space of the learned features \( \mathbf{h}_\theta \), and the NN is optimized to produce features that best approximate the leading slow modes, or in other words, that have the longest autocorrelation. This is achieved by minimizing the negative sum of the top $n$ eigenvalues :
\begin{equation}
\mathcal{L}_{\text{Deep-TICA}}(\theta) = - \sum_{i=1}^n \tilde{\lambda}_i^2
\end{equation}
obtained by solving the TICA eigenproblem on the transformed descriptors \( \mathbf{d}_\theta \).
To address the challenge of obtaining relevant data, the authors proposed to start from CVs free methods to generate initial biased trajectories, such as multicanonical sampling. Alternatively, an initial simulation may be carried out using CVs optimized via structural criteria, and Deep-TICA can then be applied to refine the CVs. In both cases, the time is rescaled according to the instantaneous acceleration induced by the bias potential \( V \), i.e., \( \Delta t' = e^{\beta V(x_{t_k})} \,\Delta t \), and the time-correlation functions are computed in this accelerated space using unevenly spaced intervals proposed in Ref.~\citenum{yang2018refining}. While being an approximation of the unbiased dynamical modes, this method identifies the sampling bottlenecks of the initial biased simulation and, by using them as CVs, can enhance sampling by orders of magnitude. 

Another nonlinear variant builds on the \emph{Variational Approach to Markov Processes} (VAMP), which generalizes the VAC principle to non-equilibrium settings. In particular, \method{VAMPnets}~\cite{mardt2018vampnets} uses a two-lobed unsupervised neural network that maps pairs of molecular configurations (separated by a lag time \( \tau \)) into a low-dimensional latent space. These outputs are then used to estimate time-lagged covariance matrices and optimize the VAMP score, which quantifies the quality of the learned dynamical model. However, because VAMPnets typically express their output in terms of soft assignments to metastable states, they are not directly suited for defining CVs in enhanced sampling.
To address this, Chen \textit{et al.} introduced a variant called \method{state-free reversible VAMPnets (SRVs)}~\cite{chen2019nonlinear}, which directly approximates the eigenfunctions \( \tilde{\psi}_i \) of the transfer operator using a siamese neural network architecture, similar in spirit to Deep-TICA (although SRVs were introduced earlier, but only for unbiased simulations).
Building on this methodology, Shmilovich \textit{et al.} proposed the \method{Girsanov reweighting enhanced sampling technique (GREST)}~\cite{shmilovich2023girsanov}, which extends SRVs to biased simulations. GREST uses the Girsanov formalism~\cite{donati2017girsanov,donati2018girsanov} to reweight biased trajectories, accounting for both thermodynamic and integrator-specific path corrections, and enables unbiased estimation of dynamical observables from biased simulations.

Furthermore, the TICA principle has been integrated into the t-distributed stochastic neighbor embedding (t-SNE) method, leading to the development of \method{time-lagged t-SNE} \cite{spiwok2020time}, a variant that emphasizes slowly evolving molecular modes over fast fluctuations. To address the requirement of out-of-sample embedding and differentiability required by enhanced sampling techniques, this approach was extended into a parametric time-lagged t-SNE, where a neural network was trained to map Cartesian coordinates to a low-dimensional latent space while preserving time-lagged similarities. The resulting coordinates were then used as CVs in metadynamics simulations \cite{hradiska2024acceleration}.

To summarize, all these methods aim to approximate the leading eigenfunctions of a dynamical operator. A central challenge common to all approaches is the need for sufficiently informative data. Since the operators being learned reflect the system’s long-time dynamics, the quality of the approximation crucially depends on whether the relevant transitions are sampled in the input trajectories. This often requires the use of enhanced sampling techniques to generate such data. However, biased simulations introduce distortions in the sampled dynamics, which must be corrected if one wishes to recover unbiased dynamical information, as exemplified by the GREST approach. For a comparison between different reweighting strategies for time-lagged data, see also  Ref.~\citenum{chen2023chasing}.

An alternative approach has recently been pursued by considering the limit of vanishing lag time, which leads to the definition of the infinitesimal generator rather than the transfer operator. When assuming that the probability density evolves toward equilibrium according to an overdamped Langevin equation, the \method{infinitesimal generator} admits an analytical expression given by the backward Kolmogorov equation \cite{schutte2023overcoming}. This analytic form enables the computation of slow modes directly from Boltzmann-weighted averages 
\cite{zhang2022solving,klus2020data,nuske2023efficient,devergne2024unbiased,devergne2025slow}, thereby facilitating the use of data obtained from biased simulations~\cite{zhang2022solving,devergne2024unbiased,devergne2025slow}. Notably, Devergne \textit{et al.} 
\cite{devergne2024unbiased,devergne2025slow} demonstrated that, even using biased simulations, it is possible to recover the time evolution of the occupation numbers of metastable states in molecular systems. Moreover, the infinitesimal generator has been used in conjunction with generative models to extract kinetic properties \cite{moqvist2025thermodynamic} (Sec.~\ref{sec:generative}).

\subsubsection{Spatial techniques}

While the methods discussed in the previous section aim to learn the spectral properties of dynamical operators directly from time-series data, another family of approaches focuses instead on deriving CVs by constructing or approximating a \emph{transition matrix} between the states. 
A common feature of these methods is that they infer dynamical information by analyzing how configurations are likely to evolve probabilistically between states. In a recent review, Gokdemir and Rydzewski~\cite{gokdemir2025machine} refer to this class as \textit{spatial techniques}, since they exploit only equilibrium or thermodynamic information to build the transition matrix as opposed to time-lagged data.

\method{Diffusion maps} \cite{belkin2001laplacian} estimate slow CVs by constructing a Markovian representation of data and diagonalizing the transition matrix. The construction of the matrix starts with computing pairwise similarities between data points using a Gaussian kernel $G_\epsilon(x_k, x_l) = \exp\left(-\frac{\|x_k - x_l\|^2}{\epsilon^2}\right)$, where the bandwidth  $\epsilon$ controls the locality of interactions. To correct for non-uniform sampling, an anisotropic kernel is often employed: $
K(x_k, x_l) = G_\epsilon(x_k, x_l)/(\rho^\alpha(x_k) \rho^\alpha(x_l))$,
where  $\rho(x_k) = \sum_l G_\epsilon(x_k, x_l)$  is the density estimate and $\alpha$  is a parameter adjusting the degree of density correction. Normalizing this kernel yields the Markov transition matrix:
\begin{equation}
M(x_k, x_l) = \frac{K(x_k, x_l)}{\sum_i K(x_k, x_i)}
\end{equation}
which defines a discrete diffusion process over the dataset. The essential step in diffusion maps is the eigenvalue decomposition of this transition matrix:
$M v_k = \lambda_k v_k$, where the eigenvalues  $\lambda_k$  provide a measure of the timescales of diffusion, and thus the leading eigenfunctions define the diffusion coordinates, which project data into a reduced space preserving slow dynamics:
$\Psi_k(x) = \lambda_k^t v_k(x)$, where $t$ is a diffusion time parameter controlling the scale of the transformation. For this reason, the diffusion coordinates could serve as effective CVs. \cite{ferguson2011integrating,zheng2013rapid,preto2014fast}

Different generalizations have been proposed to adjust the transition probabilities and extract unbiased slow CVs in the case in which the dataset comes from enhanced sampling simulations. Evans \textit{et al.} used a Mahalanobis kernel to account for position-dependent diffusion coefficients \cite{evans2023computing} and corrected the probability distribution based on the applied bias potential\cite{evans2022computing}, $
p(x) \propto p_{\text{bias}}(x) e^{\beta V_{\text{bias}}(x)}$.
Other techniques, such as \method{stochastic kinetic embedding (StKE) }\cite{zhang2018unfolding} and \method{multiscale reweighted stochastic embedding (MRSE)} \cite{rydzewski2021multiscale},  incorporate the effect of the bias potential as importance weights to construct an unbiased Markov transition matrix. A crucial aspect that distinguishes stochastic embedding methods from diffusion maps is that they are optimized by minimizing the KL divergence between the transition probabilities in feature space $p_{ij}$ and those in latent space $q_{ij}$. In this way, it is possible to learn CVs that preserve the dynamical structure of the system. 

Another family of methods, which includes \method{spectral gap optimization of order parameters (SGOOP)} \cite{tiwary2016spectral}, seeks to optimize CVs by maximizing the spectral gap of a transition matrix. The spectral gap indeed quantifies the separation between slow and fast dynamics, with a large value indicating a good choice of CVs, ensuring that metastable states are well separated. In SGOOP, a linear combination was optimized using a maximum path entropy estimate of the spectral gap. Similarly, \method{Spectral Map} \cite{rydzewski2023spectral} used a neural network mapping and optimized the spectral gap of the transition matrix in the reduced nonlinear space. Maximizing timescale separation in the spectral map induces dynamics with Markovian characteristics in the reduced space~\cite{gokdemir2025machine}, and this framework can also be extended to learn TSEs \cite{rydzewski2024spectral}.

\setlength{\arrayrulewidth}{0.0mm}
\setlength{\tabcolsep}{-1pt}
\renewcommand{\arraystretch}{1.5}

\newcolumntype{C}[1]{>{\centering\arraybackslash}p{#1}}

\begin{table*}[htbp]
\centering

\begin{tabular}{lC{5.5cm}C{2.2cm}C{2.2cm}C{2.2cm}C{2.2cm}}
& \shortstack{Notes} & \shortstack{Conformational\\Biophysics} & \shortstack{Ligand\\Binding} & \shortstack{Phase\\Transformations} \\
\hline
\rowcolor{fessa3!20} \paragraphtitle{Forecasting} &  &  &&  \\
\rowcolor{fessa3!20} TAE\cite{wehmeyer2018time} &  Time-lagged AE &\citenum{wehmeyer2018time}   &  &  \\
\rowcolor{fessa3!20} VDE\cite{hernandez2018variational} & Time-lagged VAE + autocorrelation&\citenum{sultan2018transferable}  &  &  \\
\rowcolor{fessa3!20} (S)PIB\cite{wang2019past,wang2021state} & Information bottleneck AE&\citenum{wang2021state,mehdi2022accelerating,majumder2024machine}  &  \citenum{mehdi2022accelerating} &\citenum{zou2023driving,wang2024local,meraz2024simulating,wang2025electric} \\

\hline
\hline
\rowcolor{fessa4!20} \paragraphtitle{Slow modes} &  &  & & \\
\rowcolor{fessa4!20} TICA & Linear VAC  &\citenum{m2017tica}  &  & \citenum{zhang2019improving}\\
\rowcolor{fessa4!20} Deep-TICA\cite{bonati2021deep} &  NN extension of TICA&\citenum{bonati2021deep}  &\citenum{ansari2022water,muscat2024leveraging}  &\citenum{bonati2021deep}  \\
\rowcolor{fessa4!20} SRV \cite{chen2019nonlinear}\& GREST\cite{shmilovich2023girsanov} & State-free VAMPnets &\citenum{sidky2019high,shmilovich2023girsanov} & & \\ 
\rowcolor{fessa4!20} Time-lagged t-SNE \cite{spiwok2020time} &t-SNE + TICA  &\citenum{hradiska2024acceleration} & & \\ 

\hline
\hline
\rowcolor{fessa5!20}\paragraphtitle{Spatial techniques} &  &  &  &\\
\rowcolor{fessa5!20} Diffusion maps\cite{belkin2001laplacian} & Non-linear kinetic diffusion embedding&\citenum{ferguson2011integrating,zheng2013rapid,preto2014fast} & & \\
\rowcolor{fessa5!20} StKE\cite{zhang2018unfolding} \& MRSE\cite{rydzewski2021multiscale}  &  Stochastic embedding&\citenum{zhang2018unfolding}  &  &  \\
\rowcolor{fessa5!20} SGOOP\cite{tiwary2016spectral} &Linear spectral gap optimization &\citenum{tiwary2017molecular}  &\citenum{tiwary2016wet,shekhar2022protein}  &\citenum{tsai2019reaction,zou2021toward}  \\
\rowcolor{fessa5!20} Spectral Map\cite{rydzewski2023spectral} & NN spectral gap optimization  &\citenum{rydzewski2024spectral,rydzewski2024learning}  &  &  \\

\hline
\end{tabular}

\caption{Overview of physics-based machine learning collective variables and their applications.}
\end{table*}

\setlength{\arrayrulewidth}{0.1mm}

\subsection{Physics-based approaches: leveraging the committor function}
\label{sec:mlcvs_physical_committor}

In this section, we examine another class of physics-based approaches that use the committor function to learn CVs that characterize rare transitions in complex systems. 
The committor is a central quantity in the theory of rare events and underpins many enhanced sampling techniques. 
As a result, a number of methods have been developed that either machine-learn the committor or derive CVs based on its properties.

We start by recalling its definition and some relevant properties. 
Given two metastable states, A and B, the committor function $p_B(\mathbf{R})$ denotes the probability that a trajectory initiated from configuration $\mathbf{R}$ will reach state B before A~\cite{vanden2010transition, bolhuis2002transition}. 
As a consequence, it satisfies $p_B(\mathbf{R})=0$ in basin A, $p_B(\mathbf{R})=1$ in basin B, and smoothly interpolates between these values along transition paths. 
In addition, sampled configurations for which $p_B(\mathbf{R}) \simeq 0.5$ are usually associated with the TSE, as they are equally likely to proceed to either basin. 
For these reasons, the committor is considered by many an ideal reaction coordinate for the description of rare events~\cite{ma2005automatic, peters2006obtaining, krivov2013optimal, rogal2021reaction}.

In practice, committor values for a given configuration $\tilde{\mathbf{R}}$ can be estimated by initiating a large number of unbiased trajectories from $\tilde{\mathbf{R}}$ and counting the fraction that reaches B before A. 
This empirical committor distribution can also be used to assess the quality of a CV, based on the idea that a good CV should strongly correlate with $p_B(\mathbf{R})$, i.e., configurations with the same CV value should lie on the same isocommittor surface. 
This principle can be used to guide the construction or optimization of CVs\cite{best2005reaction}.

Methods for learning the committor can be broadly divided into three classes: (1) regression approaches that fit an explicit model to empirical committor values, (2) maximum likelihood approaches used with transition path data, and (3) variational principles. In the following, we will refer to as $q(\mathbf{R})$ as the parametrizations of the committor. 

\paragraphtitle{Regression}. If a dataset of "experimental" committor values $\left(\mathbf{R}_i\rightarrow p_B(\mathbf{R}_i)\right)$ is available, one can directly optimize the function $q(\mathbf{R})$ approximate the committor function by minimizing the residual squared $|q(\mathbf{R}_i)-p_B(\mathbf{R}_i)|^2$.
In a pioneering study, Ma and Dinner~\cite{ma2005automatic} trained a neural network on structural features (e.g., distances and angles) to predict committor values directly from molecular configurations.
As shown by France-Lanord \textit{et al.}, this approach can also be seen as learning a data-driven path CV~\cite{france2024data}. 
A systematic comparison of ML models to learn the committor was carried out by Naleem \textit{et al.}~\cite{naleem2023exploration}; however, this type of supervised learning approaches requires large numbers of committor trajectories, which are computationally expensive to generate~\cite{peters2007extension}.

\paragraphtitle{Maximum likelihood}. An alternative strategy to reduce the cost of learning the committor is based on Maximum Likelihood Estimation\cite{peters2006obtaining,peters2007extension,allen2009forward}.  
The core idea behind MLE is to determine the model parameters that best describe the observed data by maximizing the likelihood function.
In the approach proposed by Peters and Trout~\cite{peters2006obtaining}, the committor is modeled as a sigmoid function of a single reaction coordinate $s(\mathbf{R})$, which is expressed as a linear combination of physical descriptors. This yields a committor model of the form:
    \begin{equation}
        q(\mathbf{R}) = q(s(\mathbf{R}))  = \frac{1}{1 + e^{-s(\mathbf{R})}}
    \end{equation}
The data are obtained from TPS, where each shooting point is labeled according to whether the resulting trajectory reaches state B or A. 
This outcome can be interpreted as an instantaneous evaluation of the committor function.  
Given trajectories shot from  configurations $\{\mathbf{R}_i\}$, the likelihood of observing their outcome is written as:
\begin{equation}
    \mathcal{L} = \prod_{i \in \mathcal{B}} q(s(\mathbf{R}_i)) \prod_{i \in \mathcal{A}} \left[1 - q(s(\mathbf{R}_i))\right]
    \label{eq:likelihood}
\end{equation}
where $\mathcal{B}$ and $\mathcal{A}$ denote the subsets of shooting points that terminate in states B and A, respectively. 
The parameters of the reaction coordinate $s(\mathbf{R})$, such as the coefficients in the linear combination, are then optimized to maximize the likelihood in Eq.~\ref{eq:likelihood}.

Several extensions and modifications of the MLE framework have since been developed. 
These include the use of nonlinear string-based coordinates~\cite{lechner2010nonlinear}, alternative path sampling techniques such as forward flux sampling~\cite{allen2009forward}, and strategies to reduce the number of committor evaluations~\cite{li2015reducing}. 

\begin{figure*}[htbp]
 \centering
 \includegraphics[width=0.85\linewidth]{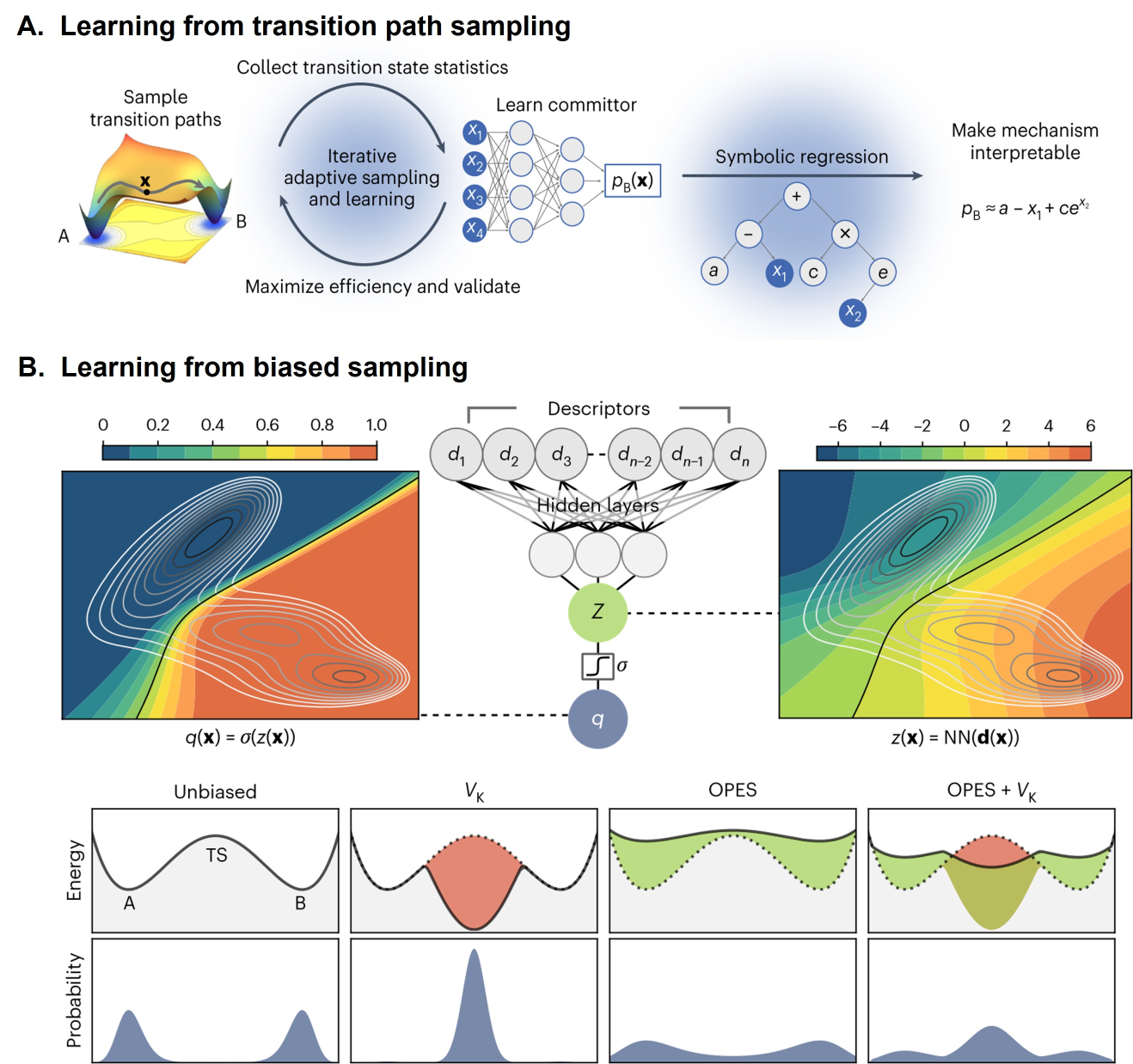}
 \caption{\justifying
 Two approaches for learning the committor function and enhancing sampling of the transition state. (A) The AIMMD method iteratively combines TPS with a neural network estimate of the committor \(p_B(x)\), which is then used to promote shooting from the transition state. At convergence, symbolic regression distills an interpretable expression for the mechanism. Image adapted from Ref.~\citenum{jung2023machine}. Copyright 2023 Springer Nature under \href{https://creativecommons.org/licenses/by/4.0/}{[CC BY 4.0 DEED]}. (B) (top) A variational approach where a neural network maps descriptors \(d(x)\) into a smooth latent space \(z\), related to the committor, and adds a bias $V_K$ to keep the system near the transition state, improving committor estimates. (bottom) The panels illustrate how the bias $V_K$ can also be integrated with standard CV-based biases such as OPES to obtain a combined effect. Image adapted from Ref.~\citenum{trizio2025everything}. Copyright 2025 Springer Nature.}
 
 \label{fig:committor-es}
\end{figure*}

More recently, neural networks have been adopted to represent the reaction coordinate in a more flexible and expressive way. 
Frassek \textit{et al.}~\cite{frassek2021extended} introduced an extended autoencoder architecture in which the latent bottleneck representation is passed to a separate predictor module trained to estimate committor values. 
Jung \textit{et al.}~\cite{jung2023machine} developed the \method{artificial intelligence for molecular mechanism discovery (AIMMD)} framework, which iteratively combines neural-network-based committor learning via MLE with adaptive selection of shooting points in TPS (see Fig.~\ref{fig:committor-es}A).
Additionally, symbolic regression is used post hoc to extract interpretable physical expressions for the learned coordinate. 
This approach was later extended by Lazzeri \textit{et al.}~\cite{lazzeri2023molecular}, who introduced reweighting schemes that allow for the recovery of free energy estimates from the TPS data.

\paragraphtitle{Variational approaches} for learning the committor function have also been proposed that bypass the need to estimate committor values explicitly. 

Krivov and coworkers exploited a variational principle based on minimizing the \method{total squared displacement} over equilibrium trajectories that start in A and end in B.~\cite{banushkin2015nonparametric, krivov2018folding} Specifically, they showed that the committor minimizes the functional
    \begin{equation}
       \Delta q^2 = \sum_k \left[q(k\Delta t_0 + \Delta t_0) - q(k\Delta t_0)\right]^2,
    \end{equation} 
where $q(t)$ is constrained to satisfy $q = 0$ in state A and $q = 1$ in state B.

Roux and coworkers formulated a variational principle based on the dynamical evolution of the system as governed by the propagator $\mathcal{P}_\tau$. 
Under appropriate assumptions~\cite{roux2021string, roux2022transition}, the committor function can be obtained by minimizing the \method{steady-state unidirectional reactive flux}:
    \begin{equation}
       J_{AB} = \frac{1}{2\tau} \left\langle \left[q(\tau) - q(0)\right]^2 \right\rangle = \frac{1}{\tau} C_{qq}(\tau)
    \end{equation}
where $C_{qq}(\tau) = \langle q(0)^2 \rangle - \langle q(0)\, q(\tau) \rangle$ is a time correlation function. 
Interestingly, this approach is somewhat akin to the one exploited in time-informed methods such as TICA~\cite{molgedey1994separation, bonati2021deep} and VAMPnets~\cite{mardt2018vampnets}. 
The variational flux principle has been applied using the string method by He \textit{et al.}~\cite{he2022committor} and siamese neural networks by Chen \textit{et al.}~\cite{chen2023discovering} and also by Megias \textit{et al.}~\cite{megias2025iterative}. In the latter, the variational approach was used to learn committor-consistent strings in a reduced CV space to be used as a path CV.

Another line of work derives a variational formulation from the \method{Kolmogorov backward equation}, which governs the committor function under overdamped Langevin dynamics. 
The corresponding function to be minimized is
\begin{equation}
  \mathcal{K}[q] = \frac{1}{Z} \int |\nabla q(\mathbf{R})|^2 e^{-\beta U(\mathbf{R})} \, \mathrm{d}\mathbf{R} = \left\langle |\nabla q(\mathbf{R})|^2 \right\rangle_{U}
\end{equation}
where $Z$ is the partition function and $\langle \cdot \rangle_U$ denotes an average over the Boltzmann distribution. 
The boundary conditions $q = 0$ in A and $q = 1$ in B are imposed.
As pointed out by Khoo \textit{et al.}~\cite{khoo2019solving}, this formulation faces two main challenges: the gradients $\nabla q$ are sharply localized in the transition region, and accurate Boltzmann-weighted sampling is required. 
To alleviate these issues, Li \textit{et al.}~\cite{li2019computing} combined high-temperature simulations or metadynamics with simple CVs. 
Rotskoff \textit{et al.}~\cite{rotskoff2022active} employed replica exchange and umbrella sampling to enrich sampling of the TSE.

To further address the sampling difficulty, Parrinello and coworkers \cite{kang2024computing} proposed a self-consistent biasing scheme (Fig.~\ref{fig:committor-es}B) that enhances sampling of the transition state region by introducing the following bias functional of the committor:
\begin{equation}
    V_\mathcal{K}(\mathbf{R}) = -\frac{\lambda}{\beta} \log\left(|\nabla q(\mathbf{R})|^2 + \epsilon\right)
\end{equation}
where $\lambda \sim 1$ controls the bias strength and $\epsilon > 0$ is a regularization term. 
This bias guides the system toward regions where the gradient norm is large, enabling efficient sampling of the TSE, thus providing the data needed to optimize the committor via the variational formulation.
  
We conclude this section with two \textit{general considerations}. 
First, learning the committor function is particularly challenging due to the difficulty of obtaining informative data. 
In rare event scenarios, the committor is nearly constant (i.e., close to 0 or 1) for the vast majority of configurations, and only exhibits non-trivial behavior in the narrow transition region. 
As a consequence, data that provide meaningful information about the committor are inherently rare and difficult to sample. 
For this reason, it is crucial to employ iterative schemes that progressively enhance the sampling of the TSE.
Although originating from different methodological frameworks, both AIMMD~\cite{jung2023machine} and the variational approach proposed by Kang \textit{et al.}~\cite{kang2024computing} follow a similar strategy: they leverage a learned approximation of the committor to guide the next generation of informative data. 
The former employs a neural network estimate of the committor to iteratively refine shooting point selection in TPS, while the latter defines a bias potential based on the committor’s gradient, effectively turning the transition state region into a free energy minimum and promoting its exploration (Fig.~\ref{fig:committor-es}).

Second, while the committor is widely regarded as an ideal reaction coordinate from a theoretical standpoint~\cite{berezhkovskii2004reaction, krivov2013optimal, roux2022transition}, its direct use as a CV in biased enhanced sampling schemes poses significant challenges.
As mentioned just above, within metastable basins, the committor is approximately constant (i.e., close to 0 or 1), leading to vanishing gradients and, consequently, ineffective biasing forces. 
In contrast, within the transition region, the committor changes rapidly over a narrow range, which can result in large and unstable gradient values. 
These features limit the stability and effectiveness of using the committor directly as a biasing variable.
To mitigate these issues, one can transform the committor using a smoothing function, for example, $\text{logit}(q) = \log\left(q / (1 -q)\right)$, or even adapt the biasing protocol itself. 
For instance, Rotskoff \textit{et al.}~\cite{rotskoff2022active} designed an umbrella sampling scheme in which the window widths are tailored to the shape of the committor. 
Another strategy, proposed by Trizio \textit{et al.}~\cite{trizio2025everything}, circumvents the use of the committor itself as a CV. 
Instead, they insert a sigmoid activation function at the final layer of the neural network and define the CV as the pre-activation output (analogous to the reaction coordinate in maximum likelihood approaches). 
This choice yields a smoothly varying variable that avoids saturation in the metastable basins while encoding the same information as the committor. 
The resulting CV can then be effectively biased, enabling stable and efficient enhanced sampling (see Fig.~\ref{fig:committor-es}B).

\subsection{Software}
  
The development of MLCVs can significantly expand the capabilities of enhanced sampling methods. However, implementing these techniques in practice requires careful handling of data preprocessing, model training, and integration with MD engines. To streamline these workflows and make MLCVs more accessible, several software packages have been created. In this section, we review prominent tools that support the construction, training, and deployment of MLCVs in molecular simulations.

\begin{figure*}[t!]
  \centering
  \includegraphics[width=0.75\linewidth]{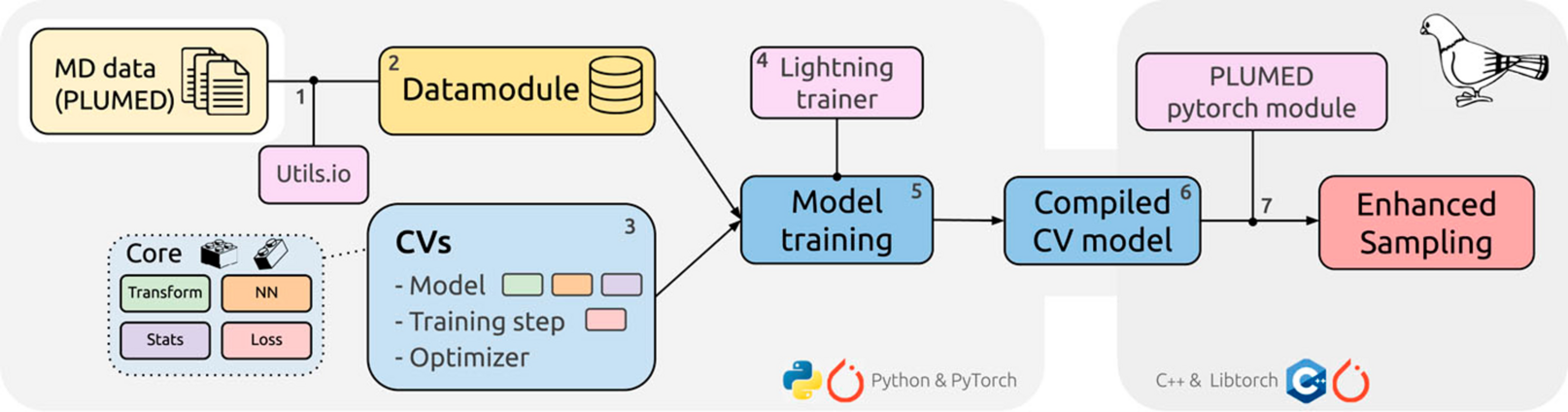}
  \caption{\justifying
  Schematic summary of the workflow for the construction of data-driven CVs in \texttt{mlcolvar}. A CV is selected from ready-to-use ones (\texttt{mlcolvar.cvs}) or built from the implemented building blocks (\texttt{mlcolvar.core}). After training, the model is compiled with the TorchScript language to be deployed to \texttt{PLUMED} for using it as a CV to enhance sampling. Image reproduced from Ref.~\citenum{bonati2023unified}. Copyright 2023 AIP Publishing LLS.}
  \label{fig:mlcolvar}
  \end{figure*}
  
\software{mlcolvar} is a \texttt{Python} package developed by Bonati \textit{et al.}\cite{bonati2023unified} to construct and deploy MLCVs via the \texttt{PLUMED}\cite{tribello2014plumed}  plugin for free energy calculations. It provides a unified interface for defining, training, and exporting a wide range of CV models. Different architectures (such as FNN, AEs, GNNs) and objective functions are available, including a multitask framework to combine multiple objectives. A typical workflow involves extracting trajectory data using \texttt{PLUMED}, training the CV with \texttt{mlcolvar}, compiling the model with Torchscript, and loading it inside \texttt{PLUMED} using the \texttt{pytorch} module (Fig.~\ref{fig:mlcolvar}). The package also supports post-processing and interpretability tools. Comprehensive documentation, together with tutorials and examples, is available at \url{https://mlcolvar.readthedocs.io/en/stable/}.

\software{MLCV}\cite{chen2021mlcv}, developed by Chipot and collaborators, integrates neural network models within the \texttt{Colvars} library\cite{fiorin2013using}. It is written in \texttt{C++}, and it provides an interface for defining and evaluating neural networks using native \texttt{Colvars} inputs.  To use MLCV, users need to extract the weights, biases, and activation functions of each layer from a \texttt{TensorFlow} neural network model into a text file using a \texttt{Python} script.  The \texttt{MLCV} module is available in the latest release of \texttt{Colvars}~\cite{fiorin2024expanded}. Source code and examples can be found at \url{https://github.com/Colvars/colvars/tree/master}.

\software{DeepCV}\cite{ketkaew2022deepcv}, developed by Ketkaew and Luber, implements the DAENN algorithm~\cite{ketkaew2022machine}. It is built on \texttt{TensorFlow} and the software is implemented in both \texttt{Python} and \texttt{C++} for efficient integration and extensibility. Documentation and tutorials are available at \url{https://lubergroup.pages.uzh.ch/deepcv/}.

\section{Applications of machine-learned CVs}
\label{sec:mlcvs_applications}
As discussed in the previous section, a wide range of ML approaches have been put forward to construct CVs, opening the door to an expanding range of applications in molecular simulations. To highlight their impact, we dedicate this section to showcasing the types of problems they can address and outlining the practical considerations involved. MLCVs have been particularly successful in tackling rare events that are beyond the reach of conventional MD, such as conformational transitions in biomolecules (e.g., protein folding), host–guest binding and unbinding, structural phase transformations, and complex chemical reactions. For each of these domains, we first review representative studies that illustrate how MLCVs have been applied to diverse systems and challenges. We then distill common methodological strategies, identify recurring limitations, and discuss open questions, aiming to provide a comprehensive perspective on the current capabilities and future directions of MLCVs in molecular simulations.

\begin{figure}[h!]
  \centering
  \includegraphics[
  width=\if\ispreprint1
    1\linewidth
\else
    0.65\linewidth
\fi]{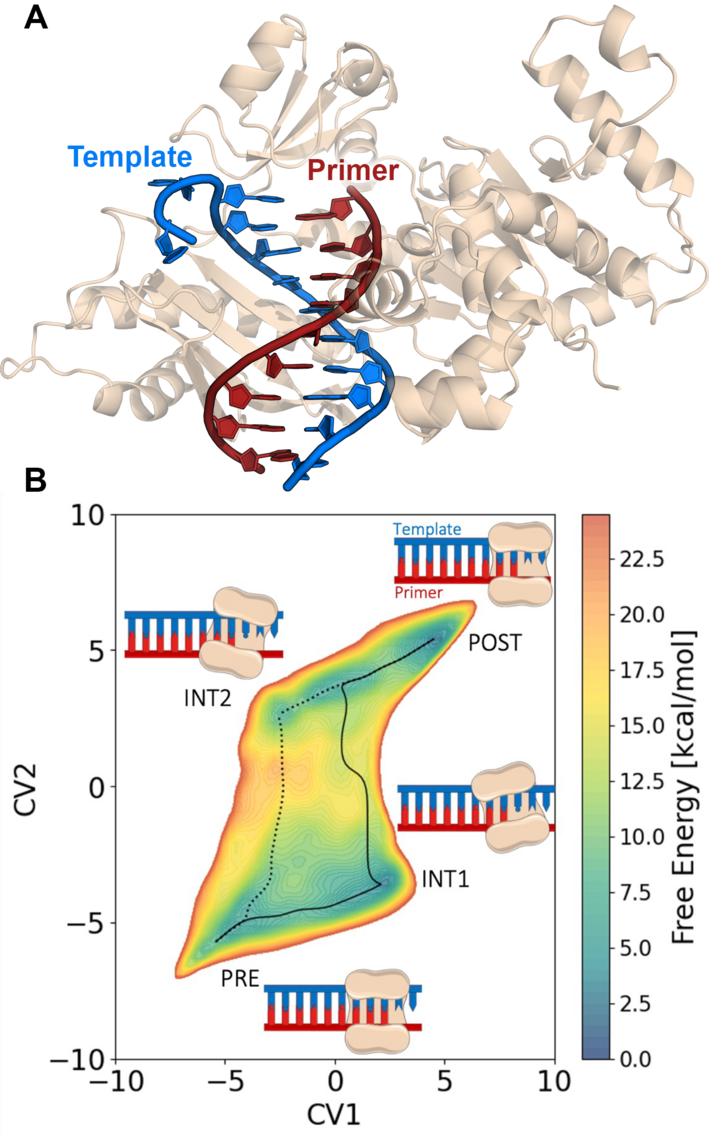}
  \caption{\justifying
    DNA translocation in the Pol$\eta$ enzyme. (A) The template strand (blue) and primer strand (red) are shown together with the enzyme (cartoon). (B) Free energy surface computed using a 2D semi-supervised multitask CV that integrates experimentally determined initial (PRE) and final (POST) states with intermediates and pathways identified from Deep-LDA-based OPES simulations. The simulations revealed two distinct translocation routes: pathway~1 (dashed line), where the primer translocates first followed by the template strand (via INT1), and pathway~2 (dotted line), where the template moves first (via INT2). This 2D representation captures the asynchronous DNA translocation mechanisms and highlights their relative free energy costs. Image A courtesy of Alessia Visigalli; image B reproduced from Ref.~\citenum{visigalli2025coordinated}. Copyright 2025 American Chemical Society under  \href{https://creativecommons.org/licenses/by/4.0/}{[CC BY 4.0 DEED]}. 
  }
  \label{fig:DNA}
\end{figure}

\subsection{Biological conformational changes}
    
Among the first and most prominent applications of MLCVs has been the study of conformational dynamics in biomolecular systems. These problems naturally involve rare transitions between metastable states and exhibit complex, high-dimensional free energy landscapes—ideal candidates for enhanced sampling aided by ML-driven dimensionality reduction. In this subsection, we focus on selected case studies where MLCVs have provided mechanistic insights and accelerated sampling in biologically relevant systems. These include protein folding, large-scale transitions in membrane transporters, the assembly of protein–protein complexes, and the impact of mutations on protein dynamics. Together, these examples showcase the versatility of ML approaches in resolving biologically meaningful motions and guiding simulation-based hypotheses.

\paragraphtitle{Protein folding} is a fundamental biological process by which an amino acid chain adopts its secondary and tertiary structure to achieve its functional three-dimensional structure. 
Many methods to construct MLCVs have been tested on simulating the folding pathways of small proteins such as chignolin and villin. \cite{mendels2018folding,pomarici2023learning, bonati2021deep, ray2023deep}. 
Also, larger proteins have been studied with similar approaches. 
For example, Belkacemi \textit{et al.} simulated the dynamics of the N-terminal domain of the heat-shock protein 90 (Hsp90) using an autoencoder-based CVs (FEBILAE) trained on clustered dihedral data, capturing transitions between known experimental conformers \cite{belkacemi2023autoencoders}.

\paragraphtitle{Membrane transporters} are proteins that mediate the movement of ions and molecules across cell membranes, often through large conformational changes. 
To study the transition between the inward-open and outward-open states of the sodium potassium–chloride cotransporter NKCC1, classifier-based CVs (Deep-LDA) have been combined with OPES sampling to reveal a rocking-bundle mechanism and highlight the membrane permeability to water.\cite{ruiz2023cation}

\paragraphtitle{DNA translocation} in polymerases is a fundamental process for the genetic transcription process.
After the addition of a new nucleotide, the forming DNA strand has to move along the enzyme to prepare for the next addition. 
Such a process has been studied by Visigalli \textit{et al.} \cite{visigalli2025coordinated} for the Pol$\eta$ enzyme, highlighting the combined action of residues at the protein-DNA interface, acting like screen wipers to favour an asynchronous translocation of the DNA strand. 
In their study, they first run OPES\cite{invernizzi2022exploration} simulations using a Deep-LDA CV\cite{bonati2020data} starting from the known crystallographic structures, identifying two possible reaction pathways with stable intermediates.
Then, they integrated this information into a 2D semi-supervised MultiTask CV\cite{bonati2023unified} to estimate the relative energetic cost of the two paths as shown in Fig.~\ref{fig:DNA}. This showcase how MLCVs can be effectively employed to combine the data coming from experiments (the initial states) with the simulations (intermediate states and pathways) into a single model.

\paragraphtitle{Protein–protein interactions} are central to many cellular processes, and their assembly or activation often involves complex and rare structural transitions. 
Majumder and Staub studied the dimerization of GpA and WALP23 transmembrane proteins, comparing the performance of classifier-based (Deep-LDA) and time-informed CVs (SPIB) using well-tempered metadynamics\cite{majumder2024machine}.

\paragraphtitle{Mutations} in protein sequences can affect stability, dynamics, or function, and understanding these effects is crucial in both basic biology and biomedical research. 
To compare the stability of three mutants of the T4 lysozyme, Smith \textit{et al.} combined data-driven descriptor selection (AMINO) and autoencoder-based MLCVs (RAVE) with metadynamics, also recovering precious insights into conformational preferences from an analysis of the learned reaction coordinates \cite{smith2020discovering}.

While these applications span a wide range of systems, they share common methodological steps and challenges. 
One key step is the initial generation of \textit{structural data }for model training. 
This often begins with experimental structures, such as those obtained from X-ray diffraction or cryo-electron-microscopy, but sequence-to-structure models (e.g., from AlphaFold2) are increasingly used to initialize simulation ensembles \cite{jumper2021highly}.  In addition, clustering of such initial conformations has also been used to define diverse starting points for short unbiased simulations, which are then used to train the CV models.
Another central challenge lies in the selection of appropriate \textit{ input features} or descriptors. 
While CV models can, in principle, operate on large sets of interatomic distances, angles, or contact functions, this high-dimensional space is often redundant and unsuitable for biasing without further filtering. 
Various strategies have been proposed to address this. 
For instance, sparse linear models such as LASSO can be used to identify a minimal set of geometric features that best discriminate between states \cite{novelli2022characterizing}. 
The AMINO method proposed by Ravindra \textit{et al.}, on the other hand, first clusters a large pool of candidate descriptors using a mutual information-based distance metric and then selects representative features from each cluster \cite{ravindra2020automatic}. 
Following a different strategy, it is also possible to first train a CV model on a full descriptor set, then perform a sensitivity analysis to identify a subset of the most relevant features, and finally use them to retrain a more compact version of the model \cite{bonati2021deep}.

\begin{figure}[h!]
  \centering
  \includegraphics[width=\if\ispreprint1
    0.85\linewidth
\else
    0.5\linewidth
\fi]{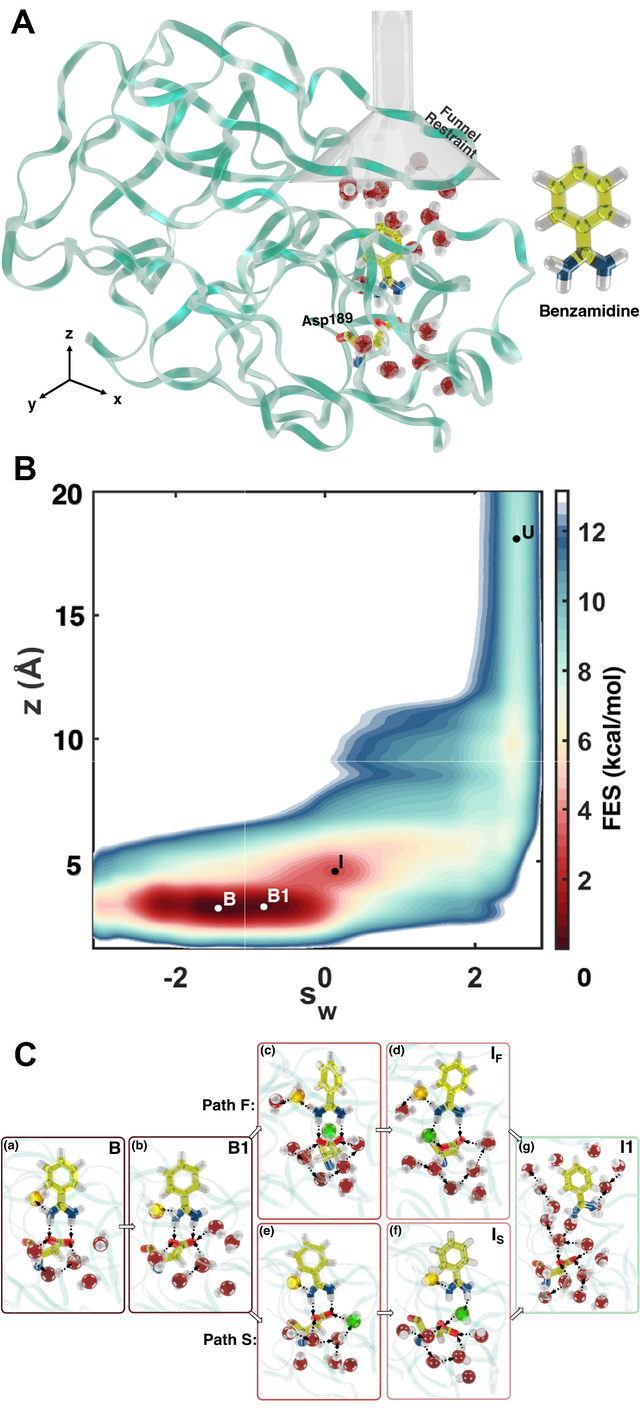}
    \caption{\justifying
    Unbinding pathways in the Trypsin–Benzamidine system. (A)  A cartoon representation of Trypsin structure with the ligand Benzamidine and the Funnel restraint.  (B) Free energy surface (FES) as a function of \(s_w\), a water-related variable learned via Deep-LDA, and \(z\), the ligand–pocket distance, highlighting bound (B/B1), intermediate (I), and unbound (U) states. (C) Two distinct ligand unbinding mechanisms identified using Deep-TICA: one faster and one slower, each characterized by specific rearrangements of water molecules in the binding pocket. Image adapted from Ref.~\citenum{ansari2022water}. Copyright 2022 Springer Nature under \href{https://creativecommons.org/licenses/by/4.0/}{[CC BY 4.0 DEED]}.}
  \label{fig:host-guset}
  \end{figure}

\subsection{Ligand binding}
\label{sec:ligand_binding}
Besides conformational transitions, MLCVs have become powerful tools for studying ligand binding processes across a broad spectrum of biological and chemical systems, spanning simplified host–guest models, pharmacologically relevant protein targets, and complex environments like RNA folds and lipid membranes.

A much-studied prototypical \paragraphtitle{host-guest system} is the set of calixarene host and small ligand guest molecules proposed in the SAMPL5 challenge, which served as benchmarks for testing several sampling strategies and CV design. For example, Rizzi \textit{et al.}~\cite{rizzi2021role} used a classifier-based (Deep-LDA) CV to systematically investigate the role of water in the (un)binding process for several combinations of molecules.
Later, classifier-based CVs were augmented by including information from the transition paths\cite{ray2023deep} in the TPI-Deep-TDA method, and insights about the transition pathways were obtained by studying the committor function \cite{trizio2025everything}.
Siddiqui \textit{et al.}~\cite{siddiqui2023application} compared different methodologies on a pharmacologically relevant drug/target complex, comprising a DNA secondary structure (G-quadruplex) and a metallodrug acting as its stabilizer. Both autoencoders and DeepLDA were found to be effective, yielding consistent results for binding modes and free energies.

In \paragraphtitle{protein–ligand systems}, ML-guided techniques have enabled detailed exploration of unbinding pathways and the computation of kinetic quantities such as residence times. 
Ribeiro and Tiwary~\cite{lamim2018toward} applied autoencoders (RAVE) to study the dissociation of benzene from T4 lysozyme, capturing transitions between metastable intermediates and achieving substantial acceleration of rare dissociation events. 
In a related study on the trypsin–benzamidine complex, a classifier (Deep-LDA) was used to generate the first CV, which was later improved using time-informed methods (Deep-TICA) to model slow solvent-driven motions and improve sampling. In particular, Ansari \textit{et al.}~\cite{ansari2022water} proposed a strategy to identify the long-lived hydration spots, which were used as input descriptors for the MLCVs. These simulations revealed how specific water molecules mediate hydrogen-bond networks that gate ligand unbinding and modulate the energy barrier~\cite{ansari2022water}. 
Classifier-based CVs have also been used to investigate substrate binding in human pancreatic $\alpha$-amylase. 
In this case, Deep-TDA was employed to train two orthogonal CVs: one to account for conformational degrees of freedom, based on nucleophile–substrate reactive contacts, and the other to capture solvation of substrates and catalytic residues~\cite{das2023and}. 
A path CV was then defined as a function of these two CVs connecting reactive and non-reactive states, revealing three distinct binding modes. 
The same framework was later extended also to substrates of different sizes but exhibiting similar binding poses~\cite{das2024correlating}.

More complex examples involve ligand binding to G-protein coupled receptors (\paragraphtitle{GPCRs}), which is associated with longer dissociation timescales. 
In a study on the $\mu$-opioid receptor, a combination of feature selection (AMINO), autoencoder CVs (RAVE), and infrequent metadynamics was used to extract unbinding kinetics and identify structural determinants of transition states, providing mechanistic insight into drug residence times~\cite{lamim2020combination}. 
Significant challenges are also associated with the study of \paragraphtitle{RNA–ligand interactions}, due to RNA's intrinsic flexibility and structural diversity. 
In this regard, Wang \textit{et al.}~\cite{wang2022interrogating} combined autoencoder-based CVs (RAVE) simulations with experimental data to study riboswitch-ligand binding, identifying distinct dissociation pathways for cognate and synthetic ligands and predicting long-range mutational effects.

\begin{figure}[h!]
  \centering
  \includegraphics[width=0.95\linewidth]{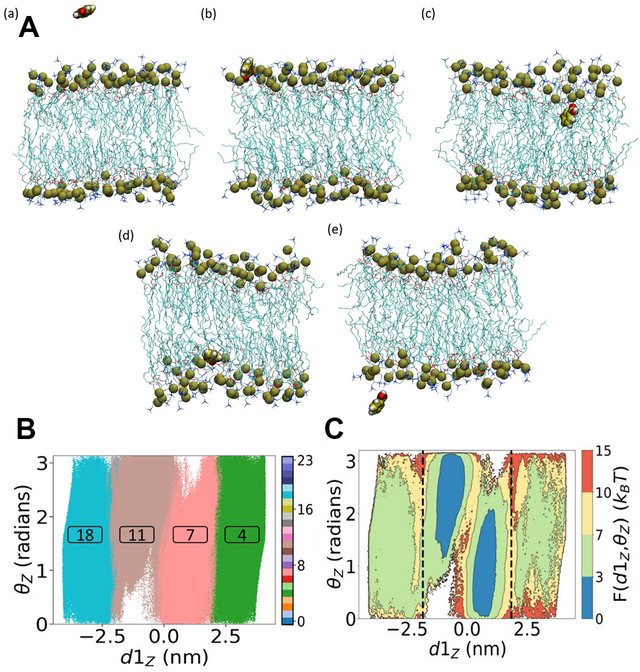}
    \caption{\justifying (A) Schematic representation of a benzoic acid molecule permeating a symmetric phospholipid bilayer, highlighting the key stages of adsorption at the membrane surface, reorientation, and translocation across the lipid core. (B) Metastable state assignments provided by the SPIB algorithm in the space of the membrane–solute distance and orientation angle, and (C) FES projected onto the same reaction coordinate space. This provides the thermodynamic barriers for membrane entry, traversal, and exit, as well as enables mechanistic insights into the role of molecular orientation and interactions with lipid headgroups.
    Adapted from Ref.~\citenum{mehdi2022accelerating}. Copyright 2022 American Chemical Society.}
  \label{fig:membrane}
  \end{figure}

While these cases involve well-defined ligand–receptor systems, similar strategies have also been applied to \paragraphtitle{membrane permeation} processes.  Mehdi \textit{et al.}\cite{mehdi2022accelerating} used the SPIB framework to investigate the permeation of benzoic acid through phospholipid bilayers. Starting from short unbiased simulations and iteratively refining the CV on biased data, they efficiently sampled permeation events between metastable states and uncovered how molecular orientation and lipid headgroup interactions shape the free energy barriers for membrane crossing (see Fig.~\ref{fig:membrane}). Similarly, Muscat \textit{et al.}\cite{muscat2024leveraging} applied Deep-TICA, initialized from a multithermal simulation, in coarse-grained models of neuron-like membranes to study the insertion of aminosterols, reconstructing the free energy landscape and identifying key metastable intermediates along the insertion pathway.

Despite their diversity, these systems share common modeling challenges. One of the most important ones is accounting for the role of water in mediating binding thermodynamics and kinetics. 
Water molecules can indeed bridge critical hydrogen bonds, occupy binding pockets or leave them empty, and even modulate energy barriers during association and dissociation (Fig.~\ref{fig:host-guset}). 
MLCVs offer a way to build water-sensitive CVs able to represent hydration shells and dynamic water networks by using permutationally invariant descriptors such as PIV~\cite{pietrucci2011graph,herringer2023permutationally} or the solvation number of relevant sites for the binding process~\cite{rizzi2021role}, e.g., close to the binding pocket or on the ligand. 
Additionally, semi-automated strategies for the identification of such hydration spots, which in complex cases may be far from trivial, have also been proposed~\cite{ansari2022water}. 
Overall, these findings highlight the need to treat water as an active component of the binding process, and not merely as a passive background.
These applications demonstrate how ML-enhanced simulations enable not just the estimation of free energies and rate constants, but also the mechanistic interpretation of molecular recognition events, provided that CVs are constructed to capture all relevant degrees of freedom.

\subsection{Structural phase transformations}

Phase transformations, including crystallization, melting, and solid–solid and liquid-liquid transitions, are rare events that span even longer timescales and involve the crossing of substantial free energy barriers. 
These processes typically begin with the formation of transient nanoscale regions of the new phase, such as nuclei or precursors, which then grow into extended domains. 
Capturing such transformations with atomistic simulations is inherently difficult, as it requires CVs capable of describing complex, system-specific structural rearrangements. 
Unlike biomolecular transitions, which often combine many simple descriptors such as distances and dihedral angles, phase transformations frequently involve more complex geometric, symmetry-based, or thermodynamic descriptors that are able to capture the changes in the ordering of the system with the additional complication of explicitly treating permutational invariance; see, for example, the recent review on crystallization by Neha \textit{et al.}~\cite{neha2022collective}.

In the study of \paragraphtitle{homogeneous crystallization}, Zhang \textit{et al.}~\cite{zhang2019improving} used X-ray diffraction (XRD) peak intensities as input features for HLDA and TICA to distinguish liquid from crystalline phases in elemental \ce{Na} and \ce{Al}. 
These descriptors enabled the resolution of metastable states and accelerated sampling of the nucleation process. 
Building on this idea, Karmakar \textit{et al.}~\cite{karmakar2021collective} employed peaks from the full three-dimensional Debye structure factor to train Deep-LDA CVs~\cite{bonati2020data}, successfully driving crystallization in \ce{NaCl} and \ce{CO2}. 
As in other domains, such CVs can serve as a starting point and be further refined, particularly in the transition region, using time-informed methods such as Deep-TICA~\cite{bonati2021deep}.

\begin{figure}[h!]
  \centering
  \includegraphics[width=0.85\linewidth]{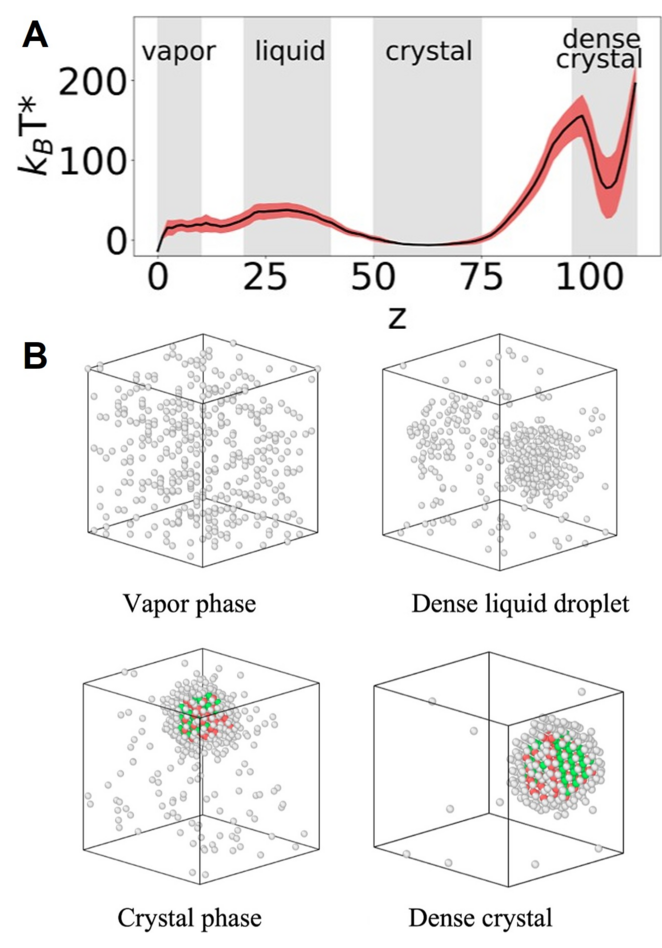}
  \caption{\justifying
  Crystal nucleation in supersaturated colloid suspensions investigated using enhanced sampling with MLCVs. (A) One-dimensional free energy profile as a function of the SPIB CV. (B) Representative structures of the four phases during the nucleation process. Image adapted from Ref.~\citenum{meraz2024simulating}. Copyright 2024 American Chemical Society.}
  \label{nucleation}
  \end{figure}
  
\begin{figure*}[ht!]
  \centering
  \includegraphics[width=0.65\linewidth]{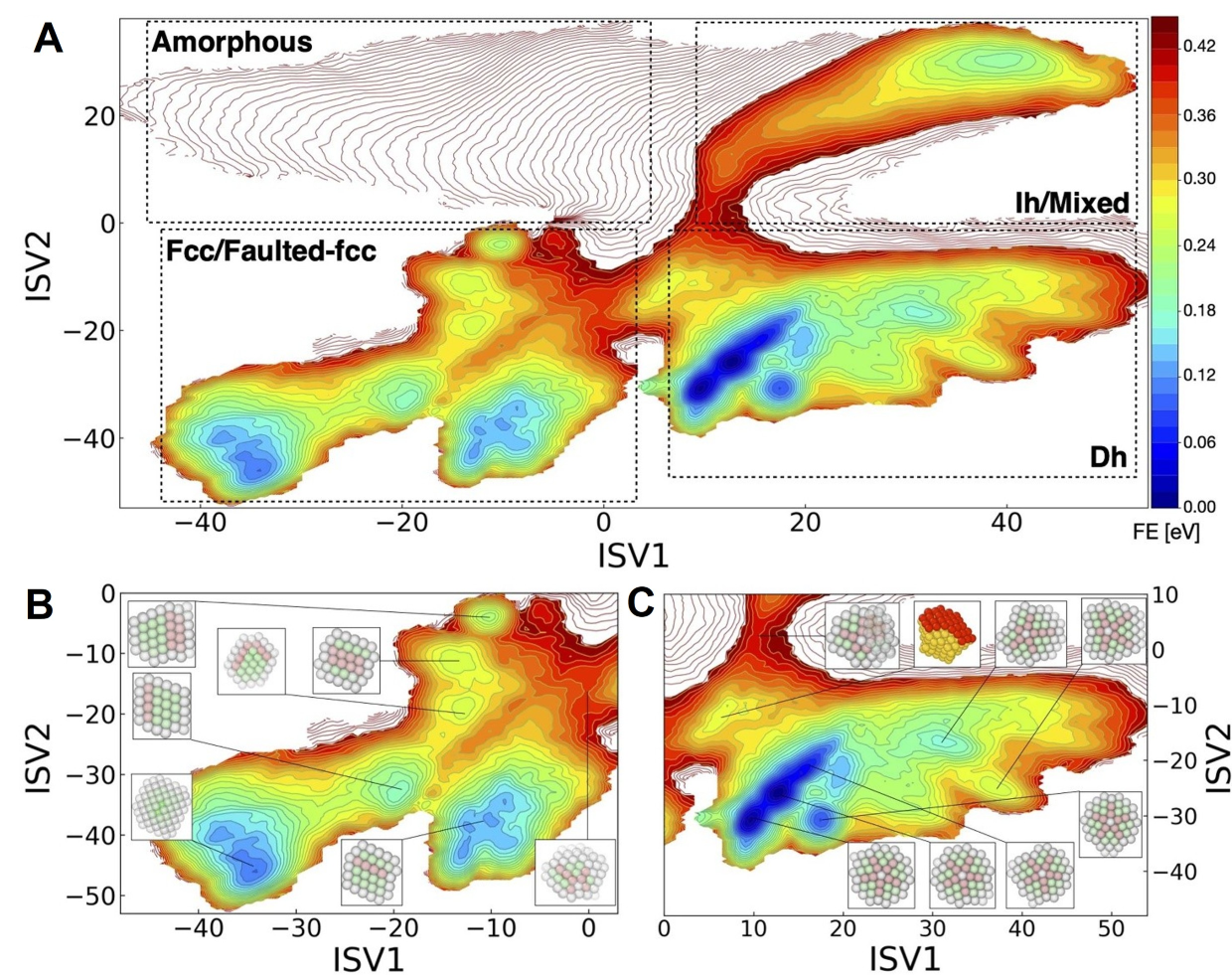}
  \caption{\justifying  Solid-solid phase transition of Au$_{147}$ at 396 K. (A) Free energy landscape obtained from umbrella sampling. The bottom left region corresponds to face-centered cubic (fcc) and faulted-fcc structures, the top right to icosahedral (Ih) and mixed structures, and the bottom right to decahedral (Dh) structures. Amorphous structures, associated with very high free energies at this temperature, are located in the top left corner. (B) Enlarged view of the free energy landscape in panel A, focusing on the fcc and faulted-fcc region and illustrating representative local minima and the bottleneck connecting this region to the Dh basin. (C) Enlarged view of the Dh region from panel A, highlighting local minima and the transition path connecting Dh to Ih and mixed structures. Atoms are colored according to their local coordination: green for fcc, pink for hcp, and white for undefined environments. Image reproduced from Ref.~\citenum{telari2025inherent}. Copyright 2025 IOP Publishing Ltd under \href{https://creativecommons.org/licenses/by/4.0/}{[CC BY 4.0 DEED]}.}
  \label{fig:nanoclusters}
  \end{figure*}
  
In the field of \paragraphtitle{nucleation}, Tiwary and collaborators applied the SPIB framework~\cite{wang2021state} to molecular and ionic systems. 
For aqueous urea and glycine~\cite{zou2023driving}, they constructed and compared CVs from a diverse set of descriptors, including coordination numbers, Steinhardt bond-order parameters~\cite{steinhardt1983bond}, intermolecular angles, orientational entropy, water structure~\cite{salvalaglio2012uncovering}, and pair entropy~\cite{piaggi2017enhancing}. 
The resulting CVs revealed that orientational descriptors, rather than cluster size alone, were critical in capturing the slow modes of nucleation, highlighting the limitations of classical nucleation theory. 
In subsequent work~\cite{wang2024local}, SPIB was used to explore \ce{NaCl} nucleation from melt and aqueous solution, showing that, while local ion density could distinguish phases, it was insufficient to drive transitions, whereas energy and local structure emerged as more effective drivers instead. 
Their recent study \cite{wang2025electric} found that removing solvent water from  Cl$^{-}$ ions on the solid precursor surface is more important than ion buildup, and that the electric field both promotes nucleation by removing water and hinders it by separating ion pairs.
A similar approach was then applied to colloidal systems~\cite{meraz2024simulating}, where a one-dimensional SPIB-derived CV, based on both local and global structural information, was trained to capture transitions among vapor, liquid, and solid states (See Fig.~\ref{nucleation}).

Other relevant transformations include \paragraphtitle{solid–solid phase transitions}. To model the A15-to-bcc transition in molybdenum, Rogal \textit{et al.}\cite{rogal2019neural} developed a neural network path CV that combines a local classifier of atomic environments (based on Behler–Parrinello symmetry functions) with a global path CV constructed from the fractions of atoms in different phases. This CV enabled the study of interface migration and the characterization of the transformation pathway. Similarly, Telari \textit{et al.}\cite{telari2025inherent} explored structural transitions in gold nanoclusters using an autoencoder-based approach. Configurations generated via replica exchange simulations were represented using the radial distribution function (RDF) as a global structural descriptor. The autoencoder, trained with a denoising-like objective, learned a latent representation capable of reconstructing the RDF associated with the inherent structures on the potential energy surface, obtained through energy minimization. This data-driven framework classified the structural diversity into three dominant families (face-centered cubic, decahedral, and icosahedral) and highlighted the role of defects in facilitating structural transformations (see Fig.~\ref{fig:nanoclusters}). By using these CVs with umbrella sampling and Markov state models, the authors reconstructed the free energy landscape, computed transition rates, and characterized the pathways connecting the different conformations.

The phase diagram of many liquids also includes \paragraphtitle{liquid-liquid phase transitions}, which present similar challenges but in a much more mobile environment.
For example, the $\lambda$-transition in liquid sulfur involves the equilibrium between a molecular phase, characterized by low viscosity and composed of eight-member crown-shaped rings, and a polymeric phase with high viscosity and composed of long linear polymeric chains.
To characterize the structures and mechanisms across such a transition, Yang \textit{et al.}\cite{yang2024structure} employed a Deep-TDA\cite{trizio2021enhanced} CV in combination with OPES, using as input descriptors for the changes in the system topology the distribution of the eigenvalues of the adjacency matrix of the system.

Overall, these studies presented several challenges, but chief among them is the selection of physically meaningful descriptors able to capture the right structural properties.
Effective CVs must indeed simultaneously capture local order and collective structural changes, and remain valid throughout the transition and guarantee permutational invariance. 
ML offers a powerful framework to handle large, heterogeneous descriptor sets and to construct low-dimensional CVs that preserve essential mechanistic features. 
Furthermore, since phase transitions often proceed through multiple intermediates, generalizable CVs must be robust across the entire reaction coordinate landscape.

\subsection{Chemical and catalytic reactions}

Traditional enhanced sampling studies of chemical reactivity often relied on biasing a few physically intuitive CVs, such as distances or angles associated with bond formation or cleavage. However, this strategy is only effective for relatively simple reactions and in cases where the surrounding environment plays a minimal role. In many realistic scenarios, especially those involving complex molecular systems, heterogeneous interfaces, or enzymatic active sites, the reaction mechanism can involve multiple steps, hidden intermediates, and collective contributions from the environment. In such cases, predefining the relevant CVs becomes exceedingly difficult. To overcome these challenges, MLCVs have been applied to chemically reactive systems, offering a data-driven route to uncover complex reaction coordinates.

In particular, a first objective is \paragraphtitle{reaction discovery}, which leverages enhanced sampling to find the possible products and pathways.
One strategy in this regard was proposed by Raucci \textit{et al.} by incorporating a first exploratory stage based on an agnostic CV from graph theory with a second stage in which, once new states were discovered, free energy calculations based on MLCVs and/or refinement of the identified structures are carried out.\cite{raucci2022discover}
This approach was first applied to simple chemical reactions, training a classifier-based CV (Deep-LDA) using atomic contacts as descriptors and using it to converge free energy profiles. Additionally, the obtained profiles, initially computed at the semi-empirical level, were also corrected to a more refined level of theory via free energy perturbation. The same approach was applied by Das \textit{et al.} to the identification of reactive conformations of substrate-enzyme complex in the sugar-degrading enzyme $\alpha$-amylase \cite{das2023and} (see also Sec.\ref{sec:ligand_binding}). A similar strategy was also used by Raucci \textit{et al.} to study the donor–acceptor Stenhouse adduct (DASA) molecular photoswitchers, which are able to undergo substantial conformational changes upon light irradiation and present a complex reaction network of multiple stable states.\cite{sanchez2021silico} In this case, after the discovery stage, static structural optimization was carried out.\cite{raucci2022enhanced} More information about part of the same reaction network was later obtained by Kang \textit{et al.}\cite{kang2024computing} by learning the corresponding committor function and using it to characterize in detail the transition state ensemble.

\begin{figure}[h!]
  \centering
  \includegraphics[width=\linewidth]{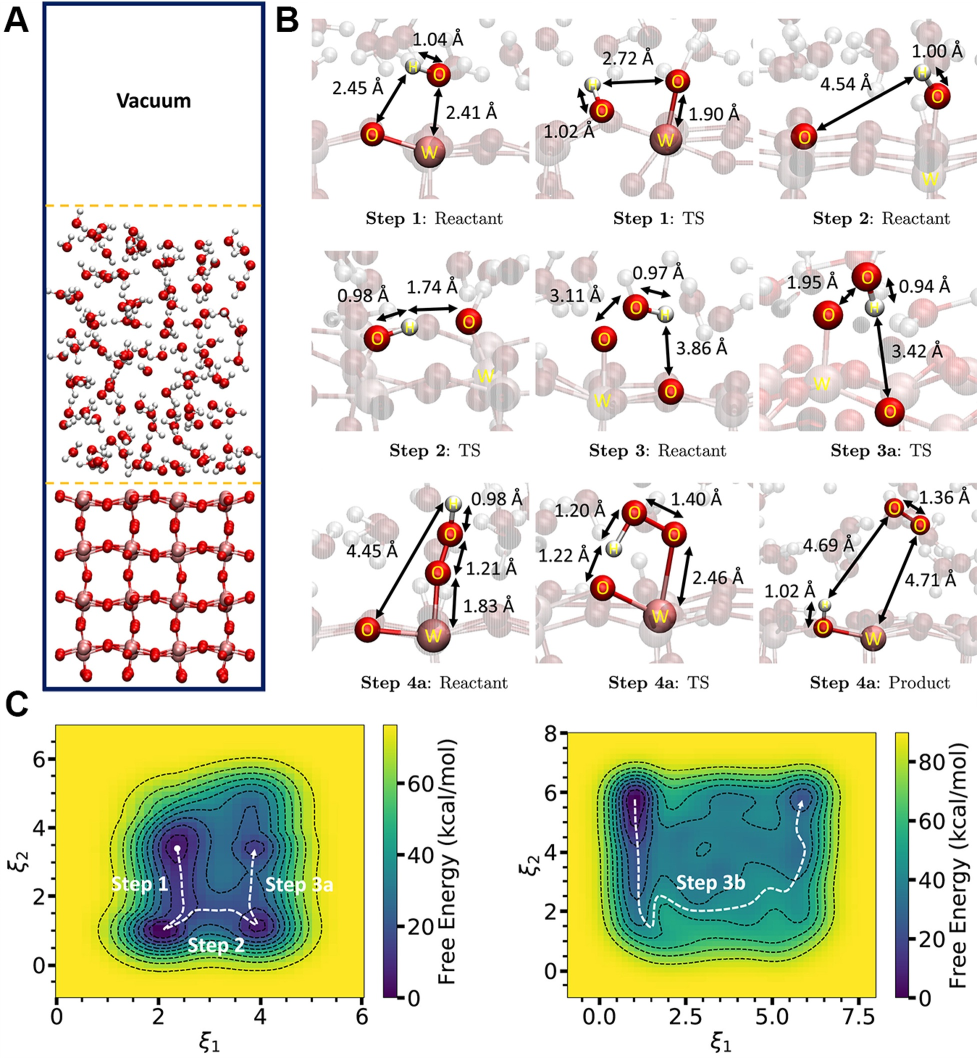}
  \caption{\justifying
   Catalytic water oxidation at a solid–liquid interface. (A) Atomistic model of the $WO_3$/water interface. (B) Representative snapshots of key intermediates and transition states along the oxygen evolution reaction (OER) pathway. (C) Free energy surfaces computed the autoencoder-based CV  (DAENN), capturing both the OER (left) and the alternative $H_2O_2$ formation pathway (right). Images reproduced Ref.~\citenum{ketkaew2024metadynamics}. Copyright 2024 Elsevier.}
  \label{fig:oer}
\end{figure}

Another crucial area of application is \paragraphtitle{heterogeneous catalysis}, which targets the reduction of energy barriers in industrially and environmentally relevant reactions. The oxygen evolution reaction at the WO$_3$/water interface (Fig.~\ref{fig:oer}) was investigated by Luber and co-workers~\cite{ketkaew2024metadynamics}, who used autoencoders (DAENN) to combine bond distances with xSPRINT descriptors~\cite{ketkaew2022machine} and drive metadynamics simulations, uncovering competing pathways such as H$_2$O$_2$ formation.
Besides biasing, MLCVs can also be used to rationalize the behavior of reactions in complex environments. For example, Bonati \textit{et al.}~\cite{bonati2023role} trained a supervised CV to capture the charge transfer during nitrogen dissociation on iron,  the first step in industrial ammonia synthesis. This CV was then used to reconstruct the free energy landscape, providing insights into the catalytic role of the surface not via structural but rather electronic descriptors.

\begin{figure}[h!]
  \centering
  \includegraphics[width=1\linewidth]{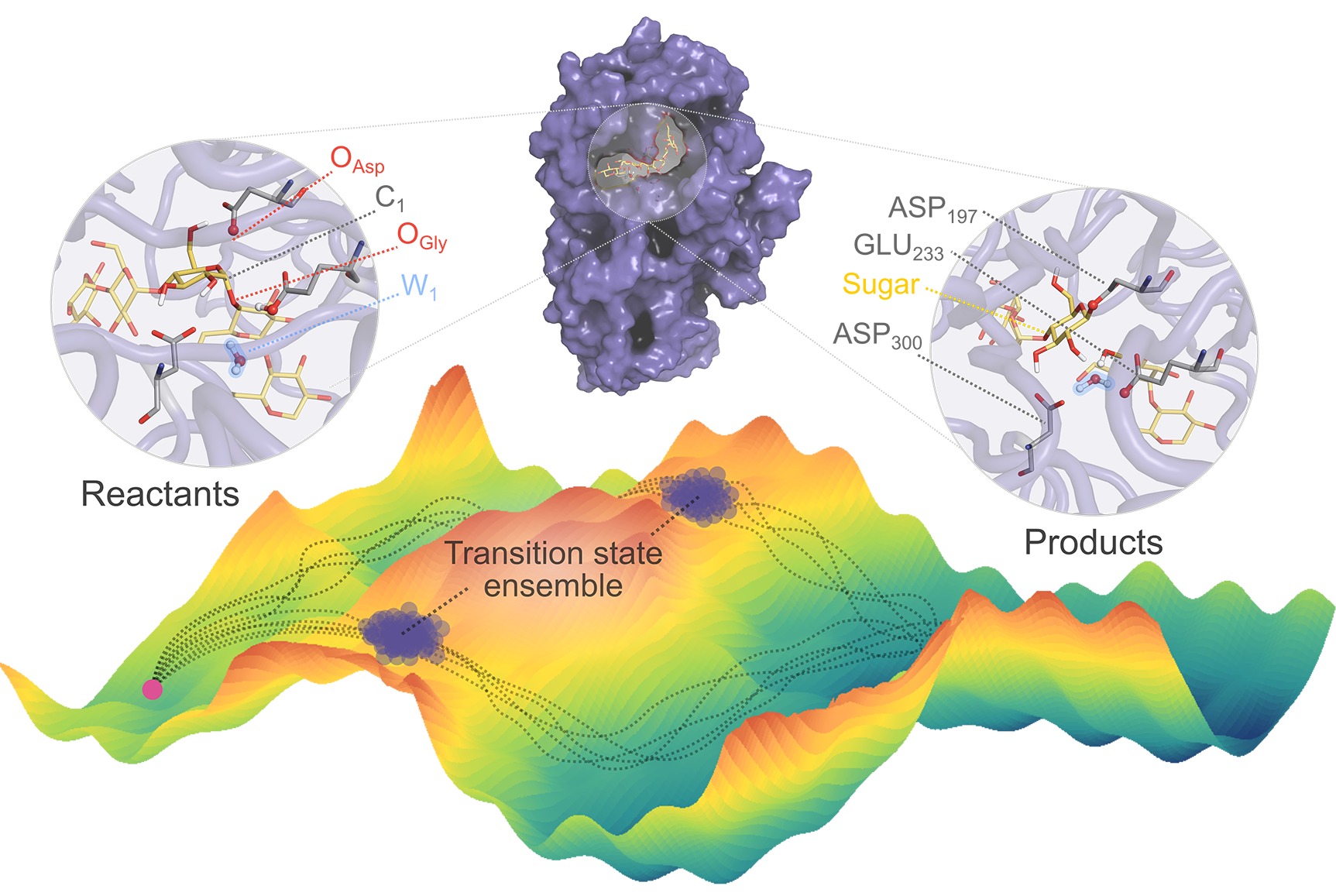}
  \caption{\justifying
  Enzymatic catalysis of substrate-bound \(\alpha\)-amylase. Schematic representation of the free energy surface connecting reactant and product states (highlighted in snapshots along with key catalytic residues in the active site). The dynamic catalytic landscape, with multiple reaction pathways, is revealed through a machine-learned committor function, enabling a probabilistic characterization of transition states.
  Image reproduced from Ref.~\citenum{das2025machine}. Copyright 2025 American Chemical Society.}
  \label{fig:catalysis}
\end{figure}

Catalytic reactions are also fundamental in biophysics, where \paragraphtitle{enzymes} efficiently accelerate biochemical reactions, thus motivating great interest in understanding their complex workings.
For example, a number of diseases are caused by enzymatic dysfunction, and enzymes are also being investigated to degrade pollutants. Recently, Das \textit{et al.}\cite{das2025machine} applied the committor-based enhanced sampling strategy \cite{kang2024computing,trizio2025everything} to the study of the glycolysis of sugars in the human pancreatic $\alpha$-amylase (Fig.~\ref{fig:catalysis}), which is important in glucose production and a drug target for type-II diabetes. 
This approach provided insights into the mechanisms and revealed the pivotal role of water molecules in competing pathways in the catalytic process.

\vspace{1em}

To conclude this section, we have seen that MLCVs have been developed and applied to address a wide variety of objectives. These range from enhancing sampling of complex landscapes and facilitating the exploration of rare events, to gaining mechanistic insights and reducing the dimensionality of high-dimensional systems. This breadth not only reflects the flexibility of MLCVs in tackling diverse challenges but also underscores that there is no single “one-size-fits-all” solution. Instead, the choice of method must be carefully aligned with the specific goals of the study and the available data.
For instance, autoencoder-based models are well suited for unsupervised exploration of high-dimensional landscapes, classifier-based CVs can be effective when metastable states are already known, and time-lagged or committor approaches typically offer deeper mechanistic insight, albeit at the cost of higher requirements in terms of quantity and quality of data.

\section{Machine learning bias potentials}
\label{sec:ml_bias}
In the previous sections, we examined approaches that employ ML to identify suitable low-dimensional representations (CVs) and to integrate them within traditional enhanced sampling methods. A complementary line of development seeks to address the inherent limitations of conventional biasing schemes themselves. Techniques such as metadynamics and umbrella sampling typically rely on bias potentials applied along a small set of carefully chosen CVs. Recent advances, by contrast, explore how ML can directly inform the design and optimization of biasing strategies, potentially bypassing these dimensionality constraints and opening new avenues for sampling complex systems.
On one hand, ML models can help overcome the limitations of low-dimensional representations by enabling the use of a larger number of CVs simultaneously, without reducing the system to just one or two dominant modes. On the other hand, they make it possible to optimize bias potentials with objectives that go beyond traditional free energy reconstruction. For instance, emerging approaches aim to generate physically meaningful, unbiased transition pathways, thereby addressing one of the longstanding shortcomings of biased sampling techniques.
In the following, we present these approaches grouped into three broad categories:
\begin{enumerate}
    \item \paragraphtitle{Representing and biasing high-dimensional FESs (Section \ref{sec:bias_high_dimensional}):}  ML models are used to represent high-dimensional free energy surfaces, which can be then used to bias the sampling.
    \item \paragraphtitle{Bias potentials optimization (Section \ref{sec:bias_neural_networks}):} Neural networks are used to represent and optimize bias potentials within existing adaptive sampling schemes (e.g. VES, ABF, GAMD).
    \item \paragraphtitle{Transition path-guided bias (Section \ref{sec:bias_transition_paths}):} These approaches aim to construct external potentials such that they can produce unbiased transition paths, often through a reinforcement learning approach.
\end{enumerate}

\subsection{Representing and biasing high-dimensional free energy surfaces}
\label{sec:bias_high_dimensional}
A key ingredient in many enhanced sampling schemes is the accurate \textit{representation of the FES} as a function of selected CVs. 
However, constructing such representations in high-dimensional spaces remains a significant challenge, due to the curse of dimensionality and the limited amount of data typically available from molecular simulations.
To address this, a variety of ML techniques, including kernel methods and neural networks, have been applied to model equilibrium probability distributions and their associated FESs. 
While differing in formalism, both approaches aim to capture complex, high-dimensional landscapes in a data-efficient manner, and their respective strengths have been systematically compared by Cendagorta \textit{et al}.~\cite{cendagorta2020comparison}.

For instance,  Csányi and collaborators proposed a \method{Gaussian process regression (GPR) of the FES} from simulation data. 
In their first work~\cite{stecher2014free}, GPR was used to model the FES obtained from umbrella sampling, using histogram-based estimates of equilibrium probabilities as training labels. 
By incorporating prior assumptions of smoothness and consistently accounting for sampling noise, the method achieved significantly improved accuracy over conventional estimators in two or more dimensions. 
Moreover, the Bayesian formulation of Gaussian processes naturally provides uncertainty estimates, enabling the quantification of confidence in the predicted free energies.
In a follow-up study~\cite{mones2016exploration}, the Authors proposed a modular approach that explicitly separates biasing, free energy gradients measurement, and free energy reconstruction to improve computational efficiency. 
In particular, they used metadynamics to guide sampling, instantaneous collective forces (akin to those used in adaptive biasing force methods) to estimate free energy gradients, and GPR to reconstruct the FES. 
This strategy led to a substantial reduction in computational cost, demonstrating that decoupling sampling from learning can be especially powerful in high-dimensional settings.

In parallel, \textit{neural networks} have been widely adopted due to their flexibility and favorable scaling with the number of data points and CVs. 
Tuckerman and collaborators~\cite{schneider2017stochastic} trained \method{neural networks to represent the FES} based on either free energy values or their derivatives, depending on the enhanced sampling method used. 
This approach facilitated both the computation of free energy differences and the evaluation of ensemble averages from the learned model. 
Sidky and Whitmer~\cite{sidky2018learning} extended this framework using Bayesian regularization to adaptively refine the FES and reduce overfitting.
In addition to direct regression of free energies, some methods rely on \textit{probability density estimation}. 
Galvelis \textit{et al.}\cite{galvelis2017neural} proposed \method{NN2B}, a hybrid approach in which a nearest neighbor density estimator (NNDE)\cite{loftsgaarden1965nonparametric} is first applied to a biased trajectory to estimate local probability densities. 
This smoothed information is then converted to free energy labels and used to train a neural network, which iteratively updates the bias potential.

Together, these techniques demonstrate how ML methods can provide accurate and scalable representations of free energy surfaces, a key ingredient for developing effective biasing strategies in high-dimensional landscapes.

\begin{figure*}[htbp]
\centering
\includegraphics[width=0.8\linewidth]{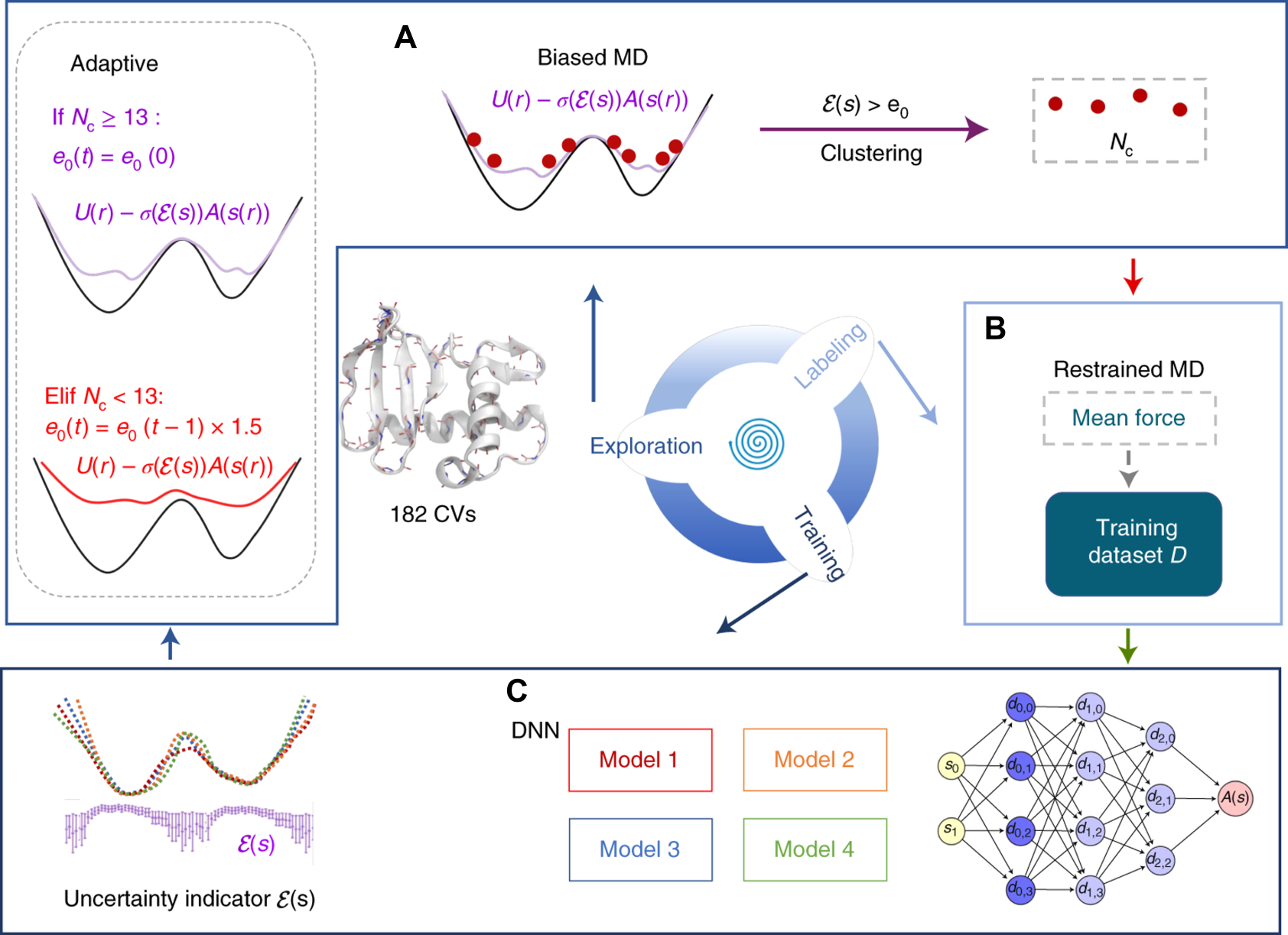}
\caption{\justifying
The workflow of adaptive RiD. (a) In the exploration step, biased MD simulations are used, and the visited CV values with the uncertainty indicators \( \mathcal{E}(s)\) larger than a certain level \(\epsilon_0\) are proposed for labeling. The proposed CVs are then clustered into \(N_c\) clusters, and one set of CV values is randomly selected from each cluster for labeling. An adaptive strategy is applied at each iteration by adjusting the uncertainty levels based on the number of clusters \(N_c\). In this case, if \(N_c\) is less than 13, the level \(\epsilon_0\) is multiplied by 1.5, and \(\epsilon_1 = \epsilon_0 + 1\). Otherwise, the same levels as the initial values are used (panel outlined by a gray dashed line). (B) The mean forces evaluated by the restrained MD simulation are used as labels to train the DNN models. (C) Four DNN models are trained using different random initial parameters, and the uncertainty indicator \(\mathcal{E}(s)\) is defined as the standard deviation of the force predictions from this ensemble of DNN models. Image reproduced from Ref.~\citenum{wang2022efficient}. Copyright 2021 Springer Nature.}
\label{fig:adaptive-rid}
\end{figure*}

Following a different strategy, Zhang \textit{et al.} introduced a reinforcement learning framework called \method{reinforced dynamics (RiD)}~\cite{zhang2018reinforced}\footnote{Available via \texttt{Rid-kit}: \url{https://github.com/deepmodeling/rid-kit}}. 
In RiD, a neural network is trained to represent the FES, and an uncertainty indicator $\mathcal{E}(s)$ is used to evaluate the reliability of the model’s predictions across the CV space. 
The uncertainty is estimated using a query-by-committee approach, in which an ensemble of $N$ neural networks predicts the mean force. 
The indicator $\mathcal{E}(s)$ is then defined as the standard deviation across the ensemble:
    \begin{equation}
        \mathcal{E}^2(s) = \langle ||f_n(s) - \bar{f}(s)||^2 \rangle
    \end{equation}
where $f_n(s)$ is the force predicted by a single model $n$, and $\bar{f}(s)$ is the average over the ensemble of models. 
A switching function $\sigma(\mathcal{E})$ is applied to modulate the force based on the model confidence, biasing the system only in regions where the uncertainty is low.
In particular, the force $f_i(R)$ acting on atom $i$ is obtained as: 
    \begin{equation}
        {f}_i(\mathbf{R}) = -\nabla_{r_i} U(\mathbf{R}) + \sigma(\mathcal{E}(s(\mathbf{R}))) \left\langle \nabla_{\mathbf{R}_i} F(s(\mathbf{R})) \right\rangle
        \label{RiD_bias}
    \end{equation}
where $U(R)$ is the physical potential, and $F(s)$ is the learned FES.

While RiD proved effective for systems involving up to 20 CVs, its performance degraded in higher-dimensional settings. 
To address this, Wang \textit{et al.}~\cite{wang2022efficient} developed an adaptive extension of RiD (see Fig.~\ref{fig:adaptive-rid}). 
In this scheme, points with high uncertainty are flagged during simulation and clustered to ensure diverse sampling.
Representative configurations are selected from each cluster, labeled via restrained MD to obtain mean forces, and used to retrain the neural network ensemble (see also Fig.~\ref{fig:adaptive-rid}). 
Furthermore, the uncertainty threshold is dynamically adjusted based on the number of clusters, balancing exploration and labeling efficiency. 
Thanks to this adaptive strategy, RiD has been successfully applied to exploratory studies involving up to 100 CVs, showcasing its potential for navigating complex free energy landscapes in high-dimensional systems.

\subsection{Enhancing biasing schemes with NNs}
\label{sec:bias_neural_networks}

In this section, we examine methods in which ML algorithms, and particularly neural networks, are employed to enhance the representation of the bias potential within established enhanced sampling frameworks. 
The expressive power and smoothness of neural networks make them well-suited for modeling complex bias potentials, especially in systems involving multiple CVs or rapidly varying free energy landscapes.

One example is the \method{variationally enhanced sampling (VES)} method~\cite{valsson2014variational}, in which the bias potential is optimized by minimizing a convex functional $\Omega[V]$, designed to drive the system toward a prescribed target distribution $p_{\mathrm{tg}}(\textbf{s})$. 
This functional is closely related to the KL divergence between the biased distribution $p_V$ and the target distribution $p_{\mathrm{tg}}$:
    \begin{equation}
        \beta \Omega[V] = D_{\mathrm{KL}}(p \| p_V) - D_{\mathrm{KL}}(p \| p_{\mathrm{tg}})
    \end{equation}
where $p$ denotes the equilibrium distribution and $\beta$ is the inverse temperature. 
In its original formulation, the VES bias potential $V(\textbf{s})$ was expressed as a linear expansion over a set of basis functions, with the expansion coefficients serving as variational parameters. 
To improve flexibility and scalability, Bonati \textit{et al.}~\cite{bonati2019neural} proposed \method{Deep-VES}, representing $V(\textbf{s})$ using a neural network.
In this formulation, the functional $\Omega[V]$ is treated as a scalar loss function, and its optimization with respect to the neural network parameters $\theta$ is performed using the gradients estimated directly from the simulation data:
    \begin{equation}
      \frac{\partial \Omega}{\partial \theta} = -\left \langle \frac{\partial V}{\partial \theta} \right \rangle _{P_V} + \left \langle \frac{\partial V}{\partial \theta} \right \rangle_{p_{\mathrm{tg}}}
    \end{equation}
where the first average is computed over the biased ensemble (via simulation) and the second over the target distribution (numerically). 
This approach leverages the representational capacity of neural networks to construct bias potentials via a principled variational framework.

A similar approach, still inspired by the variational formulation of VES, is the \method{targeted adversarial learning optimized sampling (TALOS)} method proposed by Zhang \textit{et al.}~\cite{zhang2019targeted}. 
TALOS aims to guide sampling toward a predefined target distribution using a generative adversarial learning scheme. 
The key idea is to train two neural networks simultaneously: a \textit{generator}, which defines the bias potential and modifies the sampling distribution, and a \textit{discriminator}, which learns to distinguish between samples drawn from the biased simulation and those from the desired target distribution.
A distinctive feature of TALOS is the separation between the spaces where the target and the bias are defined. 
The target distribution $p(q)$ is specified in a \textit{descriptor space} $q(\mathbf{R})$, composed of physical or structural features such as distances or angles. 
In contrast, the bias potential $V_\theta(\mathbf{R})$ is defined and acts in the full atomic coordinate space $\mathbf{R}$, not in the reduced descriptor space.
This allows TALOS to operate without requiring a traditional low-dimensional CV. 
During training, the two networks play an adversarial game: the discriminator improves its ability to tell apart sampled and target configurations, while the generator updates the bias to make the sampled distribution more closely resemble the target. 
The process converges when the two distributions match, yielding an optimized bias potential that reproduces the desired sampling behavior. 

\begin{figure}[htbp]
\centering
\includegraphics[width=1\linewidth]{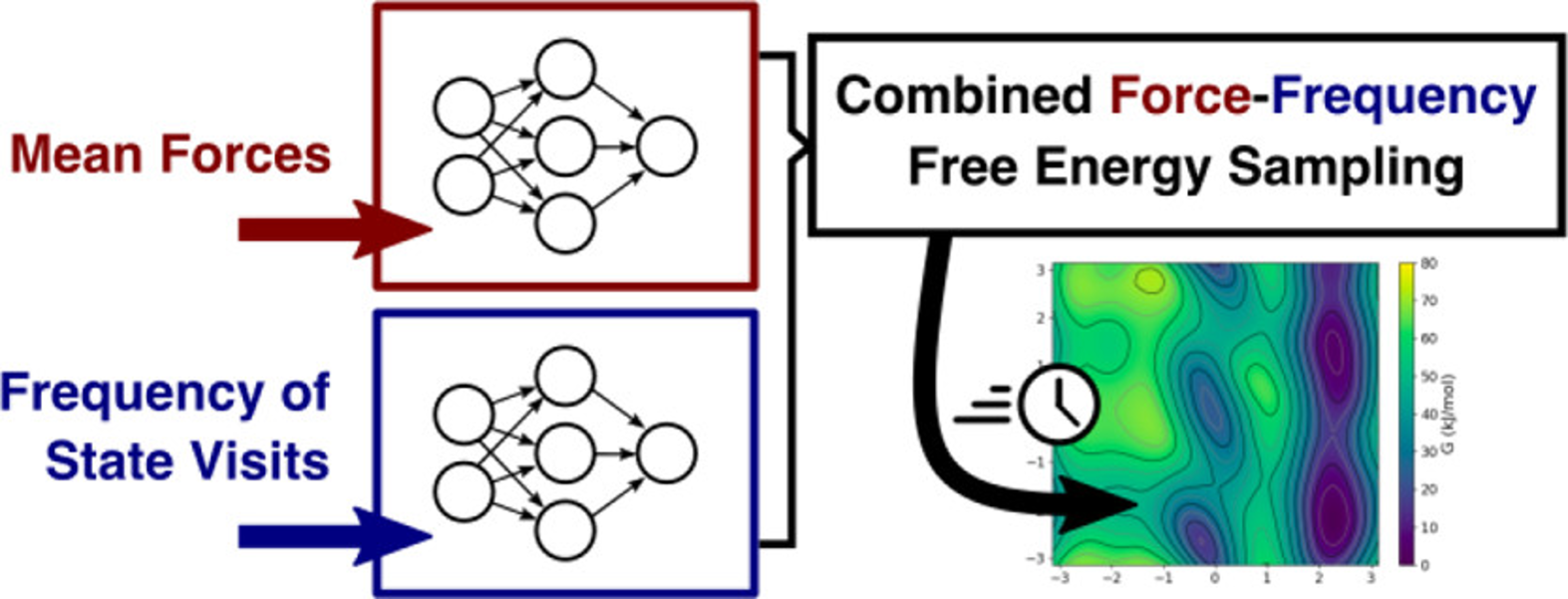}
\caption{\justifying
Schematic of the CFF method. Frequency and force data collected in CV space are used to train two neural networks: one learning the free energy from histogram frequencies and the other from its gradient estimates. Together, they provide a "combined force-frequency" free energy estimate.
Image reproduced from Ref.~\citenum{sevgen2020combined}. Copyright 2023 American Chemical Society.}
\label{fig:CFF}
\end{figure}

Another enhanced sampling method that has benefited from neural network-based representations of the bias potential is \method{adaptive biasing force (ABF)}. 
ABF aims to reconstruct the free energy landscape from its derivatives, computed as generalized mean forces, and use it to determine a biasing force. 
In traditional ABF, the mean force estimates are stored on a discrete grid, which leads to inaccuracies in poorly sampled regions and prevents generalization to unexplored areas. 
Moreover, the choice of grid resolution introduces a trade-off between accuracy and convergence speed. 
To overcome these limitations, Guo \textit{et al.} proposed the \method{force-biasing using neural networks (FUNN)} method~\cite{guo2018adaptive}, which replaces the discrete force representation with a continuous neural network model. 
This approach improves ABF by (i) providing smooth force estimates even in sparsely sampled regions, (ii) enabling force predictions in unexplored areas to avoid edge effects, and (iii) accelerating convergence by offering more accurate mean force estimates.
Building on this idea, Sevgen \textit{et al.} introduced the \method{combined force frequency (CFF)} method~\cite{sevgen2020combined}, which combines force-based and frequency-based estimators to improve free energy reconstruction (Fig.~\ref{fig:CFF}). 
CFF employs a self-integrating neural network to directly learn the free energy landscape from its derivatives, improving both robustness and accuracy over traditional approaches.
More recently, Rico \textit{et al.}~\cite{rico2025efficient} advanced this framework by incorporating \textit{Sinusoidal Representation Networks}~\cite{sitzmann2020proceedings} into the CFF methodology. 

A final example is of enhanced sampling methods boosted with ML is GaMD, which enhances sampling by applying harmonic boost potentials designed to yield a near-Gaussian energy distribution. 
However, GaMD's performance can be limited by the need for frequent updates and fine-tuning of the potential. 
To address this, Do and Miao proposed \method{deep boosted molecular dynamics (DBMD)}~\cite{do2023deep}, which leverages probabilistic Bayesian deep learning models to construct optimized boost potentials. 
DBMD first collects energy statistics from short unbiased MD runs to collect energy statistics, followed by the construction of a Gaussian-shaped boost potential that minimizes anharmonicity.

\subsection{Transition path-guided bias}
\label{sec:bias_transition_paths}

One of the limitations of enhanced sampling methods based on external bias potentials is that they typically alter the distribution of transition paths. 
As a result, approaches such as TPS, which do not perturb the system's Hamiltonian, are often employed for investigating transition mechanisms. 
However, TPS is computationally demanding due to the rarity of spontaneous transitions.

Recently, a new class of methods has been proposed that aims to preserve the statistical properties of the unbiased transition path ensemble while introducing a bias to enhance rare event sampling. 
In addition, these techniques do not rely on predefined CVs, as, instead, they introduce bias potentials that depend on both atomic positions $\mathbf{R}$ and velocities $\textbf{v}$, modifying the dynamics to generate trajectories drawn from a biased distribution. 
The central objective is then to learn a bias potential such that the resulting transition path distribution closely approximates the unbiased one.
To achieve this, several strategies have been developed using tools from reinforcement learning, stochastic optimal control, and variational inference.

In the context of \method{reinforcement learning}, the problem of sampling transition pathways is re-framed as a control task, where a neural network bias potential is trained to make rare transitions frequent by applying an optimized additional force that reshapes the dynamics while preserving correct transition statistics. 
Das \textit{et al.}\cite{das2021reinforcement} and Hua \textit{et al.}\cite{hua2023accelerated} both introduced methods in which the bias is optimized by minimizing the KL divergence between the biased and unbiased transition path distributions. 
The bias is parameterized as a neural network and trained via reinforcement learning techniques, using low-variance gradient estimators or adaptive data-driven updates to enhance convergence and sampling efficiency, as shown for a few toy model systems. 

Holdijk \textit{et al.}~\cite{holdijk2024stochastic} introduced \method{path integral path sampling (PIPS)}, which formulates the TPS problem as a stochastic optimal control problem related to the Schrödinger bridge formulation. 
PIPS learns a control force $u_\theta$ that modifies system dynamics to efficiently generate low-energy transition paths between metastable states.
This method has been validated on systems ranging from alanine dipeptide to larger biomolecules like polyproline and chignolin.

Finally, we note related approaches based on generative modeling and variational formulations. Although these methods do not use explicit biasing forces, they share the goal of enhancing sampling transition paths via learned probabilistic models.
Ahn \textit{et al.}~\cite{seong2025tpsdps} used generative flow networks for transition pathways.
Raja \textit{et al.}~\cite{raja2025action} proposed a zero-shot TPS approach, interpreting candidate transition paths as trajectories sampled from stochastic dynamics governed by a score function learned by a pre-trained generative model. Under such dynamics, identifying high-quality transition paths becomes equivalent to minimizing the Onsager-Machlup \cite{machlup1953fluctuations} functional.
Du \textit{et al.}~\cite{du2024doobs} proposed a simulation-free variational method based on Doob’s Lagrangian that directly parametrizes path distributions under boundary constraints.

\section{Generative models assist sampling}
\label{sec:generative}
Generative models have rapidly emerged as powerful tools across a broad range of scientific domains. These models learn to produce samples from complex, high-dimensional distributions and can be used to generate novel data consistent with a given statistical or physical model. Perhaps the most widely recognized success in this area is AlphaFold\cite{jumper2021highly, abramson2024accurate}, which has revolutionized structural biology by predicting the three-dimensional structures of proteins from their amino acid sequences—an achievement acknowledged by the 2024 Nobel Prize in Chemistry.

In this section, we focus on the application of generative models to the sampling problem in molecular simulations. Rather than using ML as a universal interpolator or for property prediction, the goal here is to accelerate conventional sampling procedures—or bypass them entirely. Examples of the latter include the Variational Autoregressive Network\cite{wu2019solving} and the Boltzmann Generator\cite{noe2019boltzmann}, which aim to optimize models that can be used to generate configurations distributed according to the equilibrium Boltzmann distribution.
In addition, generative models have also been employed to improve the efficiency of established simulation techniques such as free energy perturbation methods and REMD. 
In the following, we limit our focus to these types of approaches, which are closer in spirit to the enhanced sampling approaches discussed in the other sections of this Review. In particular, we leave out methods that integrate generative models with Monte Carlo algorithms. For a broader overview of generative modeling in molecular sciences, we refer the reader to recent reviews~\cite{rotskoff2024generative, aranganathan2025modeling}.

This chapter is organized as follows. Sec.~\ref{sec:generative_models} provides a brief introduction to the generative models underpinning the methods discussed later. Sec.~\ref{sec:generative_BGs} reviews the Boltzmann Generator approach, Sec.~\ref{sec:generative_LFEP} explores applications of generative models to free energy perturbation, and Sec.~\ref{sec:generative_REX} covers their integration with REMD.

\subsection{Deep generative models}
\label{sec:generative_models}

\begin{figure*}[htbp]
\centering
\includegraphics[width=0.9\linewidth]{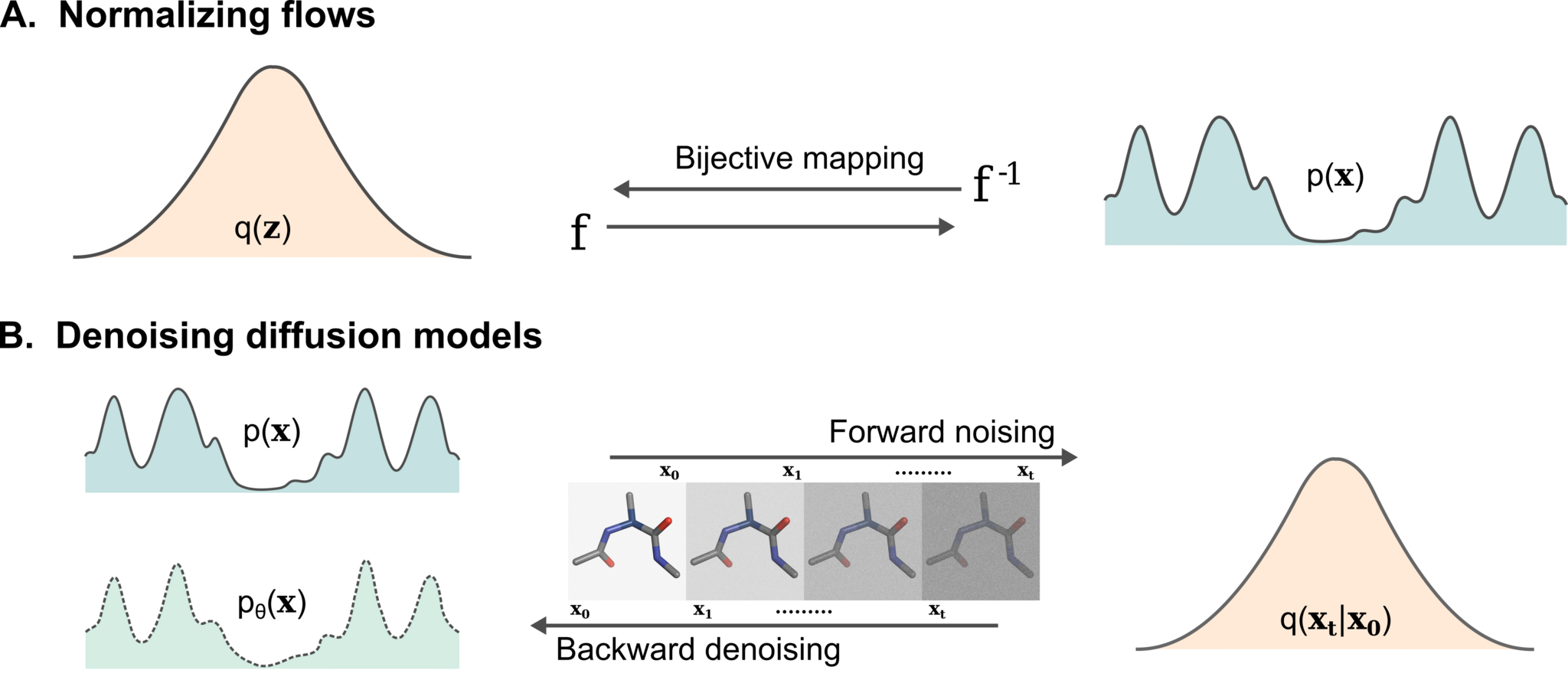}
\caption{\justifying
Two families of deep generative models. (A) Flow-based models learn a bijective mapping between a simple prior distribution and a complex data distribution, parameterized by a neural network. (B) Diffusion models learn a pair of complementary stochastic processes: a forward diffusion process that gradually transforms a data sample  $x_0 \sim p(x)$ into a noise sample $x_t \sim q(x_t \mid x_0)$ by adding Gaussian noise, and a learned reverse process that denoises $x_t$ to recover samples from a distribution $p_\theta(x)$ that approximates the original data distribution.
}
\label{fig:generative}
\end{figure*}

The general aim of generative models is to produce samples from complex target distributions by transforming samples drawn from simpler distributions.
In the following, we briefly introduce the two broad categories of such models that have shown the most relevant applications to the field of enhanced sampling, namely, normalizing flows and diffusion models, whose workings are schematically depicted in Fig.~\ref{fig:generative}.

\paragraphtitle{Normalizing flows} (NFs) are a class of deep generative models that enable exact and tractable density estimation while allowing efficient sampling. They achieve this by learning an invertible transformation mapping between arbitrary distributions, usually from a simple one (e.g., a Gaussian) into a complex target distribution of interest. This dual capability makes them especially attractive for applications in molecular simulations, where one seeks both to evaluate thermodynamic observables and generate physically meaningful configurations.

More formally, a flow-based model aim to generate samples $x$ from a target distribution \( p(\mathbf{x}) \) by transforming samples $z$ drawn from another (simpler or cheaper) distribution \( q(\mathbf{z}) \).~\cite{bishop2025deep}  
To achieve this, the flow defines a learnable invertible transformation \( f : \mathbf{z} \rightarrow \mathbf{x}\) from this space to the target one, and the corresponding inverse \( f^{-1} : \mathbf{x} \rightarrow \mathbf{z}\). The generated samples will be distributed according to the transformed distribution \( p_x(\mathbf{x})=f(q(\mathbf{z})) \), which is then optimized to match the target \( p(\mathbf{x}) \), for instance by minimizing the KL divergence.

The advantage of choosing an invertible transformation  is that we can write the relation between the two distributions as a change of variables: 
\begin{equation}
    p_x(\mathbf{x}) = q(\mathbf{z}) \left| \det(J_f(\mathbf{z})) \right|^{-1}
    \label{eq:NF_change_var}
\end{equation}
where \( J_f(\mathbf{z}) \) is the Jacobian matrix of \( f \), and \( \left| \det(J_f(\mathbf{z})) \right|^{-1} = \left| \det(J_{f^{-1}}(\mathbf{x})) \right| \).  
Hence, in order to be of practical usage, NF architectures need to be designed so that the determinant of the Jacobian is easy to compute.
A common design involves composing multiple \textit{invertible coupling layers}, where the input \( \mathbf{z} \) is split into two subsets \( \mathbf{z}_1 \) and \( \mathbf{z}_2 \). The first subset is left unchanged and used to condition the transformation of the second:
\begin{align}
    \mathbf{y}_1 &= \mathbf{z}_1 \\
    \mathbf{y}_2 &= h(\mathbf{z}_2, g_\theta(\mathbf{z}_1))
    \label{eq:NF_coupling_layer}
\end{align}
Here, \( h \) is an easily invertible \textit{coupling function}, and \( g \) is a generally non-invertible \textit{conditioning function} (typically a neural network) that depends on parameters \( \theta \). This structure leads to lower-triangular Jacobians, simplifying determinant calculations. Stacking multiple layers and alternating the roles of \( \mathbf{z}_1 \) and \( \mathbf{z}_2 \) enhances model expressivity.

Among the broad family of NFs, it is worth mentioning the \textit{conditional normalizing flows}, designed to model conditional target distributions. A conditional NF \( f(\mathbf{z} | \mathbf{c}) \) learns a transformation from the prior \( q(\mathbf{z}) \) to a conditional target \( p(\mathbf{x} | \mathbf{c}) \), where \( \mathbf{c} \) is a set of conditioning variables.~\cite{ardizzone2019conditional}  
In this case, the change-of-variables rule becomes:
\[
p_x(\mathbf{x} | \mathbf{c}) = q(\mathbf{z}) \left| \det(J_f(\mathbf{z}) | \mathbf{c}) \right|^{-1}
\]
analogous to Eq.~\ref{eq:NF_change_var} but explicitly dependent on \( \mathbf{c} \).

\paragraphtitle{Denoising diffusion models} (DDMs) are a class of stochastic generative models that construct complex distributions through a gradual, learnable denoising process. In contrast to the deterministic nature of normalizing flows, DDMs are inherently probabilistic, which grants them greater expressivity and flexibility at the expense of exact likelihood evaluation.~\cite{bishop2025deep}

The core idea behind DDMs is to define a pair of complementary stochastic processes: a forward process that gradually transforms data into noise, and a backward process that learns to reverse this transformation and recover samples from the original distribution.
The forward process, or \textit{noising} diffusion, starts from an input \( \mathbf{x}_0 \) and produces a sequence of increasingly noisy versions \( \mathbf{x}_1, \mathbf{x}_2, \dots, \mathbf{x}_T \) by adding noise in a controlled fashion. In the commonly used case of a Gaussian noise, this step takes the form:
\begin{equation}
    \mathbf{x}_t = \sqrt{1 - \beta_t} \, \mathbf{x}_{t-1} + \sqrt{\beta_t} \, \boldsymbol{\epsilon}_t
    \label{eq:diffusion_models_forward}
\end{equation}
where \( \boldsymbol{\epsilon}_t \sim \mathcal{N}(0, I) \) and \( \beta_t < 1 \) controls the noise variance at each timestep. This process transforms any structured input into pure Gaussian noise as \( t \to T \). This forward diffusion can equivalently be described using a transition kernel:
\begin{equation}
    q(\mathbf{x}_t | \mathbf{x}_{t-1}) = \mathcal{N}(\mathbf{x}_t; \sqrt{1 - \beta_t} \, \mathbf{x}_{t-1}, \beta_t I)
\end{equation}

The more complicated component is the \textit{denoising} or reverse process, which aims to reconstruct meaningful samples from noise. This is learned by parameterizing reverse transition kernels \( q'_\theta(\mathbf{x}_{t-1} | \mathbf{x}_t) \), typically using neural networks. A standard approach models the reverse step with another Gaussian distribution:
\begin{equation}
    q'_\theta(\mathbf{x}_{t-1} | \mathbf{x}_t) = \mathcal{N}(\mathbf{x}_{t-1}; \boldsymbol{\mu}_\theta(\mathbf{x}_t, t), \boldsymbol{\sigma}_\theta(\mathbf{x}_t, t))
\end{equation}
where both the mean and variance are predicted by a neural network.

To optimize the parameters, one can follow the maximum likelihood principle by training a reverse Markov chain that best explains the data. Since the exact likelihood is intractable, training typically maximizes the ELBO, whose KL terms can be computed efficiently under Gaussian assumptions for the transition kernels. Alternatively, score matching can be used, where the model learns the score function \( s(\mathbf{x}) = \nabla_{\mathbf{x}} \log p(\mathbf{x}) \) instead of directly modeling transition kernels. 

In summary, normalizing flows and diffusion models both transform simple base distributions into complex target ones, but they differ in key aspects. Flows are deterministic and enable exact likelihood evaluation with fast sampling, though their expressivity can be limited by the need for invertibility and tractable Jacobians. Diffusion models, being stochastic, are more flexible and typically perform better in high-dimensional settings, but they require iterative sampling and do not provide closed-form likelihoods. For a more in-depth comparison and analysis, see the review by John \textit{et al.}~\cite{john2025comparison}.

\subsection{Boltzmann generators}
\label{sec:generative_BGs}

\method{Boltzmann Generators (BGs)}, introduced by Noé\ \textit{et al.}~\cite{noe2019boltzmann}, represent one of the most well-known applications of generative models to (enhanced) sampling. In essence, they are designed to directly sample the equilibrium Boltzmann distribution, bypassing the need for long simulations like MD or Monte Carlo.

\begin{figure}[htbp]
\centering
  \includegraphics[
  width=\if\ispreprint1
    1\linewidth
\else
    0.7\linewidth
\fi]{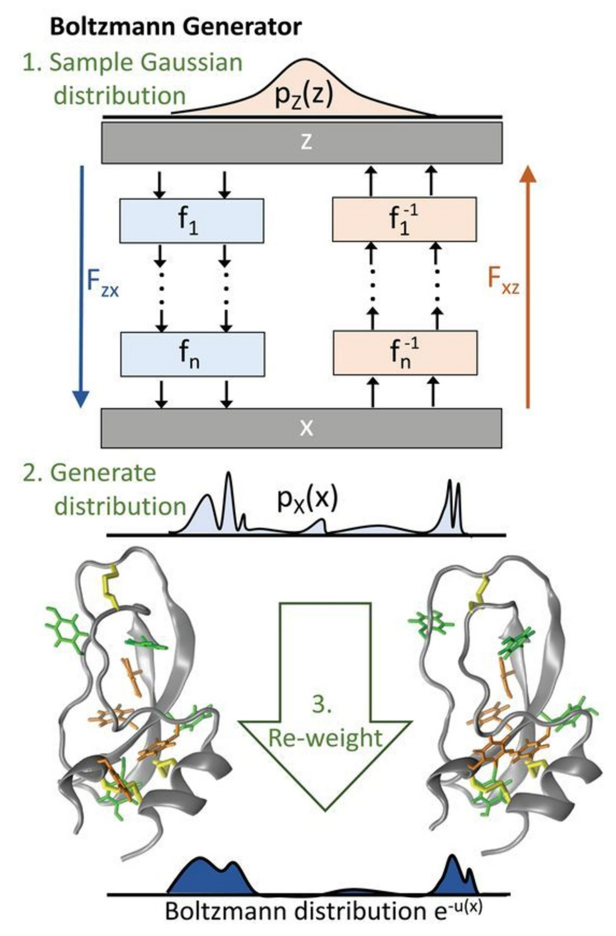}
\caption{\justifying
Boltzmann generators are optimized to minimize the discrepancy between their generated distribution and the target Boltzmann distribution. Sampling proceeds by drawing latent variables \(\textbf{z}\) from a simple prior (e.g., a Gaussian) and transforming them into molecular configurations \(\textbf{x}\). This transformation is implemented as a deep neural network \(F_{zx}\), constructed by stacking invertible layers \(f_1, \dots, f_n\), with an inverse mapping \(F_{xz}\) for efficient bidirectional sampling. To compute thermodynamic quantities, the generated samples are then reweighted to obtain the Boltzmann distribution.
Image reproduced from Ref.~\citenum{noe2019boltzmann}. Copyright 2019 American Association for the Advancement of Science.
}
\label{fig:BG}
\end{figure}

As described in Fig.~\ref{fig:BG}, the key idea is to learn an invertible transformation between a simple latent space $\textbf{z}$ with an easy-to-sample prior distribution $q(\mathbf{z}) =p_z(\mathbf{z}) $ (e.g., a standard Gaussian) and the configuration space \( \mathbf{x} \) of the physical system, distributed according to the hard-to-sample Boltzmann distribution:
\[
p(\mathbf{x})=\frac{1}{Z}e^{-u(\mathbf{x})}
\]
where $u$ is the reduced energy (divided by $k_B T$) and $Z$ the partition function. This transformation is implemented as a normalizing flow, consisting of a forward map  \( f=F_{zx} \) and its inverse \( f^{-1}=F_{xz} \). The map is optimized such that the distribution of the generated samples $p_x(\mathbf{x})$ approximates the true one $p(\textbf{x})$. Once trained, the transformation can be used to generate equilibrium samples by drawing latent variables \( \mathbf{z} \sim q(\mathbf{z}) \) and mapping them to physical configurations via \( \mathbf{x} = F_{zx}(\mathbf{z}) \).  In particular, expectation values of physical observables \(O(\mathbf{x}) \) can be computed as a weighted average over generated samples:
\[
 \langle O(x) \rangle
= \frac{\sum_i w(\mathbf{x}_i) O(\mathbf{x}_i)}{\sum_i w(\mathbf{x}_i)}
\]
where the weights account for the discrepancy between the generated distribution \( p_x(\mathbf{x}) \) and the true Boltzmann one: $w(\mathbf{x}) \propto \frac{e^{-u(\mathbf{x})}}{p_x(\mathbf{x})}$.

The standard training procedure combines two main learning objectives, corresponding to the directions of the invertible transformation. The primary component is \textit{training by energy}, which encourages the generation of Boltzmann-distributed samples in the transformed space. Unlike conventional ML models trained on fixed datasets, here the parameters are optimized using configurations generated by the model itself. Specifically, latent variables \( \mathbf{z} \sim p_z \) are sampled from the prior and mapped to configurations via \( \mathbf{x} = F_{zx}(\mathbf{z}) \). The generative map is then optimized by minimizing the KL divergence between the generated distribution and the target Boltzmann one:
\begin{equation}
  \mathcal{L}_{\text{KL}} = \left\langle u(F_{zx}(\mathbf{z})) - \log \det J_{zx}(\mathbf{z}) \right\rangle
  \label{eq:boltzmann_generator_loss_KL}
\end{equation}
where \( J_{zx} \) is the Jacobian of the generative transformation.

While effective, this energy-based training alone can lead to \textit{mode collapse}, in which the model learns only the most probable thermodynamic state, failing to capture the full diversity of the distribution. To avoid this, a complementary objective is introduced: \textit{training by example}. In this approach, reference configurations \( \tilde{\mathbf{x}} \) (e.g., representative structures from different metastable states) are encoded into the latent space via \( \tilde{\mathbf{z}} = F_{xz}(\tilde{\mathbf{x}}) \), and their likelihood under the prior is maximized:
\begin{equation}
  \mathcal{L}_{\text{ML}} = \left\langle \frac{1}{2} \| F_{xz}(\mathbf{x}) \|^2 - \log \det J_{xz}(\mathbf{x}) \right\rangle
  \label{eq:boltzmann_generator_loss_ML}
\end{equation}
where \( J_{xz} \) is the Jacobian of the encoding transformation.

It is important to note that the two training modes described above do not rely on the identification of reaction coordinates or CVs. However, if such coordinates are known, they can be incorporated into the training via auxiliary loss functions that encourage exploration outside of the metastable basins, for instance, by explicitly targeting transition-state configurations. This enhances the generation of low-probability states and enables the computation of continuous free energy profiles and realistic transition pathways.

Despite their conceptual appeal, the application of BGs to complex systems remained so far limited by several challenges. A primary difficulty stems from the intrinsic complexity of the Boltzmann distribution itself, which makes learning an accurate generative map highly demanding, even for relatively simple systems. 
For example, modeling systems with explicit solvent is particularly problematic due to the dramatic increase in dimensionality. Likewise, long-range interactions, which are common in biological and charged systems, pose further difficulties for accurately capturing the distribution. Another critical limitation arises from the invertibility constraint imposed by the normalizing flow architecture, which restricts the model’s expressivity unless a large number of transformation layers are employed. This, in turn, increases the computational cost associated with training. 

To address these issues, several technical improvements have been proposed. These include stochastic normalizing flows~\cite{wu2020stochastic}, equivariant flows~\cite{kohler2020equivariant}, and smooth flows~\cite{kohler2021smooth}, all designed to enhance flexibility and scalability. Notably, the introduction of equivariant flow matching~\cite{klein2024equivariant} has improved sampling efficiency and enabled the first transferable BGs~\cite{klein2024transferable}.
Beyond architectural improvements, some efforts have aimed to extend the physical applicability of BGs to a wider range of thermodynamic transformations.
For example, temperature-steerable flows, introduced by Dibak \textit{et al.}\cite{dibak2022steerable}, generalize the BG framework to sample across a family of thermodynamic states parameterized by temperature. 
Moqvist \textit{et al.} introduced a thermodynamic interpolation method \cite{moqvist2025thermodynamic} to generate sampling statistics in a range of temperatures either by learning direct mapping between thermodynamic states in the configurational space, or by passing through a latent space. 
In a similar direction, Van Leeuwen \textit{et al.} proposed a prototypical BG for the isothermal-isobaric ensemble, which can be used to predict fluctuations of the particle positions but also of the box itself.\cite{vanleeuwen2023iso}
Finally, Schebek \textit{et al.}\cite{schebek2024efficient} presented a BG-based method that combines conditioning on temperature and pressure with elements of free energy perturbation (see Sec.~\ref{sec:generative_LFEP}) to compute phase diagrams across a continuous range of thermodynamic conditions.

\subsection{Learned free energy perturbation}
\label{sec:generative_LFEP}
Generative models have also been applied to extend the capabilities of free energy perturbation (FEP) methods. The classical FEP method, introduced by Zwanzig,~\cite{zwanzig1954high} aims to estimate the free energy difference \( \Delta f_{AB} \) between two thermodynamic states \( A \) (reference) and \( B \) (target), characterized by reduced potentials \( u_A(\mathbf{x}) \) and \( u_B(\mathbf{x}) \), using the identity:
\begin{equation}
    \left\langle e^{-\Delta u_{AB}} \right\rangle_{A} = e^{-\beta \Delta f_{AB}}
    \label{eq:FEP_identity}
\end{equation}
where \( \Delta u_{AB} = u_B(\mathbf{x}) - u_A(\mathbf{x}) \). Two key factors govern the accuracy of FEP: sufficient sampling of the reference distribution \( A \), and sufficient overlap between the probability distributions of states \( A \) and \( B \) in configuration space.~\cite{pohorille2010good,kaus2015deal,wang2012achieving,mobley2007confine}  
The former often requires enhanced sampling techniques, while the latter is typically addressed using a multi-stage mapping. That is, one defines a set of intermediate states, decomposing the transformation into smaller steps and bridging the gap between poorly overlapping endpoints.

An alternative approach, particularly suited for generative models, is targeted free energy perturbation (TFEP), proposed by Jarzynski.~\cite{jarzynski2002targeted} TFEP introduces an invertible transformation \( M \) that maps configurations from state \( A \) to a modified distribution \( A' \) with increased overlap with state \( B \). Being this transformation invertible, its effect on the free energy is captured through the map work:
\begin{equation}
    w[M](\mathbf{x}) = u_B(M(\mathbf{x})) - \log |\det J_M(\mathbf{x})| - u_A(\mathbf{x})
    \label{eq:TFEP_map_work}
\end{equation}
which leads to the modified identity:
\begin{equation}
    \beta \Delta f_{AB} = -\log \left\langle e^{-w[M](\mathbf{x})} \right\rangle_{A}
    \label{eq:TFEP_identity}
\end{equation}
This approach improves convergence by enhancing overlap, but hinges on the ability to design a suitable transformation \( M \), which is a nontrivial task.

\begin{figure}[htbp]
\centering
\includegraphics[width=0.9\linewidth]{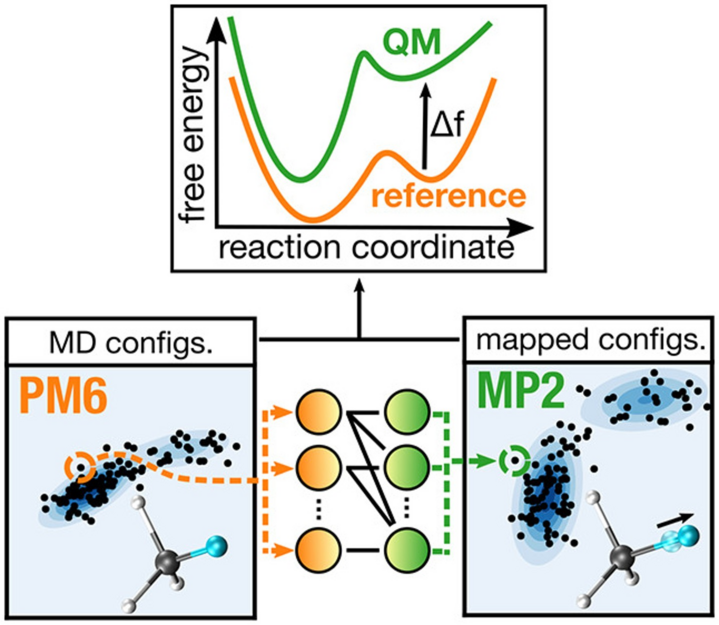}
\caption{\justifying
Schematic of the extended TFEP framework. The goal is to compute free energy differences and profiles at the quantum mechanical level starting from a cheaper reference potential. This is achieved by training a normalizing flow to map between the reference and target distributions, enhancing overlap and enabling efficient reweighting. Image reproduced from Ref.~\citenum{rizzi2021targeted}. Copyright 2021 American Chemical Society under \href{https://creativecommons.org/licenses/by/4.0/}{[CC BY 4.0 DEED]}.}
\label{fig:TFEP}
\end{figure}

To address this, Wirnsberger \textit{et al.}~\cite{wirnsberger2020targeted} proposed \method{learned free energy perturbation (LFEP)}, where the transformation \( M \) is represented by a normalizing flow, trained to minimize the expected map work:
\begin{equation}
    \mathcal{L}_{\text{LFEP}} = \langle w \rangle_{A}
    \label{eq:LFEP_loss}
\end{equation}
This avoids the need to know \( \Delta f_{AB} \), as it only contributes a constant to the KL divergence used for training. The model is designed to be permutation equivariant and consistent with periodic boundary conditions, making it applicable to atomistic systems.
Besides this unidirectional training, the authors also introduced a bidirectional scheme called \method{learned bennett acceptance ratio (LBAR)}. This method optimizes both the forward map \( M: A \rightarrow A' \) and its inverse \( M^{-1}: B \rightarrow B' \), leading to the combined loss:
\begin{equation}
    \mathcal{L}_{\text{LBAR}} = \langle w_M \rangle_{A} + \langle w_{M^{-1}} \rangle_{B}
    \label{eq:LBAR_loss}
\end{equation}
where \( w_{M^{-1}} \) is analogous to Eq.~\ref{eq:TFEP_map_work}, computed on samples from state \( B \).

LFEP was later applied by Rizzi \textit{et al.}~\cite{rizzi2021targeted} to \method{reference potential methods}, where FEP is used to reweight configurations generated with a cheaper Hamiltonian to a more accurate one (Fig.~\ref{fig:TFEP}). In this setting, bidirectional training is often infeasible due to the high cost of generating samples according to the target potential. 
They introduced several improvements to the unidirectional training, such as using an independent test dataset to evaluate \( \Delta f_{AB} \), to eliminate the bias that arises when evaluating on the same dataset used for training. Furthermore, they  extended the method to allow the computation of the free profile as a function of a general CV $f(\textbf{s})$, for which a sufficient condition is that the transformation $M:A \rightarrow A'$ satisfies the condition $\textbf{s}(M(\textbf{x})) = \textbf{s}(\textbf{x})$
which prevents the map from moving probability density along $\textbf{s}$, thus transforming only degrees of freedom orthogonal to $\textbf{s}$.
This makes it possible to employ CV-based enhanced sampling methods to gather the training points, extending the coverage of the reference phase space.
This work was further improved with a \method{multimap TFEP} formulation\cite{rizzi2023free}, which addresses two key inefficiencies: (i) the cost of the energy calculations at the expensive target potential, which are needed to compute the loss but are then discarded to avoid the systematic error, and (ii) the risk of overfitting, difficult to monitor due to the cost of the loss function. Their solution combined one-epoch training (so that each sample used is only once) with a multi-map ensemble approach, which computes the free energy difference from a collection of $N_m$ independent maps $\{ M^m \}_{m=1}^{N_m}$:
\begin{equation}
    \Delta f_{AB} = - \log \frac{1}{N_m} \sum_{m=1}^{N_m} \left\langle e^{-w[M^m](\mathbf{x})} \right\rangle_{A}
\end{equation}
The main advantage of this reformulation is to allow using the full dataset for both training and evaluating $\Delta f_{AB}$, rather than discarding the data generated during the training. The power of the approach was demonstrated by computing the free energy correction between a reference force field and a semiempirical potential across the HiPen dataset of drug-like molecules.

\subsection{Integrations with replica exchange}
\label{sec:generative_REX}

\begin{figure}[htbp]
\centering
\includegraphics[width=1\linewidth]{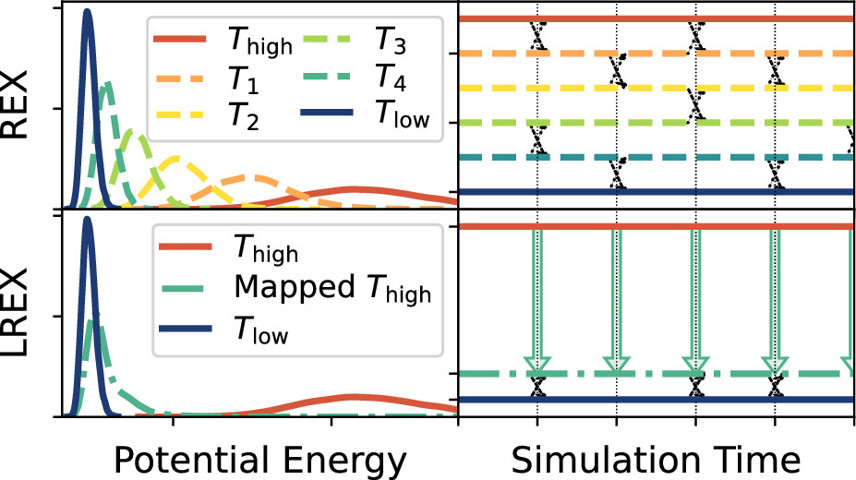}
\caption{\justifying
Scheme of the learned replica exchange (LREX). In LREX, a normalizing flow is trained to map the configurations of the prior replica to those of the target replica, allowing direct exchanges between the two without the need to simulate intermediate replicas. Image reproduced from Ref.~\citenum{invernizzi2022skipping}. Copyright 2022 American Chemical Society.} 
\label{fig:LREX}
\end{figure}

Replica exchange (REX), also known as parallel tempering, is a widely used enhanced sampling technique designed to improve sampling across complex energy landscapes by simulating multiple replicas of the system in parallel at different thermodynamic conditions.\cite{abrams2013enhanced,abel2017advancing}  
The replicas, typically arranged along a ladder of temperatures or other control parameters, are periodically allowed to exchange configurations, with an acceptance probability that depend on the difference in reduced energy between the two distributions \( \Delta u_{ij}(\mathbf{x}) = u_i(\mathbf{x}) - u_j(\mathbf{x}) \):
\begin{align}
\alpha_{\text{REX}} 
&= \min \left\{ 
    1, 
    \frac{p_j(\mathbf{x}_i)}{p_i(\mathbf{x}_i)} \cdot 
    \frac{p_i(\mathbf{x}_j)}{p_j(\mathbf{x}_j)} 
\right\} \nonumber \\
&= \min \left\{ 
    1, 
    e^{\Delta u_{ij}(\mathbf{x}_i) - \Delta u_{ij}(\mathbf{x}_j)} 
\right\}
\end{align}
The overall goal is to connect a hard-to-sample target distribution (such as a low-temperature Boltzmann distribution) with an easy-to-sample one (such as a high-temperature distribution) by enabling information flow across replicas.
A well-known limitation of REX is that energy is an extensive quantity, so a large number of intermediate replicas is often required to ensure sufficient overlap between neighboring distributions.  

To bypass this limitation, Invernizzi \textit{et al.} introduced the \method{learned replica exchange (LREX)} method,~\cite{invernizzi2022skipping} which uses a normalizing flow to learn a transformation between the prior and target distributions.  
This transformation is optimized to ensure sufficient overlap so that direct exchanges can be attempted between only two replicas, eliminating the need for a full ladder and drastically reducing computational cost. In practice, a short MD simulation is first run to sample configurations from the prior distribution \( q(\mathbf{x}) \).  These are used to train a normalizing flow \( f \) using an energy-based loss, similar to that used in BGs.  
Training convergence can be monitored using the Kish effective sample size~\cite{kish1965sampling}, which also provides an estimate of the expected exchange acceptance rate.

After training, the system is simulated at both prior and target conditions, and exchanges between the two are proposed with an acceptance probability:
\begin{equation}
    \alpha_{\text{LREX}} = \min \left\{ 1, \frac{p(\mathbf{x}_q')}{q'(\mathbf{x}_q')} \cdot \frac{q'(\mathbf{x}_p)}{p(\mathbf{x}_p)} \right\}
\end{equation}
where \( \mathbf{x}_p \) and \( \mathbf{x}_q \) are the current configurations of the target and prior replicas, respectively.  
Importantly, the learned transformation does not need to be exact—only sufficient to induce overlap—since the correct target statistics can be recovered by reweighting with the importance weights:
\begin{equation}
    w_f(\mathbf{x}) = e^{u_q(\mathbf{x}) - u_p(f(\mathbf{x}))} + \log |\det J_f(\mathbf{x})|
\end{equation}

A different point of view was adopted by Wang \textit{et al.} in combining REX with generative models, as they proposed to use them as a \method{postprocessing tool} to improve the sampling of the low-temperature replica.
\cite{wang2022data}
They noted that configurations sampled across replicas can be viewed as drawn from a joint distribution \( p(\mathbf{x}, \mathcal{T}) \), rather than from independent temperature-specific ensembles.  
Here, \( \mathcal{T} \) denotes the instantaneous kinetic temperature, whose ensemble average equals the heat bath temperature \( T \).
Based on this insight, they trained a denoising diffusion probabilistic model to learn \( p(\mathbf{x}, \mathcal{T}) \), using REX-generated data.  
The trained model was then used to generate new samples at low temperatures, improving sampling of rare configurations, and even to extrapolate to temperatures not included in the original REX ladder.  
This approach was successfully applied to small peptides and RNA strands, demonstrating how generative models can augment traditional replica exchange schemes.

\section{Conclusions}
Enhanced sampling methods have evolved over the past five decades into indispensable tools for exploring rare events and complex free energy landscapes in molecular simulations. In recent years, ML has transformed this field, enabling innovative solutions to challenges posed by the high dimensionality of molecular systems and the inherent sampling problem. In this review, we have surveyed the interplay between ML and enhanced sampling, highlighting both their synergies and their limitations.

Among the areas of integration, the most substantial and widespread advances have occurred in the construction of CVs. The challenge of identifying low-dimensional yet expressive representations of molecular systems aligns naturally with the strengths of ML. Unlike other tasks, such as learning the potential energy surface, constructing CVs does not require perfect coordinates: substantial (and often sufficient) acceleration could be achieved even with approximate variables. This has led to two major consequences.

On the one hand, it has enabled the development and application of a wide variety of strategies and learning objectives. These range from physics-based CVs, such as those informed by the committor function or dynamical operators, to pragmatic proxies based on structural information, such as preserving information content or distinguishing between metastable states. Importantly, the choice of learning objective is tightly coupled to the availability and quality of data. This interdependence gives rise to a fundamental “chicken-and-egg” paradox: identifying high-quality CVs requires access to relevant configurations, yet efficiently sampling those configurations depends on already knowing the right CVs. Many successful approaches have addressed this challenge through iterative workflows, alternating between data collection (e.g., via biased simulations) and CV refinement, often with progressively more sophisticated methods.

On the other hand, the absence of a single, well-defined objective has contributed to the proliferation of methodological variants, often distinguished by minor technical differences, that result in only incremental improvements without meaningfully advancing the field. Compounding this issue, many methods have been validated only on toy models or overly simplified systems, which fail to capture the complexity and challenges of realistic applications. To overcome these limitations, the community must embrace higher standards, including the establishment of rigorous benchmark systems and well-defined baselines, to enable systematic comparisons and ensure that new methods address problems of genuine practical relevance.

Beyond CV construction, ML has contributed to enhanced sampling at multiple levels, including representing bias potentials, optimizing free energy perturbation schemes, and guiding replica exchange protocols. More ambitious efforts to replace biasing schemes entirely with ML-driven algorithms, or even to replace conventional sampling with generative models, are emerging but they are still in their infancy. While promising, these approaches still face substantial hurdles before they can deliver general-purpose solutions, especially for large, realistic systems with many degrees of freedom (e.g., solvent molecules). 

In addition to surveying methodological developments, we aimed to provide a systematic perspective on their applications, particularly those enabled by advances in the construction of collective variables. These span diverse domains, from protein folding and ligand binding to phase transformations and catalytic reactions, each presenting its own unique challenges. Across these disparate areas, ML–enhanced sampling methods have been shown to be able not only to facilitate efficient exploration of complex landscapes but also to uncover mechanistic insights into the key degrees of freedom driving rare events. 

Yet, scaling these approaches to larger and more heterogeneous systems such as intrinsically disordered proteins, biomolecular assemblies, or realistic catalytic environments remains a formidable challenge. A key reason for this is that deploying these methods is not yet a fully automated process: substantial chemical intuition is often required to select initial conditions, define suitable representations, and identify processes of interest. Closing this gap and moving toward fully automated enhanced sampling will require advances on several fronts.

First, progress in \textit{representation learning} is essential. For large and complex systems, constructing suitable descriptors remains a major bottleneck, often demanding extensive domain expertise. Promising developments in geometric deep learning, such as equivariant graph neural networks, offer the ability to naturally encode all the system's degrees of freedom while preserving the required symmetries\cite{dietrich2023machine,zou2024enhanced,zhang2024descriptor,zou2025graph}. These approaches, however, are still computationally demanding and are currently more suited to \textit{ab initio }simulations or systems driven by ML potentials than to classical force-field-based studies. Transfer learning~\cite{sipka2023constructing} and self-supervised~\cite{pengmei2025using,turri2025self} paradigms offer complementary solutions: the former by enabling the reuse of pre-trained representations across related systems and tasks, and the latter by learning generalizable representations directly from data, thus reducing reliance on extensive simulations.

A particularly promising direction is the unification of CV and bias potential learning within a single, \textit{end-to-end framework}. Traditionally treated as separate steps, coupling the identification of low-dimensional representations with the adaptive construction of bias potentials could yield fully integrated workflows, automating both exploration and convergence.
In parallel, there is growing potential in combining traditionally distinct methodologies, such as TPS and CV-based enhanced sampling, to harness them as complementary sources of information and objectives~\cite{zhang2024combining,mouaffac2023optimal,falkner2024enhanced}. Integrating these paradigms could provide richer datasets and more accurate models of complex molecular processes.

As these methodologies grow in complexity and expressiveness, \textit{interpretability} becomes an equally pressing concern. Understanding what a model has learned and explaining its predictions are critical for extracting meaningful physical and chemical insights. Different approaches have been used, especially in the field of CV discovery, ranging from sensitivity analysis~\cite{bonati2020data} and symbolic regression~\cite{jung2023machine} to surrogate models ~\cite{ribeiro2018reweighted,zhang2024descriptor,chatterjee2025acceleration} and local explanation techniques~\cite{kikutsuji2022explaining,mehdi2024thermodynamics}. In general, this aspect will require a tighter integration with the field of explainable AI to ensure that these tools remain transparent, interpretable, and accessible to practitioners.

Achieving these advances will also require a closer integration of enhanced sampling and \textit{ML potentials}, which have already transformed chemical reaction modeling and materials science. The development of accurate ML potentials relies on datasets that span thermodynamically relevant configurations, a task where enhanced sampling plays a crucial role, particularly for modeling rare events~\cite{yang2024machine,yang2022using,perego2024data} but not limited to~\cite{van2023hyperactive, kulichenko2023uncertainty,zaverkin2024uncertainty}. Bringing these two domains closer together offers exciting opportunities for delivering highly accurate, ab initio-level simulations.

To realize this potential, the development of \textit{unified software ecosystems} will be essential. Such frameworks should seamlessly integrate all stages of the workflow: from representation learning and CV construction to biasing schemes, ML potentials, and post-processing analysis tools and interpretation. Providing modular and interoperable components would significantly lower the barrier to adoption and enable the widespread application of ML–enhanced sampling across diverse scientific domains.

Together, these advances will transform molecular dynamics into a true “computational microscope”, capable of providing atomistic insights into the structure, dynamics, and reactivity of complex physical, chemical, and biological systems over extended time and length-scales. 

\section*{Acknowledgments}
\vspace{-1em}

{\small

We are grateful to Timothee Devergne, Florian Dietrich, Ioannis Galdadas, Umberto Raucci,  Andrea Rizzi, Giorgia Rossi, Alice Triveri, Yihang Wang, Hui Zhang, for carefully reading this manuscript and providing feedback, and to Alessia Visigalli for providing a figure of the DNA translocation application. 

This study was supported by the National Key Research and Development Program of China (2024YFA1300051), the National Natural Science Foundation of China (22220102001), and the European Union - NextGenerationEU initiative and the Italian National Recovery and Resilience Plan (PNRR) from the Ministry of University and Research (MUR), under Project PE0000013 CUP J53C22003010006 ``Future Artificial Intelligence Research (FAIR)''.
}
\section*{Author Contributions}
\vspace{-1em}
{\small 
\textbf{Kai Zhu} and \textbf{Enrico Trizio} contributed to the conceptualization, investigation, and writing of the original draft.
 \textbf{Jintu Zhang} participated in the investigation and revision of the manuscript, particularly Chapter 2. \textbf{Renling Hu} contributed to the investigation and revision, with an emphasis on Chapter 6. \textbf{Linlong Jiang}, Kai Zhu, and Enrico Trizio prepared the original figures. \textbf{Tingjun Hou} provided a comprehensive review of the manuscript. \textbf{Luigi Bonati} contributed to the conceptualization, investigation, writing of the original draft, and supervision of the entire project.
}
\section*{Biographies}
\vspace{-1em}
{\small
\textbf{Kai Zhu} received his B.S. in Pharmaceutics from Wenzhou Medical University and is currently pursuing a Ph.D. in Pharmaceutics at Zhejiang University under Prof. Tingjun Hou and an incoming visiting student at the Italian Institute of Technology, Genova. His research focuses on machine learning-based enhanced sampling methods and their applications to atomic systems.

\textbf{Enrico Trizio} received his Ph.D. in Materials Science and Nanotechnology from the University of Milano-Bicocca in Milan, Italy, jointly with the Italian Institute of Technology in Genoa, Italy, under the supervision of Prof. Michele Parrinello.
His research focuses on combining machine learning techniques with molecular dynamics and enhanced sampling simulations.

\textbf{Jintu Zhang} received his Ph.D. degree in Pharmaceutics from Zhejiang University in 2025. His research focuses on machine learning-based methods for molecular dynamics simulations.

\textbf{Renling Hu} received her B.S. degree in Chemistry from Sichuan University. Since 2022, she has been pursuing a Ph.D. degree in Pharmaceutics at Zhejiang University under the advisement of Prof. Tingjun Hou. Her primary research interest is integrating machine learning, enhanced sampling, and quantum mechanics with free energy calculations.

\textbf{Linlong Jiang} received his B.S. degree in Pharmaceutics from Zhejiang University in 2023 and is currently pursuing an M.S. degree in Pharmaceutics at Zhejiang University under the supervision of Prof. Tingjun Hou. His research focuses on machine learning-based methods for protein–protein docking and conformational sampling.

\textbf{Tingjun Hou} received his Ph.D. in Computational Chemistry from Peking University in 2002. Currently, he is a full professor in the College of Pharmaceutical Sciences at Zhejiang University. His research focuses on molecular modeling and AI-driven drug discovery, including the development of structure-based virtual screening methodologies, theoretical predictions of pharmacokinetic properties and drug-likeness, discovery of small molecule inhibitors towards important drug targets, and multiscale simulations of target-ligand recognition. Prof. Hou has co-authored more than 600 publications in peer-reviewed journals with an h-index of 93.

\textbf{Luigi Bonati} received his Ph.D. in Physics from ETH Zurich, Switzerland, under the supervision of Prof. Michele Parrinello. He is currently a researcher at the Italian Institute of Technology in Genoa, Italy. His research focuses on the integration of machine learning with enhanced sampling simulations, with contributions ranging from the development of machine learning potentials for rare events to the data-driven discovery of collective variables and the design of novel algorithms. He applies these methodologies to study rare events in physics, biology, and catalysis.
}

\section{Bibliography}
\vspace{-1em}
\bibliographystyle{achemso}
\bibliography{main}
\end{document}